\documentclass[preprint]{aastex}
\usepackage{amsmath,amssymb,epsfig}
\usepackage{wrapfig}
\usepackage[english]{babel}
\usepackage{graphics}
\usepackage{subcaption}
\usepackage{threeparttable}
\usepackage{standalone}
\usepackage{booktabs, dcolumn}
\newcommand{\be}{\begin{equation}}
\newcommand{\ee}{\end{equation}}

\newcommand{\partialgamma}{{\partial\over \partial \gamma}}

\shorttitle{Radio-Gamma Correlation In Starburst Galaxies}
\shortauthors{Eichmann \& Tjus}

\begin{document}


\title{The Radio-Gamma Correlation In Starburst Galaxies}


\author{B. Eichmann\altaffilmark{1} and J. Becker Tjus\altaffilmark{1}}
\affil{Institut f\"ur Theoretische Physik, Lehrstuhl IV: Plasma-Astroteilchenphysik, Ruhr-Universit\"at Bochum, 44780 Bochum, Germany}


\altaffiltext{1}{Institut f\"ur Theoretische Physik, Lehrstuhl IV: Weltraum- und Astrophysik, Ruhr-Universit\"at Bochum, 44780 Bochum, Germany}


\begin{abstract}
We present a systematic study of the non-thermal electron-proton plasma and its emission processes in starburst galaxies in order to explain the correlation between the luminosity in the radio band and the recently observed gamma luminosity. 
In doing so, a steady state description of the cosmic ray electrons and protons within the spatially homogeneous starburst is considered where continuous momentum losses are included as well as catastrophic losses due to diffusion and advection. 
The primary source of the relativistic cosmic rays, e.g.\ supernova remnants, provides a quasi-neutral plasma with a power law spectrum in momentum where we account for rigidity dependent differences between the electron and proton spectrum.
We examine the resulting leptonic and hadronic radiation processes by synchrotron radiation, inverse Compton scattering, Bremsstrahlung and hadronic pion production.
\\
Finally, the observations of NGC 253, M82, NGC 4945 and NGC 1068 in the radio and gamma-ray band as well as the observed supernova rate are used to constrain a best-fit model.
In the case of NGC 253, M82, NGC 4945 our model is able to accurately describe the data, showing that: \\
(i) Supernovae are the dominant particle accelerators for NGC 253, M82 and NGC 4945, but not in the case of NGC 1068. \\
(ii) All considered starburst galaxies are poor proton calorimeters in which for NGC 253 the escape is predominantly driven by the galactic wind, whereas the diffusive escape dominates in NGC 4945 and M82 (at energies $> 1\,\text{TeV}$).\\
(iii) Secondary electrons from hadronic pion production are important to model the radio flux, but the associated neutrino flux is below the current observation limit. 
\end{abstract}



\section{Introduction}
The far infrared (FIR) emission is a common indicator of the star formation rate (SFR) in galaxies. 
Here, an almost constant ratio between the FIR and the radio emission has been established (\citealt{1985ApJ...298L...7H}) in galaxies with a high SFR, so-called starburst galaxies.
In the following, several photometric SFR estimators have been suggested like the radio continuum luminosity at 1.4 GHz (\citealt{2001ApJ...554..803Y}). 
To date, four starburst galaxies (M82, NGC 253, NGC 1068 and NGC 4945) have also been detected in the $\gamma-$ray by the Fermi LAT (\citealt{2010ApJ...709L.152A}, \citealt{2010A&A...524A..72L}) without showing indications of gamma-ray variability (\citealt{2012ApJ...755..164A}).
These galaxies show evidence that there is also a quasi-linear scaling relation between the gamma-ray luminosity and the radio continuum luminosity (\citealt{2012ApJ...755..164A}).
However, only M82 and NGC 253 (with some problems modeling the radio data of the latter) have been described by a self-consistent model (\citealt{0004-637X-768-1-53}, \citealt{2014ApJ...780..137Y}) so far. \\
The FIR emission is supposed to mainly represent star light absorbed and re-radiated by dust. 
The radio emission is well explained by synchrotron radiation of relativistic electrons, so that the FIR--radio correlation mainly depends on the average photon and magnetic field energy density $U_{rad}$ and $U_B$, respectively, according to $1+U_{rad}/U_B$ (\citealt{1989A&A...218...67V}).
In contrast to the FIR and radio emission, is the origin of the $\gamma-$ray emission still under debate. 
On the one hand, leptonic scenarios like inverse Compton (IC) or non-thermal Bremsstrahlung are anticipated as vital processes. 
But on the other hand, a hadronic scenario like hadronic pion production is mostly favored as the dominant $\gamma-$ray emission process.
The basic requirement in order to obtain gamma-ray emission is an efficient particle acceleration mechanism within the starburst.
A common explanation for the high energetic cosmic rays (CRs) within the galaxy is the acceleration in supernova remnant shock fronts.
However, an other possible source of the accelerated particles in starburst galaxies with an active galactic nucleus (indicated by jet-like structures, as observed in NGC 1068 and NGC 4945) could correspond to the central engine.
Without knowing the details of the acceleration process, we still take the rigidity dependent differences between electrons and protons into account.
\\
Galactic winds with wind speeds of at least a few $100\,\text{km}\,\text{s}^{-1}$ (e.\ g.\ \citealt{2011MNRAS.414.3719W}, \citealt{2003MNRAS.343L..47S}, \citealt{1998ApJ...493..129S}) are often supposed to be another important constituent of starburst galaxies. 
Due to the advection of CRs out of the starburst, these winds can have an important influence on the emitted radiation as exposed in previous calculations (e.g.\ \citealt{0004-637X-768-1-53}, \citealt{2014ApJ...780..137Y}). 
However, we also take the diffusion of CRs into account, as the corresponding loss timescale at high energies is most probably shorter than the loss timescale due to the galactic wind.\\
In this paper we present a theoretical model that describes the emission of the spherical symmetric and spatially homogeneous part of a starburst galaxy. 
This is an appropriate assumption, since we have no a priori knowledge about the spatial and temporal distribution of the particle accelerators, the magnetic fields or the thermal particles in starburst galaxies.
Furthermore, possible small-scale inhomogeneities vanish on scales much larger than the gyroradius $r_L$ of the particles. 
We like to stress that our semi-analytical model can easily be adjusted to the given astrophysical properties and it marks a solid starting point for future extensions in the case of complexer physical situations. \\
In contrast to previous calculations (e.g.\ \citealt{0004-637X-768-1-53}, \citealt{2014ApJ...780..137Y}) we take the impact of continuous momentum losses as well as advection and diffusion on the relativistic particle spectrum into account.
In particular, the flattening of the continuous energy loss rate of hadronic pion production at proton energies $>200$ GeV (\citealt{2015ApJ...802..114K}) is included.
Compared to previous investigations (e.g.\ \citealt{0004-637X-768-1-53}, \citealt{2009ApJ...698.1054D}, \citealt{2012ApJ...755..106P} or \citealt{2011ApJ...734..107L}) we also keep the total source of the high energy CRs as general as possible, but account for possible spectral differences due to the different cooling behavior of relativistic electrons and protons.  
In doing so, we are able to constrain the efficiency of the accelerators within the starburst. \\
Hence, a reasonable physical model of the transport of CRs within the starburst is the basic assumption and all astrophysical properties are used afterward.
\\
First of all, Section \ref{sec:rel_particle_description} presents the solution of the transport equation of relativistic electrons and protons within the starburst. 
In Section \ref{sec:syn_rad} the relativistic electrons are used to determine the emergent synchrotron radiation and in Section \ref{sec:gamma_rad} the resulting $\gamma$-radiation from hadronic pion production, Inverse Compton scattering and Bremsstrahlung are determined.
Furthermore, the secondary neutrinos and electrons that result from the $\pi^\pm-$decay are calculated.
The observed supernova rates as well as the data of the gamma-ray and radio flux of four starburst galaxies that have recently been observed in the $\gamma$-ray band is used in Section \ref{sec:cons_to_obs}.
Here, we constrain the cooling and escape parameters as well as the initial source spectrum and draw consequences on the supernova explosions, the ratio of primary to secondary electrons, the calorimetric behavior and the expected neutrino flux.
\section{Steady state particle transport equation}
\label{sec:rel_particle_description}
Starburst galaxies show no hints of temporal variability (\citealt{2012ApJ...755..164A}), so that the relativistic electrons and protons within the starburst can be considered in their steady state.
Furthermore, we suppose them to be spatially homogeneous and isotropic due to the mathematical convenience and the lack of information. 
Thus, spatial diffusion effects can be simplified by the leaky-box approach, where only the retarding effect on the escape time of the electrons is considered. 
Additionally, the effects of galactic winds are considered due to the advection of particles and magnetic fields.
Furthermore, the relativistic CRs suffer continuous momentum losses (with a loss rate $|\dot{\gamma}|_{e,p}$). 
Hence, we differentiate between continuous and catastrophic loss processes, in contrast to previous approaches (e.g.\ \citealt{2009ApJ...698.1054D}, \citealt{2012ApJ...755..106P} or \citealt{2011ApJ...734..107L}) where all loss processes are described by a catastrophic loss timescale.
\\
Consequently, the differential CR density of relativistic electrons and protons $n_{e,p}$ can be described by
\begin{equation}
0=\partialgamma\left(|\dot{\gamma}|_{e,p}\,n_{e,p}\right)-{n_{e,p}\over \tau^{e,p}_{diff}(\gamma)}-{n_{e,p}\over \tau_{adv}}+q_{e,p}(\gamma)\,,
\label{teq0}
\end{equation}
where $q_{e,p}(\gamma)$ takes an energy dependent source rate of electrons and protons, respectively, into account. 
The CR acceleration is supposed to be decoupled from the considered radiation region of the starburst galaxy.
Since the energy distribution of the source rate is a major ingredient in matters of the resulting radiation spectrum, a straight power-law spectrum can be oversimplified.
Diffusive shock acceleration is the favored acceleration mechanism of supernova remnants which are usually assumed to be the main source of the relativistic particles.
The resulting power-law spectrum according to diffusive shock acceleration has a slope of around $-2$ (dependent on the compression ratio of the shocked gas).
However, the maximum energy of the resulting energy spectrum depends on the rigidity of the particle and furthermore, ultrarelativistic electrons are more susceptible to continuous momentum losses than protons.
Latter is supported by recent particle-in-cell simulations (\citealt{2015PhRvL.114h5003P}) showing that the energy gain of electrons is faster than the energy gain of protons and hence, the electrons faster reach an energy where continuous momentum losses become significant.
In addition, it is later shown that the dominant cooling mechanism of the ultrarelativistic protons is hadronic pion production, which hardly changes the spectral shape of the energy distribution.
\\ 
The total number of accelerated protons and electrons is supposed to be equal in order to provide a quasi-neutral electron-proton plasma.
\\
Besides the continuous energy losses in the transport eq.\ (\ref{teq0}) there are catastrophic losses according to diffusion and advection of particles considered.
In doing so, the diffusion timescale is approximated by (e.g.\ \citealt{2004pese.book.....P})
\be
\tau_{diff}(\gamma)\simeq {R^2\over 3\,D(\gamma)}\simeq {R^2\over c\,l_{e,p}}\,\gamma^{-\beta}\,,
\label{diff-timescale2}
\ee
where a diffusion coefficient $D(\gamma)=c\,l_{e,p}\,\gamma^\beta/3\propto r_L^\beta$ has been adopted with a corresponding mean free path $l_{e,p}\,\gamma^\beta$ of CR electrons and protons, respectively.
According to the mass dependent Larmor radius $r_L$ we obtain $l_p=l_e\,(m_p/m_e)^{\beta}$ (\citealt{schlickeiser02:book}) and $\beta=1/3$ for Kolmogorov diffusion, whereas $\beta=1$ in the case of Bohm diffusion.
The mean free path $l_e$ generally depends on the turbulence spectrum of the plasma waves that scatter the particles as well as the square of the ratio between background magnetic field strength and magnetic turbulence. 
As these pieces of information are observationally inaccessible, there is at least an order of magnitude of uncertainty on the diffusion coefficient of starburst galaxies. 
However, gamma-ray observations suggest a diffusion timescale in the order of $\sim 10\,\text{Myr}$ at GeV energies (\citealt{2009ApJ...698.1054D}, \citealt{2012ApJ...755..106P}), so that $l_e \sim 3\cdot 10^{15\pm 1}\,R_{200}^2\,\text{cm}$. 
Here as well as in the following a spherical symmetric starburst with a radius $R=200\,R_{200}\,\text{pc}$ is supposed.
Furthermore, we use a Kolmogorov diffusion spectrum as observed in our Galaxy (\citealt{1990acr..book.....B}), i.e. $\beta=1/3$.
Thus, a weaker energy dependence of the diffusion coefficient as in previous approaches by \citealt{2009ApJ...698.1054D}, \citealt{2012ApJ...755..106P} or \citealt{2011ApJ...734..107L} is supposed.
\\
The timescale of advection $\tau_{adv}$ is determined by the galactic wind speed $v_{adv}$ which is supposed to be a constant in the order of a few hundred km/s, so that $\tau_{adv}\simeq R/v_{adv}$.
\begin{figure}[h!]
\centering
    \includegraphics[width=0.6\textwidth]{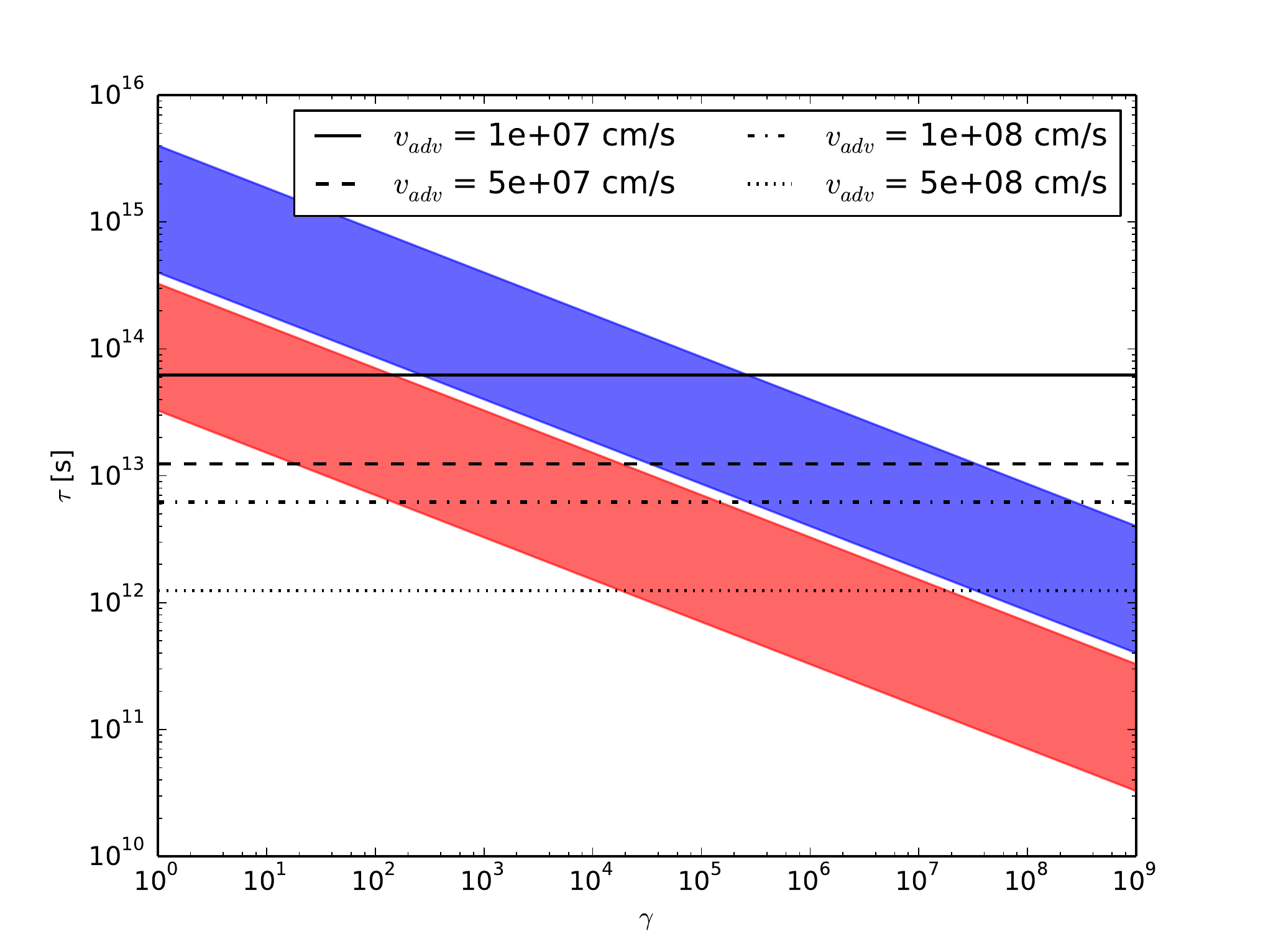}
  \caption{Range of the diffusive loss timescale $\tau_{diff}$ of electrons (blue) and protons (red) as well as the timescales of advection $\tau_{adv}$ (black) for four different galactic wind speeds.}
\label{Adv-Diff-timescale}
\end{figure} \\
Consequently, in the case of a galactic wind speed of
\be
v_{adv}\leq {3\,D(E)\over R} = 2.0\cdot 10^{6\pm 1}\,R_{200}^{-1}\,\left( E\over 1\,\text{GeV} \right)^{1/3}\,\text{cm}/\,\text{s}
\ee
the relativistic particles are predominantly affected by the diffusion process and not by the galactic wind (see fig.\ \ref{Adv-Diff-timescale}). 
In general, diffusion and advection losses need to be taken into account in order to determine accurately the whole energy spectrum of the relativistic CRs.
\\
Starburst galaxies are generally extremely bright at infrared wavelengths with $L_{IR}=3.85\cdot 10^{43}\,L_{43}\,\text{erg}\,\text{s}^{-1}$, where the dimensionless IR luminosity parameter is given by $L_{43}\sim 1-10$ (\citealt{1980ApJ...235..392T}) for most starburst galaxies.
Thus, the appropriate energy density yields $U_{IR}= 2.7\cdot 10^{-10}\,L_{43}\,R_{200}^{-2}\,\text{erg}\,\text{cm}^{-3}$. 
Furthermore, the high SFR in starburst galaxies also implicates a high density of target particles $N_t\gg 1\,\text{cm}^{-3}$ as well as a magnetic field strength $B$ in the order of a few $10\,\mu \text{G}$ up to a few $\text{mG}$.
\\
It is shown in the Appendix \ref{app:pte_sol}, that the general solution of the transport equation (\ref{teq0}) is given by
\be
n_{e,p}(\gamma)=\exp\left(  \int {\text{d}\gamma \over |\dot{\gamma}|_{e,p}\tau_{adv}}  \right)\text{e}^{-\zeta^{e,p}_0(\gamma)}\,\int_{\gamma}^{\gamma_{max}} d\gamma'\,\,{q_{e,p}(\gamma') \over  |\dot{\gamma'}|_{e,p}}\,\text{e}^{\zeta^{e,p}_0(\gamma')}\,\exp\left( - \int {\text{d}\gamma' \over |\dot{\gamma'}|_{e,p}\tau_{adv}}  \right)\,,
\label{gen_sol_part_dens_ii}
\ee
where
\be
\zeta^{e,p}_0(\gamma)=\int d\gamma\,\left(  {1\over |\dot{\gamma}|_{e,p}}\,{\partial |\dot{\gamma}|_{e,p}\over \partial \gamma}-{1\over \tau^{e,p}_{diff}(\gamma)\,|\dot{\gamma}|_{e,p}}\right) \,.
\label{zeta0}
\ee
In the following, the general solution (\ref{gen_sol_part_dens_ii}) is specified by the individual interactions of relativistic electrons and protons with their environment.

\subsection{Solution of the electron transport equation}
\begin{figure}[h!]
\centering
    \includegraphics[width=0.6\textwidth]{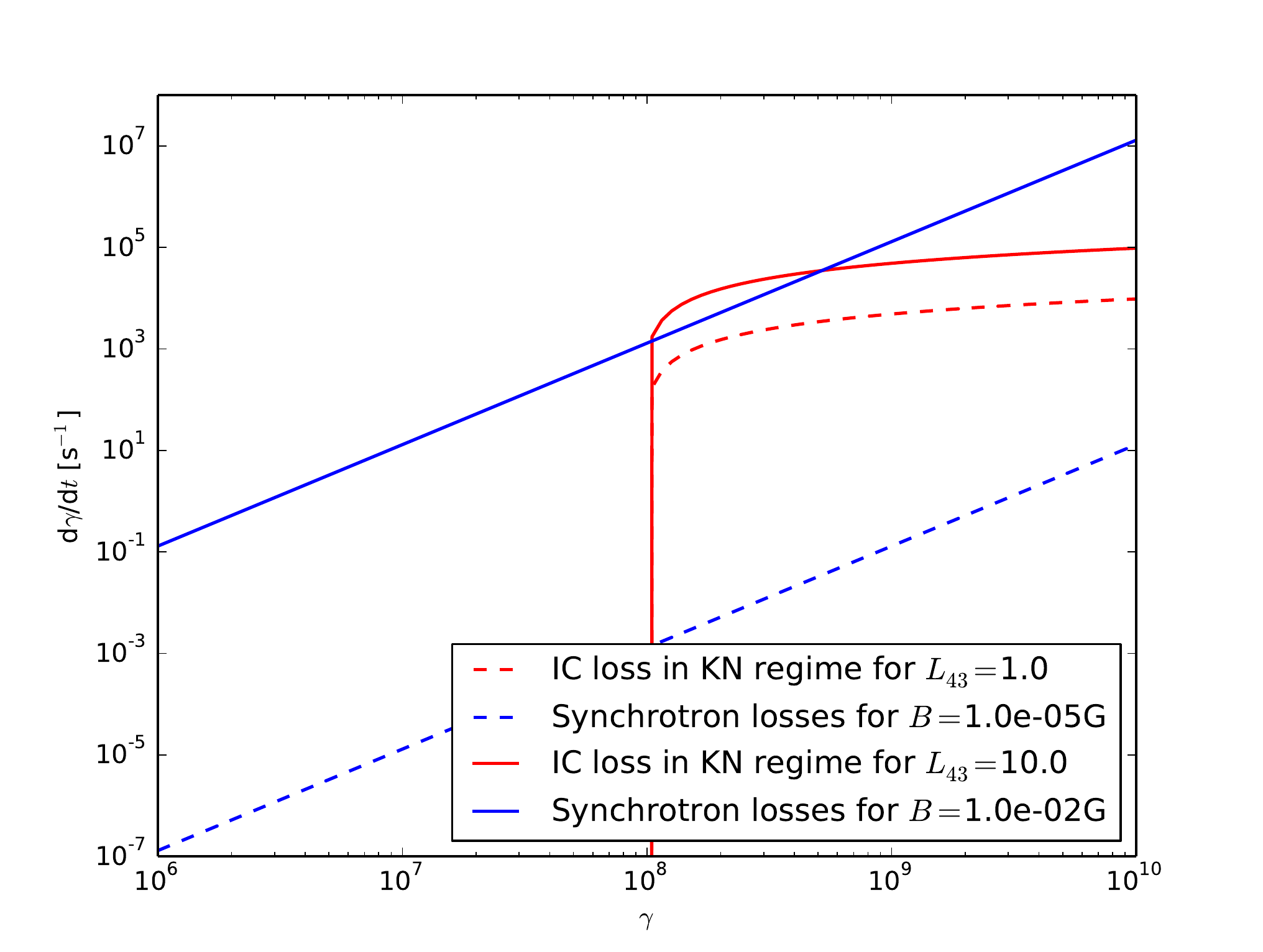}
  \caption{Loss rate of relativistic electrons according to synchrotron radiation as well as IC collisions in the KN regime (\citealt{schlickeiser02:book}).}
\label{IC_KN-Syn-lossRate}
\end{figure}

In the case of relativistic electrons, the continuous momentum losses are determined by synchrotron, inverse Compton (IC), Bremsstrahlung and ionization losses.
We consider the IC losses only in the Thomson limit, where the energy $\epsilon$ of the scattered infrared photon satisfies $\epsilon\,\gamma\ll m_ec^2$. 
In general, ultra-relativistic electrons with $\gamma > m_ec^2/\epsilon$ are in the Klein-Nishina (KN) regime, where the electrons lose a sizeable fraction of their energy by a Compton collision.
Although, only a small amount of the injected power-law distributed electrons are within the KN regime, we consider a maximal initial Lorentz factor of $\gamma_{max}^e=10^8$, caused by the IC losses in the KN regime (see fig. \ref{IC_KN-Syn-lossRate}).
Hence, the maximal Lorentz factor of electrons is not given by the CR accelerator but the onset of IC losses in the KN regime.
 \\
Thus, the total loss rate of relativistic electrons is determined by
\be
|\dot{\gamma}|_e\simeq(\Lambda_{syn}+\Lambda_{IC})\,\gamma^2\,+\Lambda_{Br}\,\gamma+\Lambda_{io,e}\,,
\label{tot_loss_e}
\ee 
where 
\be
\Lambda_{syn} \simeq 1.3\cdot 10^{-9}\left( B\over \text{Gauss} \right)^2\,\text{s}^{-1}\,,
\label{syn_loss_rate}
\ee
\be
\Lambda_{IC} \simeq 8.6\cdot 10^{-18}\,L_{43}\,R_{200}^{-2}\,\text{s}^{-1}\,,
\label{ic_loss_rate}
\ee
\be
\Lambda_{Br} \simeq 10^{-15}\left( N_t\over \text{cm}^{-3} \right)\,\text{s}^{-1}\,,
\label{brems_loss_rate}
\ee
and
\be
\Lambda_{io,e} \simeq 7.2\cdot 10^{-13}\left( N_t\over \text{cm}^{-3} \right)\,\text{s}^{-1}\,.
\label{io_loss_rate_e}
\ee
Here, $N_t$ denotes the particle density of the target plasma and $B$ refers to the magnetic field strength.
\\
In the following, the diffusion timescale (\ref{diff-timescale2}) and the total relativistic energy loss term (\ref{tot_loss_e}) is used, so that the $\zeta^e_0$-term (\ref{zeta0}) yields
\be
\zeta^e_0(\gamma) = \ln\left( {(\Lambda_{syn}+\Lambda_{IC})\,\gamma^2+\Lambda_{Br}\,\gamma+\Lambda_{io} \over \Lambda_{syn}+\Lambda_{IC}}  \right)-I^e_{diff}(\gamma)\,,
\label{zeta0_e_def}
\ee
with 
\be
\begin{split}
I^e_{diff}(\gamma) & ={c\,l_e \over R^2\,(\Lambda_{syn}+\Lambda_{IC})}\,\int \text{d}\gamma\,\,{\gamma^{\beta} \over \gamma^2+\Lambda_1\,\gamma+\Lambda_2} \\
& = {2^{-\beta}\,c\,l_e \over R^2\,(\Lambda_{syn}+\Lambda_{IC})}\, \left( \beta\,\sqrt{\Lambda_{1}^2-4\,\Lambda_{2}} \right)^{-1}\\
&\quad\times\Biggl[ \left({ \gamma \over \Lambda_{1} + 2\,\gamma-\sqrt{\Lambda_{1}^2-4\,\Lambda_{2}} }\right)^{-\beta} \, {}_2F_1\left(-\beta,\,-\beta;\,1-\beta;\, { \Lambda_{1} -\sqrt{\Lambda_{1}^2-4\,\Lambda_{2}}  \over \Lambda_{1} + 2\,\gamma-\sqrt{\Lambda_{1}^2-4\,\Lambda_{2}} } \right) \\
&\qquad- \left({ \gamma \over \Lambda_{1} + 2\,\gamma+\sqrt{\Lambda_{1}^2-4\,\Lambda_{2}} }\right)^{-\beta} \,{}_2F_1\left(-\beta,\,-\beta;\,1-\beta;\, { \Lambda_{1} +\sqrt{\Lambda_{1}^2-4\,\Lambda_{2}}  \over \Lambda_{1} + 2\,\gamma+\sqrt{\Lambda_{1}^2-4\,\Lambda_{2}} } \right) \Biggr]\,,
\end{split}
\ee
where 
\begin{equation}
 \Lambda_1 = {\Lambda_{Br} \over (\Lambda_{syn}+\Lambda_{IC})}\quad \text{and}\quad \Lambda_2 = {\Lambda_{io} \over (\Lambda_{syn}+\Lambda_{IC})}\,,
\end{equation}
and ${}_2F_1(a,b;c;x)$ denotes the hypergeometric function.
\\
In order to determine the relativistic electron density (\ref{gen_sol_part_dens_ii}), the source term $q_e(\gamma)$ needs to be specified:
\\
First, there are primary electrons related to a source rate $q_{e_1}(\gamma)$ due to the particle accelerators that are supposed to be homogeneously distributed within the starburst. 
Thus, in steady state we obtain a power-law distributed spectrum in momentum space with a spectral index $\alpha$ between $p_{min}=m_e\,c\,\sqrt{\gamma_{min}^2-1}$ and $p_{max}=m_e\,c\,\sqrt{\gamma_{max}^2-1}$.
Based on the dense IR photon fields within the starburst galaxy, the primary electron source spectrum is already effected by IC losses when $| \dot{\gamma} |_{IC} \geq | \dot{\gamma} |_{acc}$.
We suppose a strong shock, so that the ratio of the upstream to downstream plasma velocity yields $u_u/u_d = 4$. 
Furthermore, the diffusion coefficient on both sides of the shock is assumed to equal the diffusion coefficient $D(\gamma)=c\,l_{e,p}\gamma^\beta/3$ within the starburst galaxy.
Hence, the timescale $\tau_{cycle}$ for one back-and forth encounter yields (\citealt{1990cup..book.....G})
\be
\tau_{cycle}={20 \over 3}\,{l_{e,p}\,\gamma^\beta \over u_u}\,.
\ee
With an average fractional energy gain per encounter of $\xi=u_u/c$ according to the first order Fermi acceleration, the continuous energy gain is determined by
\be
| \dot{\gamma} |_{acc} = {\xi\,\gamma \over \tau_{cycle}} = {3\over 20}\,{u_u^2 \over c}\,l_{e,p}^{-1}\,\gamma^{1-\beta}\,.
\label{acc-rate}
\ee
In the lab frame, the upstream velocity $u_u$ equals the shock velocity $u_s$, which is given by $u_s = 5\cdot10^8\,u_5\,\text{cm/s}$.
Thus, the primary electron source spectrum is dominated by IC losses for
\be
\gamma > \gamma_B \equiv \left[ 1.5\cdot 10^{23}\,\left( {l_e \over 1\,\text{cm}}\right)^{-1} u_5^2\,L_{43}^{-1}\,R_{200}^2 \right]^{{1\over 1+\beta}}\,.
\label{initial_spec-break}
\ee
Here, the energy spectrum with a spectral index of $\alpha$ steepens to $\alpha-1$ due to the IC losses (see fig.\ \ref{e-p-distr}).
Note, that the consideration of the continuous synchrotron losses instead of the IC losses lead to the same results for $B=81\,L_{43}^{1/2}\,R_{200}^{-1}\,\mu\text{G}$.
But it needs to be taken into account that an increasing magnetic field strength is also supposed to increase the acceleration rate (\ref{acc-rate}) since $l_e\propto (r_L/\gamma)^\beta \propto B^{-\beta}$.  
As the details of the acceleration mechanism are still unknown and several assumptions have been made in order to determine eq.\ (\ref{initial_spec-break}), we will subsequently treat $\gamma_B$ as a free parameter.
The resulting $\gamma$-ray flux contribution of the accelerator to the total gamma radiation can be neglected as the lifetime $T_A\sim 1000\, \text{yrs}$ of the accelerator (\citealt{1990cup..book.....G}) is much smaller than the duration of stay within the starburst galaxy (see fig.\ \ref{Adv-Diff-timescale}).
\\
Secondly, there is a contribution by secondary electrons with a rate $q_{e2}(\gamma)$ due to the hadronic pion production of the relativistic protons as shown in Section \ref{sec:secondaries}.
Hence, the relativistic electron density also depends on the relativistic proton density. 
\\
In total, the source rate of relativistic electrons yields
\be
q_e(\gamma)=q_{e_{1}}(\gamma)+q_{e_{2}}(\gamma)=q^e_0\,\gamma\,[\gamma^2-1]^{-{\alpha+1 \over 2}}(1+\gamma/\gamma_B)^{-1}\,H[\gamma-\gamma_{min}]\,H[\gamma_{max}^e-\gamma]\,+\,q_{e_2}(\gamma)\,,
\label{primary_electron_source_rate}
\ee
where we account for the steepening of the spectrum at $\gamma_B$. 
Note, that the break Lorentz factor $\gamma_B$ strongly depends on the acceleration mechanism. 
\\
Using the sum of the primary and secondary source rate, the loss term (\ref{tot_loss_e}), as well as the eq.\ (\ref{zeta0_e_def}) we obtain the relativistic electron density
\be
\begin{split}
n_e(\gamma)&=\left( (\Lambda_{syn}+\Lambda_{IC})\,\gamma^2+\Lambda_{Br}\,\gamma+\Lambda_{io} \right)^{-1}\,\exp(I^e_{diff}(\gamma)+F_{adv}^e(\gamma))\,\\
&\quad\times\,\int_{\gamma_{l}}^{\gamma_{max}^e} d\gamma'\,\,\left(q^e_0 \,\gamma'\,[\gamma'^2-1]^{-{\alpha+1 \over 2}}+q_{e_2}(\gamma')\right)\,\exp(-(I^e_{diff}(\gamma')+F_{adv}^e(\gamma')))\\
\end{split}
\label{electron_dens_distr}
\ee
where the function
\be
F^e_{adv}(\gamma)={2 \over \tau_{adv}\,\sqrt{4(\Lambda_{syn}+\Lambda_{IC})\Lambda_{io}-\Lambda_{Br}^2}}\,\arctan\left( {2(\Lambda_{syn}+\Lambda_{IC})\gamma+\Lambda_{Br} \over \sqrt{4(\Lambda_{syn}+\Lambda_{IC})\Lambda_{io}-\Lambda_{Br}^2}}  \right)
\ee
determines the effect of particle losses due to advection in the same manner as $I^e_{diff}(\gamma)$ expresses the particle losses caused by diffusion.
The lower integration limit is given by
\be
\gamma_{l}=\begin{cases} \gamma\,,\,&\text{ for }\gamma_{min}< \gamma < \gamma_{max}^e\,,\\
            \gamma_{min}\,,\,&\text{ for } \gamma\leq \gamma_{min}\,.
           \end{cases}
\ee
In the following, the minimal Lorentz factor is supposed to be $\gamma_{min}=1$, whereas the maximal Lorentz factor is derived from the onset of KN loss yielding $\gamma_{max}^e = 10^8$. 
\subsection{Solution of the proton transport equation}
In the case of relativistic protons, the hadronic pion production is the dominant loss mechanism as well as the dominant source of $\gamma-$rays as later shown. 
An alternative hadronic gamma-ray production scenario is the photohadronic pion production. 
Here, a minimal Lorentz factor of 
\be
\gamma^{\gamma p}_{min}\simeq {m_\pi c^2 \over 2 E_{IR}}\,\left( 1+{m_\pi \over 2 m_p}  \right)   =      7.2\cdot 10^7\,\left({E_{IR}\over \text{eV}}\right)^{-1}\,.
\ee
is needed to generate a pion by a head-on collision of a proton with an infrared photon target in the energy range $E_{IR}\sim [1\,\text{meV};\,\,1\,\text{eV}]$.\\
The appropriate loss rate of a proton with $\gamma\simeq \gamma^{\gamma p}_{min}$ can be approximated by (\citealt{1994A&A...286..983M})
\be 
|\dot{\gamma}|_{p\gamma}\simeq 19\,\left(  {U_{IR} \over \text{erg}\,\text{cm}^{-3}} \right)\,\left( {E_{IR} \over \text{eV}}  \right)^{-2}\,\text{s}^{-1}  = 8.2\cdot(10^{-10}-10^{-9})\,\left( {E_{IR} \over \text{eV}}  \right)^{-2}\,\text{s}^{-1}\,.
\ee
On the other hand, hadronic pion production can be approximated by (\citealt{2015ApJ...802..114K})
\be
|\dot{\gamma}|_{pp} \simeq 4.4\cdot 10^{-16}\left( N_t\over \text{cm}^{-3} \right)\,\gamma^{1.28}\,\left( \gamma + 187.6  \right)^{-0.2}\,\text{s}^{-1}\,,
\label{had_pion_prod_loss_rate}
\ee
so that (at $\gamma\simeq \gamma^{\gamma p}_{min}$) photohadronic interactions are negligible if $|\dot{\gamma}|_{pp} \gg |\dot{\gamma}|_{p\gamma}$, i.e.
\be
N_t\gg 6.1\cdot (10^{-3}-10^{-2})\,\left(E_{IR}\over \text{eV} \right)^{-0.72}\,\left( \left(E_{IR}\over \text{eV} \right)^{-1} + 2.6\cdot 10^{-6} \right)^{-0.2}\,\text{cm}^{-3}\,.
\label{phohadron-hadron-pion-limit}
\ee
In starburst galaxies the target densities usually satisfy the previous condition (\ref{phohadron-hadron-pion-limit}) and hence, the continuous momentum losses of relativistic protons can be considered to be dominated by hadronic pion production as well as ionization losses at lower energies.
In total, we obtain
\be
|\dot{\gamma}|_p\simeq\Lambda_{pp}\,\gamma^{1.28}\,\left( \gamma + 187.6  \right)^{-0.2}+\Lambda_{io,p}\,,
\label{p_loss}
\ee 
with 
\be
\Lambda_{pp} \simeq 4.4\cdot 10^{-16}\left( N_t\over \text{cm}^{-3} \right)\,\text{s}^{-1}\,,
\label{pion_prod_loss_rate}
\ee
and 
\be
\Lambda_{io,p} \simeq 1.9\cdot 10^{-16}\left( N_t\over \text{cm}^{-3} \right)\,\text{s}^{-1}\,.
\label{pion_prod_loss_rate}
\ee
Note, that at energies $\gamma\gg 1$ the ionization losses are negligible.
Subsequently, the diffusion timescale (\ref{diff-timescale2}) and the hadronic loss term (\ref{p_loss}) is used, so that the $\zeta^p_0$-term yields
\be
\zeta^p_0(\gamma) = \ln\left(\Lambda_{pp}\, \gamma^{1.28}\,\left( \gamma + 187.6  \right)^{-0.2} + {\Lambda_{io}}  \right)-I^p_{diff}(\gamma)\,,
\label{zeta0_p_def}
\ee
with 
\be
\begin{split}
I^p_{diff}(\gamma) & ={c\,l_p \over R^2\,\Lambda_{pp}}\,\int \text{d}\gamma\,\,{\gamma^{\beta} \over \gamma^{1.28}\left( \gamma + 187.6  \right)^{-0.2}+{\Lambda_{io} \over \Lambda_{pp}}} \\
&\simeq {c\,l_p \over R^2\,\Lambda_{pp}}\, \,{\left( \gamma + 187.6 \right)^{0.2}\,\gamma^{\beta-0.28} \over (\beta-0.28)\,(0.00533\,\gamma+1)^{0.2}}\,\,{}_2F_1\left(-0.2,\,\beta-0.28;\,\beta+0.72;\,-0.00533\,\gamma  \right) \,.
\end{split}
\ee
In the latter ionization losses are neglected and ${}_2F_1(a,b;c;x)$ denotes the hypergeometric function.
The source rate of relativistic protons is also determined by the accelerator, however, the continuous momentum losses of ultrarelativistic protons are too slow to have an effect on the accelerated energy spectrum.
Hence, the source rate of the relativistic protons of Lorentz factor $\gamma_{min}\leq \gamma \leq \gamma_{max}^p$ yields
\be
q_p(\gamma)=q^p_0\,\gamma\,[\gamma^2-1]^{-{\alpha+1 \over 2}}\,H[\gamma-\gamma_{min}]\,H[\gamma_{max}^p-\gamma]\,.
\label{primary_proton_source_rate}
\ee
When supernovae (SNe) represent the dominant source of relativistic protons, the initial proton source rate is related to the SN rate $\nu_{SN}$ according to (e.g.\ \citealt{2010ApJ...709L.152A})
\be
{4\over 3}\,\pi\,R^3\,\int_{\gamma_{rel}}^{\gamma_{max}}\text{d}\gamma\,\,\gamma\,q_p(\gamma) = \nu_{SN}\,{E_{SN}\over m_p\,c^2}\,\eta\,.
\label{protonToSNrate}
\ee
Here, $\gamma_{rel}=\left( {1\,\text{GeV} \over m_pc^2}  \right)+1$ denotes the minimal Lorentz factor of relativistic protons, $E_{SN}$ denotes the kinetic energy ($>1\,\text{GeV}$) released per SN and $\eta$ is the acceleration efficiency of the CR protons.
\\
Consequently, with the loss term (\ref{p_loss}) and the eq.\ (\ref{zeta0_p_def}) the differential proton density yields
\be
\begin{split}
n_p(\gamma)=&{q^p_0}\,\left( \Lambda_{pp}\, \gamma^{1.28}\,\left( \gamma + 187.6  \right)^{-0.2} + {\Lambda_{io}}  \right)^{-1} \,\exp(F^p_{adv}(\gamma)+I^p_{diff}(\gamma))\\
&\times\,\int_{\gamma_{l}}^{\gamma_{max}^p} d\gamma'\,\,\gamma'\,[\gamma'^2-1]^{-{\alpha+1 \over 2}}\,\,\exp(-(F^p_{adv}(\gamma')+I^p_{diff}(\gamma')))\,,
\end{split}
\label{proton_dens_distr}
\ee
with
\be
\begin{split}
F^p_{adv}(\gamma)&=\int d\gamma\,\, [\tau_{adv}\,(\Lambda_{pp}\, \gamma^{1.28}\,\left( \gamma + 187.6  \right)^{-0.2} + \Lambda_{io})]^{-1} \\
&\simeq -{1 \over \Lambda_{pp}\,\tau_{adv}} \,{3.571\,\left( \gamma + 187.6 \right)^{0.2} \over (0.00533\,\gamma+1)^{0.2}\,\gamma^{0.28}}\,\,{}_2F_1\left(-0.28,\,-0.2;\,0.72;\,-0.00533\,\gamma  \right) \,.
\end{split}
\ee
The lower integration limit is given by
\be
\gamma_{l}=\begin{cases} \gamma\,,\,&\text{ for }\gamma_{min}< \gamma < \gamma_{max}^p\,,\\
            \gamma_{min}\,,\,&\text{ for } \gamma\leq \gamma_{min}\,.
           \end{cases}
\ee
Here, the relativistic protons are supposed to have the same minimal Lorentz factor as the electrons, whereas the maximal Lorentz factor of the relativistic proton is constrained by the accelerator.
Considering supernova remnants with a lifetime $T_A \sim 1000\, \text{yrs}$ according to \citealt{1990cup..book.....G} we obtain $\gamma_{max}^p = 10^{10}\,(B/1\,\text{G})$.
\newpage
\subsection{Electron--to--Proton Ratio}
\begin{figure}[h!]
\centering
    \includegraphics[width=0.8\textwidth]{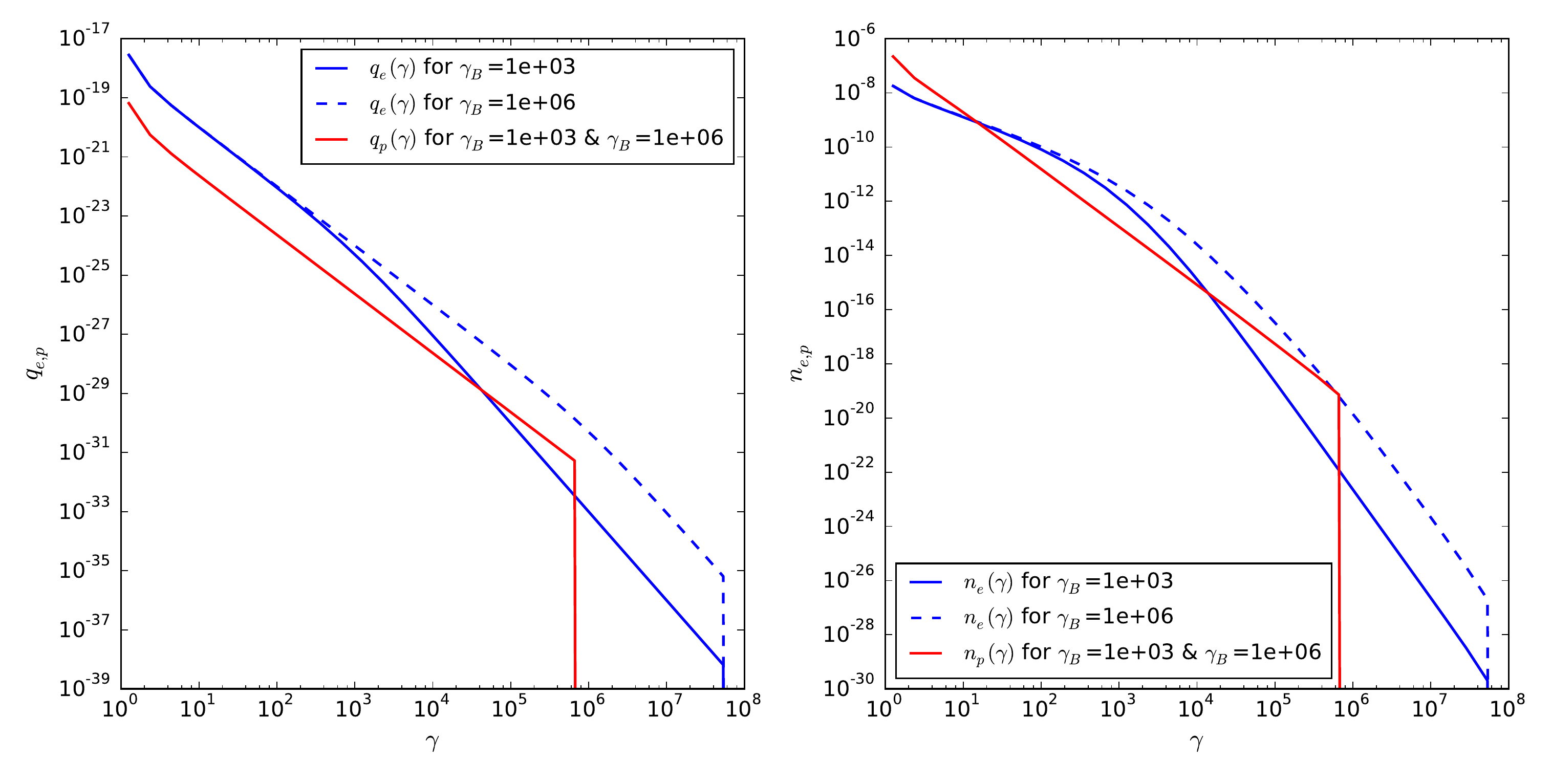}
  \caption{Quasi-neutral initial source rates of primary electrons and protons (left) as well as the resulting density distributions (right) for different $\gamma_B$. Here we used $v_{adv}=5\cdot 10^7\,$cm/s, $l_e=5\cdot 10^{15}\,$cm, $B=100\,\mu$G and $N_t=100\,\text{cm}^{-3}$.}
\label{e-p-distr}
\end{figure}

If there exists a steady state description of relativistic electrons and protons within the starburst galaxy where no large scale electric fields occur (due to charge imbalance), the total number of injected relativistic electrons and protons will need to be the same. 
Hence, we demand
\be
N_0=T_A\,\int_{\gamma_0^e}^\infty\,\text{d}\gamma\,\,q_{e_{1}}(\gamma)=T_A\,\int_{\gamma_0^p}^\infty\,\text{d}\gamma\,\,q_p(\gamma)
\label{same_total_particle_number}
\ee
where $\gamma_0^{e,p}$ denotes the minimal Lorentz factor of non-relativistic protons and electrons, respectively, that are accelerated to CR energies $>1\,\text{GeV}$. 
The initial kinetic energy of the particles needs to exceed $E_{kin}^{min}=4\,(1/2\,m_p\,u_s^2)$ in order to suffer effective particle acceleration (\citealt{1978MNRAS.182..147B}, \citealt{1978MNRAS.182..443B}). 
Hence, we obtain for a shock velocity $u_s\sim 700\,\text{km/s}$ a minimal kinetic energy of $10\,\text{keV}$ (e.g.\ \citealt{schlickeiser02:book}), i.e. $\gamma_0^{e,p}\simeq \left({10\,\text{keV} \over m_{e,p}c^2 } \right) + 1$. 
Since electrons and protons are supposed to be accelerated by the same cosmic accelerator, the lifetime $T_A$ of the accelerator also needs be the same.
\\
Thus, the relation (\ref{same_total_particle_number}) enables to fix the $q$-ratio between the electron and proton source rate according to
\be
\begin{split}
{q_0^e \over q_0^p}={ [(\gamma_{max}^p)^2-1]^{-{\alpha-1 \over 2}} - [(\gamma_{0}^p)^2-1]^{-{\alpha-1 \over 2}}  \over 1-\alpha  } \,\left[ \int_{\gamma_0^e}^{\gamma_{max}^e}\,\text{d}\gamma\, \gamma\,[\gamma^2-1]^{-{\alpha+1 \over 2}}(1+\gamma/\gamma_B)^{-1}  \right]^{-1}\,.
\label{gen_source-ratio}
\end{split}
\ee
This ratio represents a generalization of previous results (e.g.\ \citealt{schlickeiser02:book}, or \citealt{1993A&A...270...91P}) where in the case of an unbroken power-law distribution of electrons and protons a constant ratio of 
\be
{q_0^e \over q_0^p} \simeq \left(  {m_p \over m_e}  \right)^{(\alpha-1)/2}
\label{q-ratio_conv}
\ee
is obtained. Note that it is more common to determine the ratio of the differential number of particles depend on its energy (or momentum) which results in the inverse of eq.\ (\ref{q-ratio_conv}).
\begin{figure}[h!]
\centering
    \includegraphics[width=0.6\textwidth]{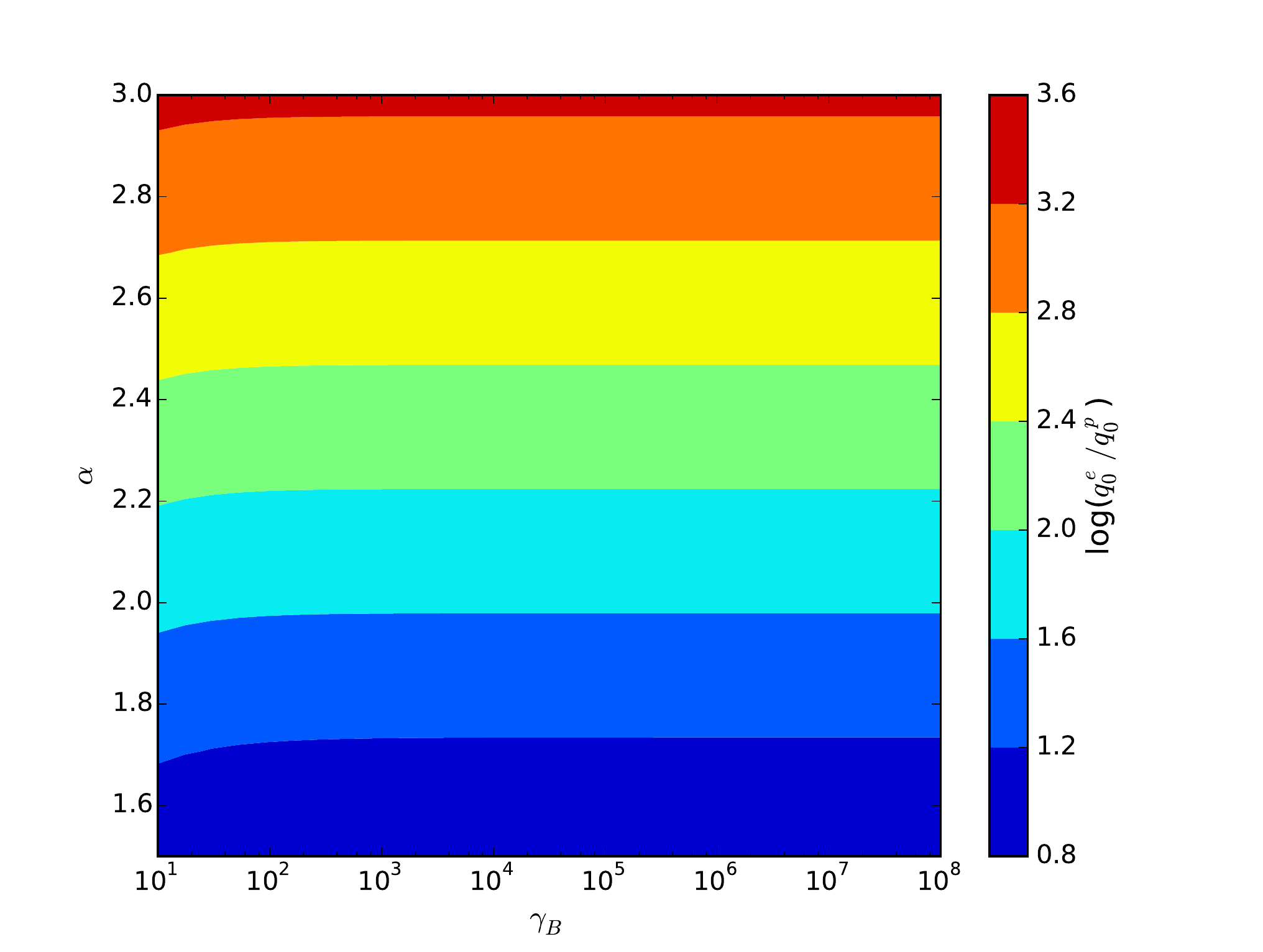}
  \caption{Source rate ratio $q_0^e/ q_0^p$ dependent on $\alpha$ and $\gamma_B$.}
\label{Q_ratio}
\end{figure} \\
The Fig.\ \ref{Q_ratio} shows that the spectral index $\alpha$ has a significant effect on source rate ratio as already exposed by the ordinary ratio (\ref{q-ratio_conv}). 
The steepening of the initial electron source rate at $\gamma_B$ only effects the source rate ratio in the case of $\gamma_B\ll 10^2$ since the total number of particles in the starburst region is dominated by the particles at the lowest energies. \\
Consequently, the source rate ratio (\ref{gen_source-ratio}) enables to determine the relativistic proton density dependent on the electron density for a given set of cooling and escape parameters (see fig.\ \ref{e-p-distr}).
 \\
\newpage
%
\section{Radio spectrum}
\label{sec:syn_rad}
The radio spectrum of starburst galaxies around a few GHz show a slope according to $\nu^{\rho}$ with $\rho \lesssim -0.5$. 
This is commonly explained by synchrotron radiation of relativistic electrons in the optically thin regime as here the relation between the electron and the radio spectrum is given by $\rho = - (\alpha -1)/2$. 
However, most radio spectra flatten at smaller frequencies due to free-free absorption by the ionized gas within the starburst.
In addition, the corresponding free-free emission can become dominant at the upper end of the radio spectrum which results in a flattening of the spectrum to $\rho\sim-0.1$ (\citealt{schlickeiser02:book}).
Hence, the properties of the ionized gas needs to be taken into account in order to describe the influence of free-free radiation. 
But, to keep the number of free parameters as small as possible, we confine our model to the steep part of the radio spectrum between $1\,\text{GHz}\lesssim \nu \lesssim 10\,\text{GHz}$ which needs to be dominated by synchrotron radiation.
The influence of synchrotron self absorption is negligible, as free-free absorption dominates the spectrum in the optically thick regime below $\sim 1\,\text{GHz}$.
The spectral synchrotron power is well described by
\be
P_s(\nu,\,\gamma)=P_0\,\left({\nu \over \nu_s\,\gamma^2}\right)^{1/3}\,\exp\left( - {\nu \over \nu_s\,\gamma^2} \right)\,, 
\ee
with $P_0=2.65\times10^{-10}(B/1\,\text{G})\,\text{eV}\,\text{s}^{-1}\,\text{Hz}^{-1}$ and a characteristic frequency $\nu_s=4.2\times10^6\,(B/1\,\text{G})\,\text{Hz}$.
Hence, the isotropic, spontaneous synchrotron emission coefficient of the relativistic electron distribution (\ref{electron_dens_distr}) is given by
\be
j_{syn}(\nu)={1\over 4\pi}\,\int_1^\infty d\gamma\,\,n_{e}(\gamma)\,P_{s}(\nu,\,\gamma) = {P_0\over 4\pi}\,\left( {\nu \over \nu_s} \right)^{1\over 3}  \int_{1}^\infty d\gamma\,\,\gamma^{-{2\over 3}}\,\exp\left( - {\nu \over \nu_s\,\gamma^2}  \right)\,n_{e}(\gamma)
\label{sy_em}
\ee
The emergent synchrotron intensity (in units of eV\,cm$^{-2}$\,ster$^{-1}$\,s$^{-1}$\,Hz$^{-1}$) in the optically thin case is determined by $I_{syn}(\nu,\,R) = j_{syn}(\nu)\,R$, so that the corresponding synchrotron flux (in units of eV\,cm$^{-2}$\,s$^{-1}$\,Hz$^{-1}$) at a distance $d$ yields
\be
\Phi_{syn}(\nu) = \pi\,{R^2\over d^2}\,I_{syn}(\nu,\,R) = {P_0\,R^3\over d^2}\,\left( {\nu \over \nu_s} \right)^{1\over 3}  \int_{1}^\infty d\gamma\,\,\gamma^{-{2\over 3}}\,\exp\left( - {\nu \over \nu_s\,\gamma^2}  \right)\,n_{e}(\gamma)\,.                                                 
\label{syn-flux}
\ee
\newpage
\section{Gamma radiation and secondary particles}
\label{sec:gamma_rad}
The most promising $\gamma$-radiation mechanisms in starburst galaxies are non-thermal Bremsstrahlung, IC radiation and $\pi^0-$ decay. 
However, the significance of each process depends pretty much on the magnetic field strength $B$ and the target density $N_t$. 
Furthermore, pionic $\gamma$-rays must be accompanied by secondary electrons as well as neutrinos. 
We suppose that the starburst galaxies are opaque for gamma radiation, so that $\gamma-\gamma$ interactions are negligible and the secondary particles are predominantly generated by hadronic pion production. 
Thereby, each process yields a certain source function $q_i(E_i)$ of $\gamma-$rays ($i=\gamma$) and secondaries ($i=\{e,\,\nu_e,\,\nu_\mu\}$), respectively, in units of $\text{cm}^{-3}\,\text{s}^{-1}\,\text{eV}^{-1}$.
Assuming the starburst galaxy to be opaque for $\gamma-$rays and neutrinos, 
the corresponding differential flux (in units of cm$^{-2}$\,s$^{-1}$\,eV$^{-1}$)  
at a distance $d$ is determined by
\be
\Phi_{i}(E_i)=\pi\,{R^2\over d^2}\,{R\,q_{i}(E_i) \over 4\pi}\,.
 \label{gammaFlux_general}
 \ee
In the following, the source functions of each process are discussed in detail.
\subsection{Gamma radiation from non-thermal Bremsstrahlung}
In addition to synchrotron radiation, the relativistic electrons are able to generate $\gamma-$rays by non-thermal Bremsstrahlung. 
According to \citealt{1971NASSP.249.....S}, the source function $q_{\gamma,Br}(E_\gamma)$ for $\gamma-$rays (in units of $\text{cm}^{-3}\,\text{s}^{-1}\,\text{eV}^{-1}$) is well approximated by
\be
q_{\gamma,Br}(E_\gamma)={c\,N_t\,\sigma_{Br} \over E_\gamma}\,\int_{E_\gamma/(m_ec^2)}^\infty \text{d}\gamma\,\, n_e(\gamma)\,,
\ee
where $\sigma_{Br}\simeq 3.38 \cdot 10^{-26}\,\text{cm}^{2}$. \\
%
%
%
%
%
%
%
%
%
%
%
%
%
%
%
%
%
%
%
%
%
%
%
%
%
%
%
%
%
%
%
%
%
%
%
%
%
%
\subsection{Gamma radiation from IC collisions}
Another leptonic process that generates gamma-radiation is IC collisions of relativistic electrons with a photon field.
In the case of starburst galaxies, the starlight that has been absorbed and re-radiated by dust in the infrared can be considered as the dominant photon field.
Hence, the differential photon density $n_{IR}(\epsilon)$ can be described by isotropic, diluted modified blackbody radiation according to the dust temperature $T_d$ and the spatial dilution factor $C_{dil}$, so that
\be
n_{IR}(\epsilon) = {C_{dil} \over \pi^2\,(\hbar c)^3}\,{\epsilon^2 \over \exp(\epsilon/(k_B\,T_d))-1}\,\left( {\epsilon \over \epsilon_0}  \right)\,,
\label{diff_IR-photon-dens}
\ee
where $T_d=45\,\text{K}$ and $\epsilon_0=8.2\,\text{meV}$ (\citealt{2008A&A...486..143P}). 
The referring energy density is given by $U_{IR}=\int\text{d} \epsilon\,\,\epsilon\,n_{IR}(\epsilon)=5.55\cdot 10^{-8}\,C_{dil}\,\text{erg}\,\text{cm}^{-3}$.
Thus, the dilution factor $C_{dil}$ is determined by conversion to the observed energy density $U_{IR}=2.7\cdot 10^{-10}\,L_{43}\,R_{200}^{-2}\,\text{erg}\,\text{cm}^{-3}$, which yields $C_{dil} = 4.8\cdot 10^{-3}\,L_{43}\,R_{200}^{-2}$. 
A more recent and accurate approach to describe the infrared photon spectrum has been proposed by \citealt{2012MNRAS.425.3094C}, where a coupled modified greybody plus a mid-infrared power law has been used. 
However, the effect on the IC process is small (and in addition section \ref{sec:cons_to_obs} shows that the contribution of $\gamma-$photons from IC scattering to the total $\gamma-$ray flux is negligible), so that we rely on the simplified approach given by eq.\ (\ref{diff_IR-photon-dens}). \\
The resulting source function $q_{\gamma,IC}(E_\gamma)$ for $\gamma-$rays is determined by (\citealt{1979rpa..book.....R})
\be
q_{\gamma,IC}(E_\gamma)={3\,c\,\sigma_T \over 4}\,\int_0^\infty \text{d}\epsilon {n_{IR}(\epsilon) \over \epsilon}\,\int_{\bar{\gamma}_{min}}^\infty \text{d}\gamma\,\, {n_e(\gamma)\over \gamma^2}\,F(q,\Gamma)\,,
\ee
where $\sigma_T$ denotes the Thomson cross-section.
The lower integration limit $\bar{\gamma}_{min}$ is given by (\citealt{schlickeiser02:book})
\be
\bar{\gamma}_{min} = {E_\gamma \over 2m_ec^2}\,\left[ 1 + \left(1+{m_e^2c^4 \over \epsilon\,E_\gamma}\right)^{1/2}   \right]\,,
\ee
as well as the Klein-Nishina related function
\be
F(q,\Gamma)= 2q\ln(q) + 1+q-2q^2 + {(\Gamma q)^2\,(1-q) \over 2\,(1+\Gamma q)}\,,
\ee
where
\be
\Gamma = {4\epsilon \gamma \over m_e\,c^2}\,\quad\text{and}\quad q={E_\gamma \over \Gamma(\gamma m_e c^2-E_\gamma)}\,.
\ee
\subsection{Gamma radiation from $\pi^0-\,$decay}
The dominant gamma ray flux component in most starburst galaxies is supposed to be a result of the inelastic interactions of relativistic protons with the ambient medium due to hadronic pion production.
Foremost, this interaction generates $\pi^0-\,$mesons which quasi-instantaneously decay into two gamma-rays.
The corresponding energy spectra $F_\gamma(x,\,E_p)$ of the resulting $\gamma-$rays with $x=E_\gamma/E_p$ are given by \citealt{2006PhRvD..74c4018K}.
Using the relativistic proton density (\ref{proton_dens_distr}) in the relativistic limit where $n_p(E_p)=n_p(\gamma=E_p/(m_pc^2))/(m_pc^2)$, the gamma ray production rate is determined by
\be
q_{\gamma,\pi^{0}}(E_\gamma)=c\,N_t\,\int_{E_\gamma}^\infty  {\text{d}E_p \over E_p}\,\,\sigma_{inel}(E_p)\,n_p(E_p)\,F_\gamma(x,\,E_p)\,.
\ee
Here, $\sigma_{inel}(E_p)$ denotes the inelastic part of the total cross-section of p-p interactions which is given by
\be
\sigma_{inel}(E_p)=\begin{cases} (34.3+1.88\,L+0.25\,L^2)\,\text{mb}\,,&\text{ for }E_p>0.1\,\text{TeV}\,,\\
                    (34.3+1.88\,L+0.25\,L^2)\cdot\left( 1-\left({E_{th}\over E_p}\right)^4 \right)^2\,\text{mb}\,,&\text{ for }E_{th}\leq E_p\leq0.1\,\text{TeV}\,,
                   \end{cases}
\label{tot_inel_cross-section}
\ee
with $L=\ln(E_p/1\,\text{TeV})$ and the threshold energy $E_{th}=m_pc^2+2m_\pi c^2+(m_\pi c^2)^2/(2m_p c^2)$ of the production of $\pi^0-\,$mesons.
\subsection{Neutrinos and secondary electrons from $\pi^\pm-\,$decay}
\label{sec:secondaries}
\begin{figure}[h!]
\centering
\begin{subfigure}[b]{0.48\textwidth}
     \includegraphics[width=1.05\textwidth]{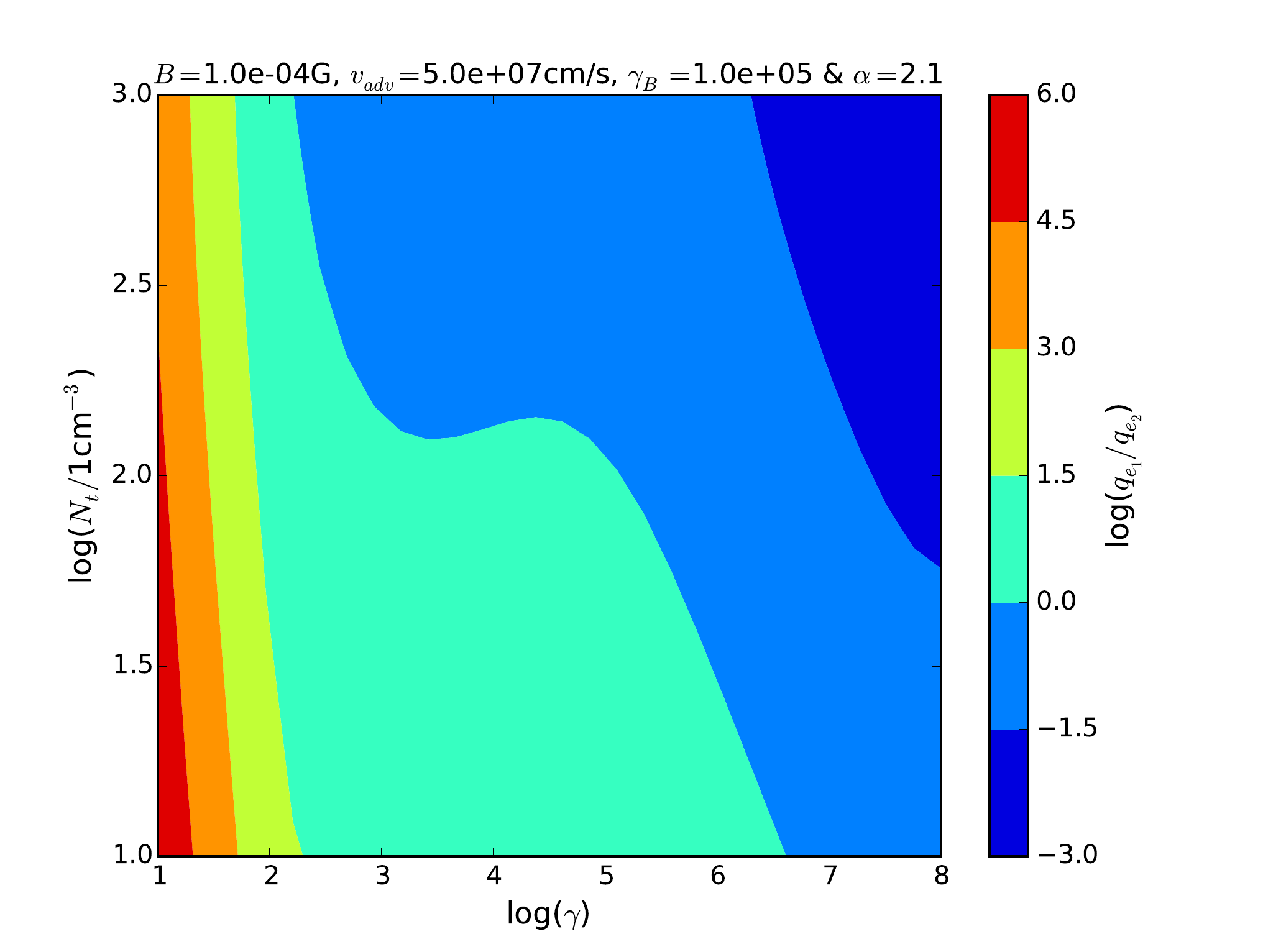}
 \label{PS-ratio_1}
 \end{subfigure}
 \begin{subfigure}[b]{0.48\textwidth}
     \includegraphics[width=1.05\textwidth]{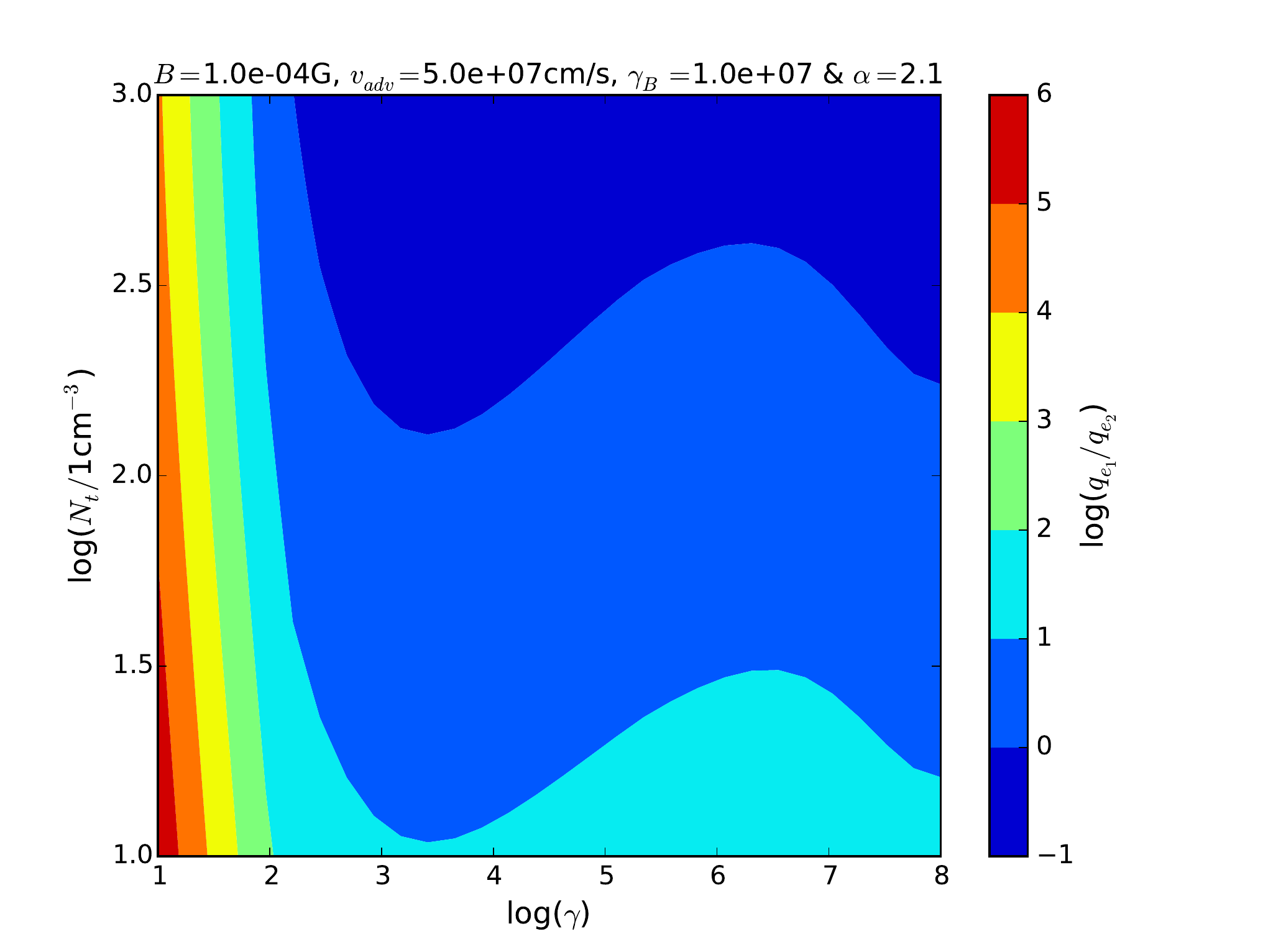}
 \label{PS-ratio_2}
 \end{subfigure}
    \caption{Ratio of primary to secondary electrons for $\gamma_B=10^{5}$ (left) and $\gamma_B=10^{7}$ (right).}
\label{Prim-Sec-Ratio}
\end{figure}

Additionally, hadronic pion production generates charged pions which quasi instantaneously decay according to $\pi^\pm \rightarrow \mu^\pm + \nu_\mu (\bar{\nu}_\mu)$ and the generated muon decays on similar timescales in a 3-body process according to $\mu^\pm \rightarrow e^\pm + \nu_e (\bar{\nu}_e) + \bar{\nu}_\mu (\nu_\mu)$.
\citealt{2006PhRvD..74c4018K} have also approximated the energy spectra of the resulting leptons.
The total muon neutrino spectrum is determined by the sum of the first muon neutrino (directly from the pion) with an energy spectrum $F_{\nu_{\mu}^{(1)}}(x,\, E_p)$ and the second muon neutrino (from the muon decay) corresponding to $F_{\nu_{\mu}^{(2)}}(x,\, E_p)$.
The energy spectrum of the decay products of the muon are almost the same, so that i.e.\ $F_{\nu_{\mu}^{(2)}}(x,\, E_p)\simeq F_{\nu_{e}}(x,\, E_p) \simeq F_{e}(x,\, E_p)$. 
\\
Hence, the source function of muon and electron neutrinos is given by
\be
q_{\nu_{\mu}}(E_\nu)=c\,N_t\,\int_{E_\nu}^\infty  {\text{d}E_p \over E_p}\,\,\sigma_{inel}(E_p)\,n_p(E_p)\,\left(F_{\nu_{\mu}^{(1)}}(x,\, E_p)+F_{e}(x,\, E_p)\right)\,.
\label{mu-neutrino-source-rate}
\ee
and
\be
q_{\nu_{e}}(E_\nu)=c\,N_t\,\int_{E_\nu}^\infty  {\text{d}E_p \over E_p}\,\,\sigma_{inel}(E_p)\,n_p(E_p)\,F_{e}(x,\, E_p)\,,
\label{e-neutrino-source-rate}
\ee
respectively, with the total inelastic cross-section $\sigma_{inel}(E_p)$ according to eq.\ (\ref{tot_inel_cross-section}) and the dimensionless neutrino energy $x=E_\nu/E_p$.
\\
In the relativistic limit, the (Lorentz factor dependent) source rate of secondary electrons that is needed in order to determine the relativistic electron density (\ref{electron_dens_distr}) is determined by
\be
q_{e_2}(\gamma)=q_{e_2}(E_e)\,m_e\,c^2=c\,N_t\,m_e\,c^2\,\int_{E_e}^\infty  {\text{d}E_p \over E_p}\,\,\sigma_{inel}(E_p)\,n_p(E_p)\,F_{e}(x,\, E_p)\,.
\ee
The fig.\ \ref{Prim-Sec-Ratio} shows that the significance of secondary electrons depends pretty much on the influence of IC losses on the primary electron spectrum according to $\gamma_B$. 
In general, secondary electrons are not relevant at energies below about 100 MeV, but at higher energies their relevance increases with decreasing $\gamma_B$. 
For $\gamma_B\sim 10^{7}$ the secondary electrons only become significant in the case of target densities above a few hundred particles per cubic centimeter, whereas for $\gamma_B\sim 10^{5}$ way lower target densities already result in a dominant contribution of secodary electrons at high energies.  
%
%
%
%
%
%
%
%
%
%
%
%
%
%
%
%
%
%
%
%
%
%
%
%
%
%
%
%
%
%
%
%
%
%
%
%
%
%
%
%
%
%
%
%
%
%
%
%
%
%
%
%
%
%
%
%
%
%
%
%
%
%
%
%
%
%
%
%
%
%
\newpage
\section{Consequences on the observations}
\label{sec:cons_to_obs}
The recently published Fermi LAT 4-Year Point Source Catalog (3FGL) measured the gamma-ray flux of four different galaxies, namely NGC 253, M82, NGC 4945 and NGC 1068.
In addition, NGC 253 and M82 have also been observed at TeV energies by the HESS (\citealt{2012ApJ...757..158A}) and VERITAS (\citealt{2009Natur.462..770V}) telescopes, respectively.
Here, the spectral shape does not change significantly, so that all four galaxies show a slope in the $\gamma$-ray band that is steeper than $E_\gamma^{-2}$. 
In addition, all four starburst galaxies have been observed in the radio band with a slope according to $\rho\lesssim -0.5$ between $1.4\,\text{GHz}\leq \nu \lesssim 10\,\text{GHz}$. 
NGC 253 and M82 show a flattening of the spectrum at $\nu\lesssim 1\,\text{GHz}$ which is supposed to refer to free-free absorption by a thermal gas.
However, we do not account for this spectral behavior as it has already been shown (\citealt{2014ApJ...780..137Y}; \citealt{0004-637X-768-1-53}) that the free-free absorption of an additional warm ionized gas component is able to model the spectrum below 1 GHz. 
In contrast to the gamma-ray observations do the radio observations enable to resolve the sub-kpc structure of the starburst galaxies.
Hence, there are radio observations of different spatial structures up to sub-arcsecond scales of the previously mentioned galaxies showing different flux behaviors.
Especially the Seyfert 2 galaxies NGC 4945 and NGC 1068 show jet like structures that could be the driving force of the non-thermal emission.
In general, the spatial boundary of the so-called starburst region is hard to define (in particular for NGC 253, NGC 4945 and NGC 1068), but as the gamma-ray flux on sub-arcsecond scales is unknown we need to consider the total radio flux of a few arcseconds.
Therefore, we suppose the radio and gamma-ray flux from the innermost region of 400 pc of diameter, i.e.\ $R_{200}=1$ to be related to the starburst region. 
Furthermore, the relativistic particles are supposed to be homogeneously distributed in space which does not hold in the case of jet-driven emission.
However, the effect of a complexer geometric structures as well as spatial inhomogeneities shall be addressed in future examinations. 
Thus, only the radio data that represents the whole starburst region is chosen, so that mostly older radio data is used where the spatial substructures could not be resolved. 
\\
In order to obtain the best-fit model we take into account that some of the radio data of different observations indicate not to be in agreement with each other. 
Thus, we exclude some observations due to computational limitations (in relation to the data by \citealt{2010ApJ...710.1462W}) as well as discrepancies between different observations (in relation to NGC 253 and NGC 4945).
Though all radio data show a range of agreement around a few GHz, so that the used data is chosen in order to obtain the most reasonable continuation to lower and higher frequencies.
\\ 
The table \ref{starburst_data_table} displays the distance $d$, the observed supernova rate $\nu_{SN,obs}$ and the luminosity $L_{IR}=3.85\cdot 10^{43}\,L_{43}\,\text{erg}\,\text{s}^{-1}$ of the IR background between 8 and 1000 $\mu$m which is related to the dominant photon field for IC interactions.
\begin{table}[h!]
\centering
  \caption{Observed Model Parameters}
  \begin{tabular}{ l || c | c | c | c }
    Physical parameters & NGC 253 & M82 & NGC 4945 & NGC 1068 \\ \hline \hline
    $d$ [Mpc] & 3.5$^{[0]}$ & 3.4 & 3.7 & 16.7 \\ \hline
    $L_{43}$ & 2.1 & 4.6 & 2.6 & 28  \\ \hline
    $\nu_{SN,obs}$ [yr$^{-1}$] & 0.11$^{[0]}$ & 0.07$^{[1]}$ & 0.12$^{[2]}$ & 0.07$^{[3]}$  \\ \hline
  \end{tabular} 
  \begin{tablenotes}
  \item\emph{References}: $^{[0]}$ - \citealt{2006AJ....132.1333L}; $^{[1]}$ - \citealt{2003RMxAC..15..303P}; $^{[2]}$ - \citealt{2009AJ....137..537L}; \\$^{[3]}$ - \citealt{2012ApJ...755...87S}; non-indexed data taken from \citealt{2012ApJ...755..164A}.
  \end{tablenotes}
  \label{starburst_data_table}
\end{table}
\subsection{The best-fit model}
\begin{figure}
  \centering
\begin{subfigure}[b]{0.4\textwidth}
    \includegraphics[width=0.9\textwidth]{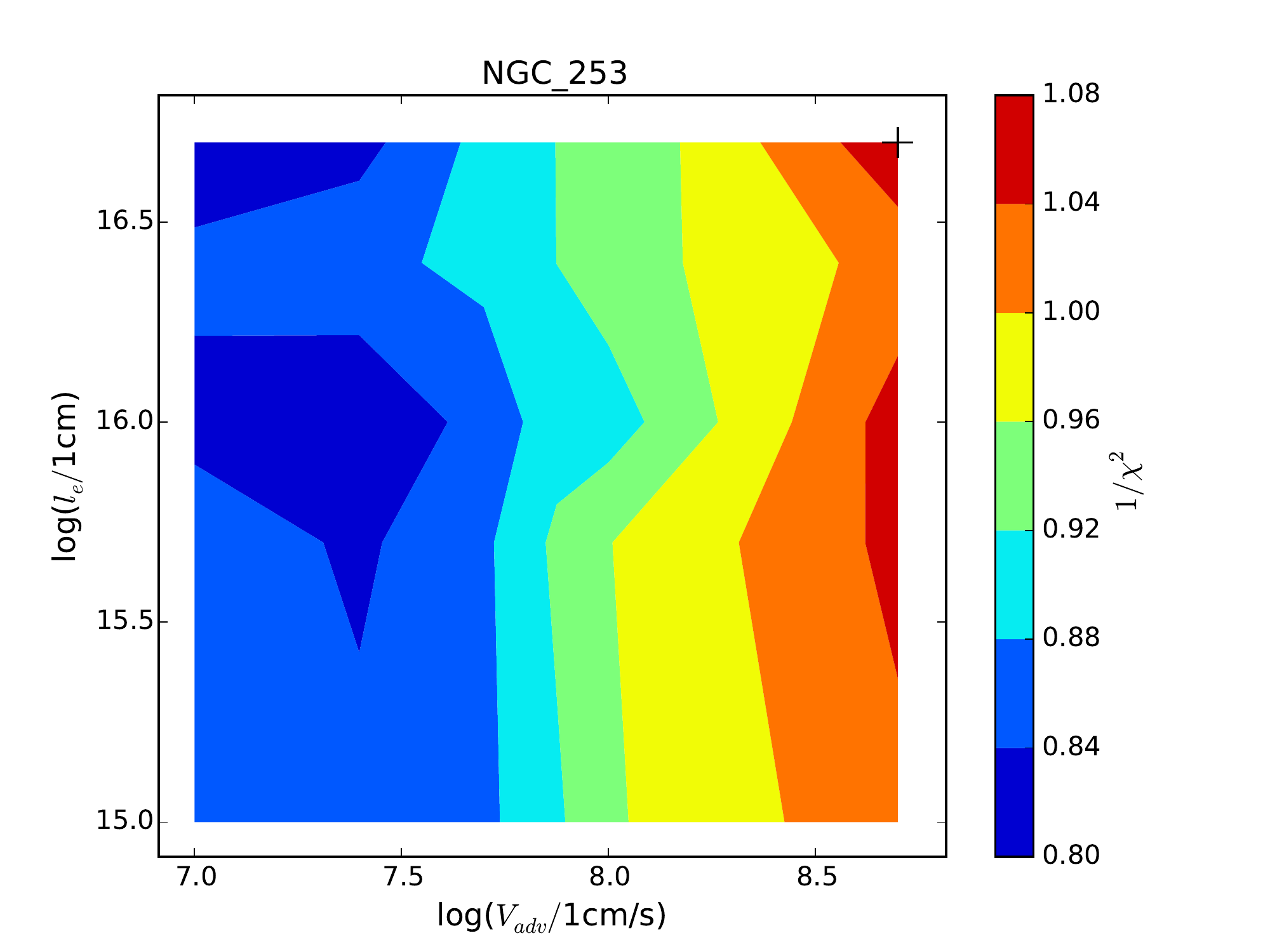}
\end{subfigure}
\begin{subfigure}[b]{0.4\textwidth}
    \includegraphics[width=0.9\textwidth]{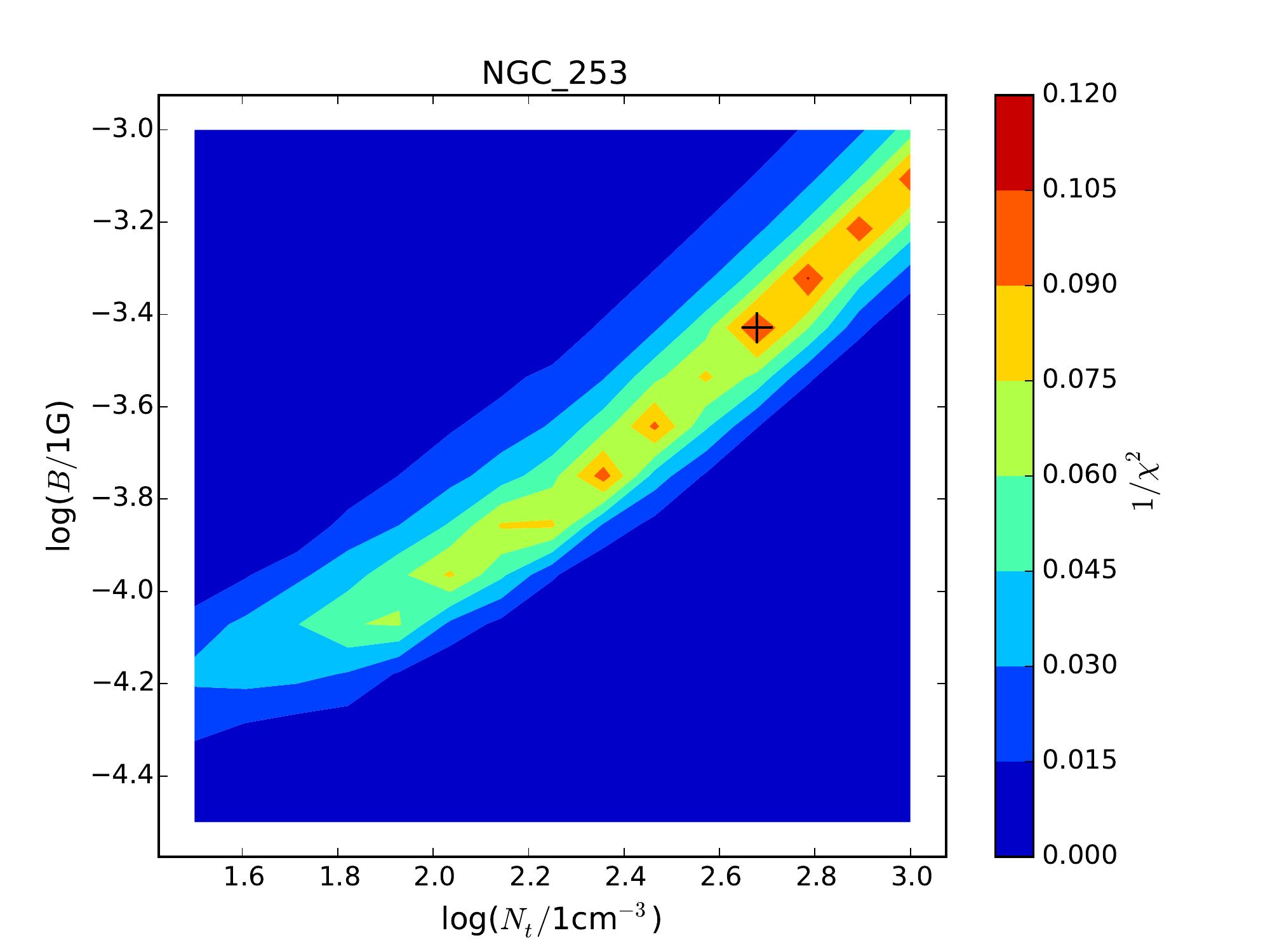}
\end{subfigure}
\begin{subfigure}[b]{0.4\textwidth}
    \includegraphics[width=0.9\textwidth]{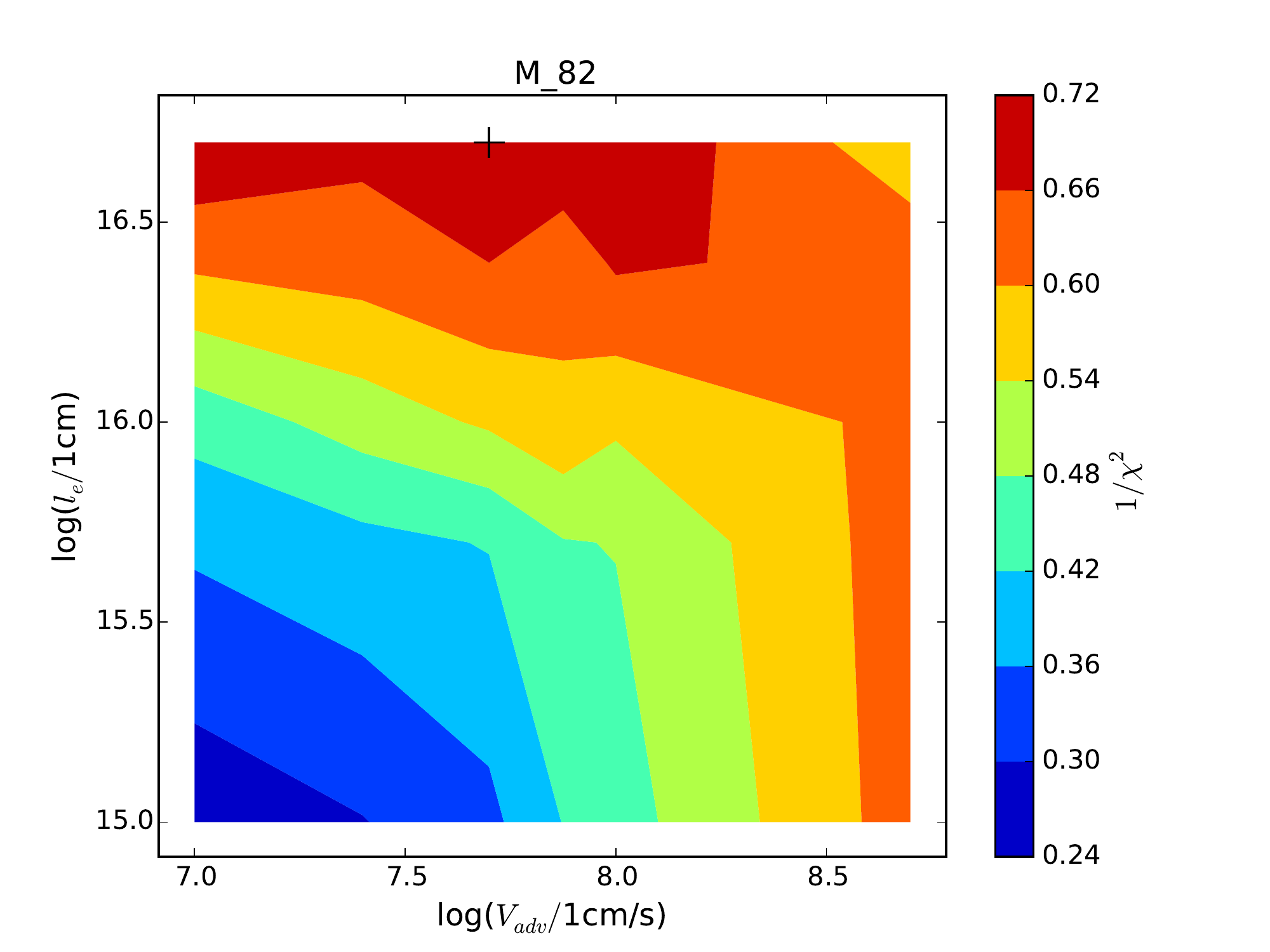}
\end{subfigure}
\begin{subfigure}[b]{0.4\textwidth}
    \includegraphics[width=0.9\textwidth]{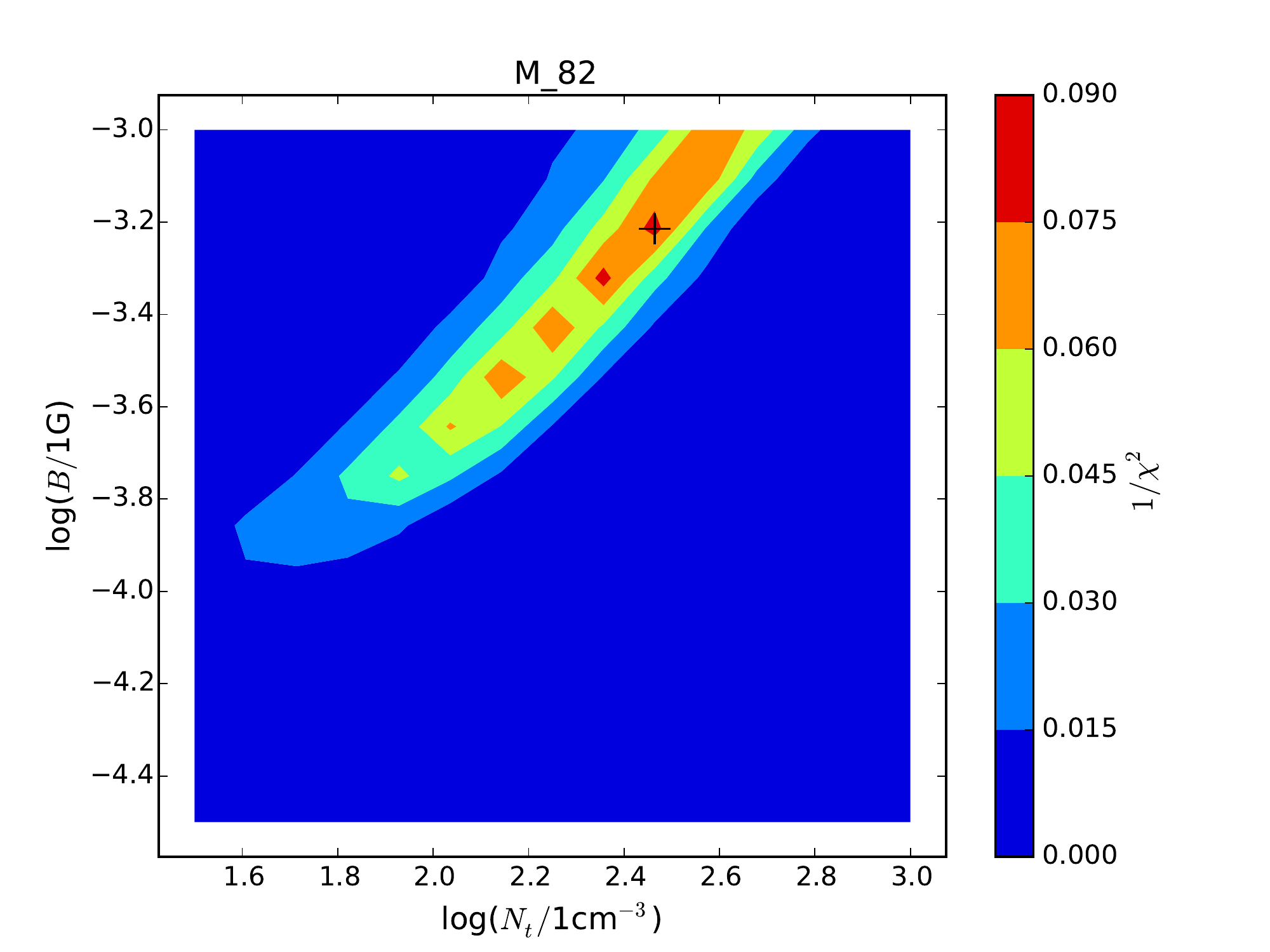}
\end{subfigure}
\begin{subfigure}[b]{0.4\textwidth}
    \includegraphics[width=0.9\textwidth]{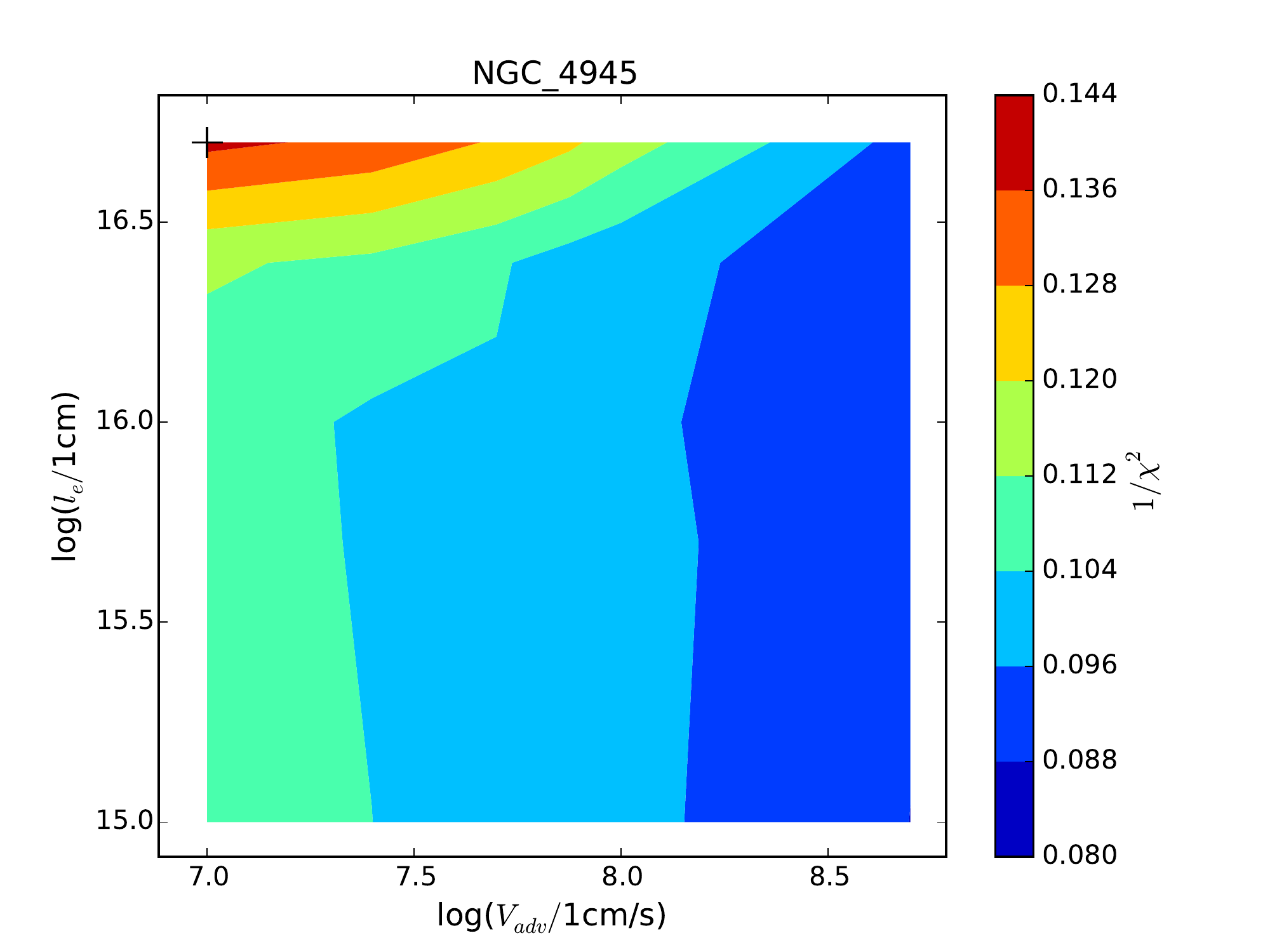}
\end{subfigure}
\begin{subfigure}[b]{0.4\textwidth}
    \includegraphics[width=0.9\textwidth]{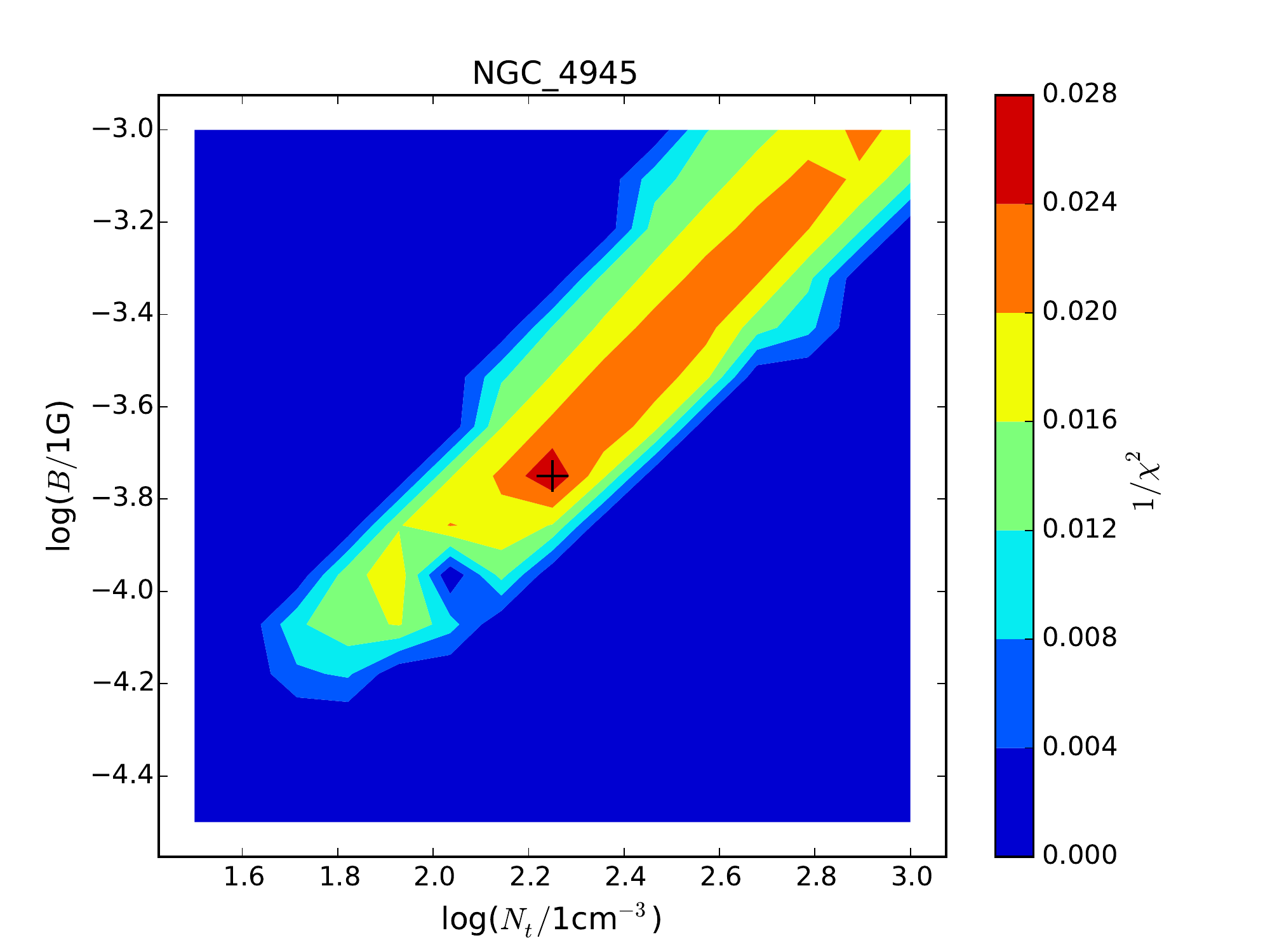}
\end{subfigure}
\begin{subfigure}[b]{0.4\textwidth}
    \includegraphics[width=0.9\textwidth]{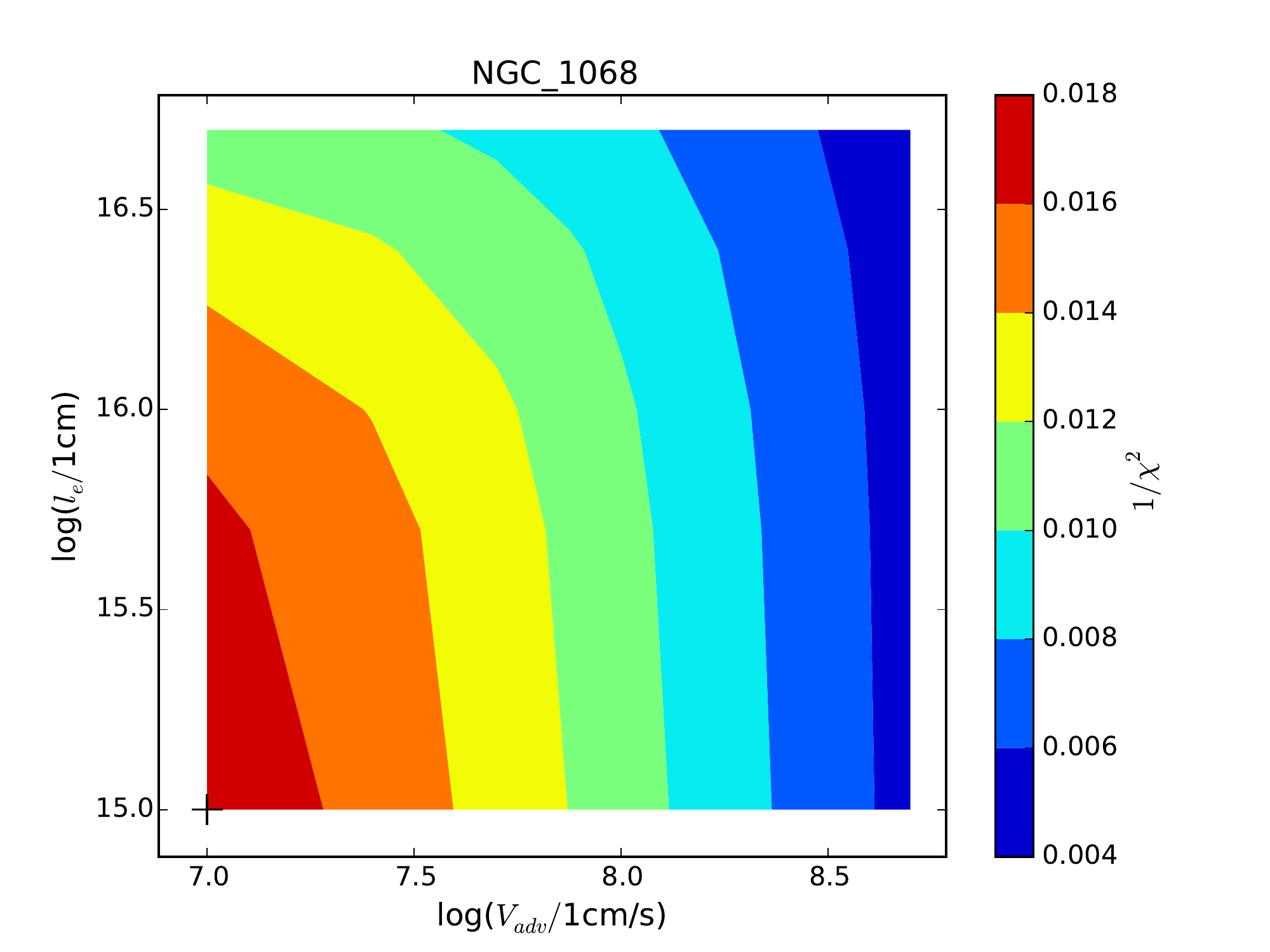}
\end{subfigure}
\begin{subfigure}[b]{0.4\textwidth}
    \includegraphics[width=0.9\textwidth]{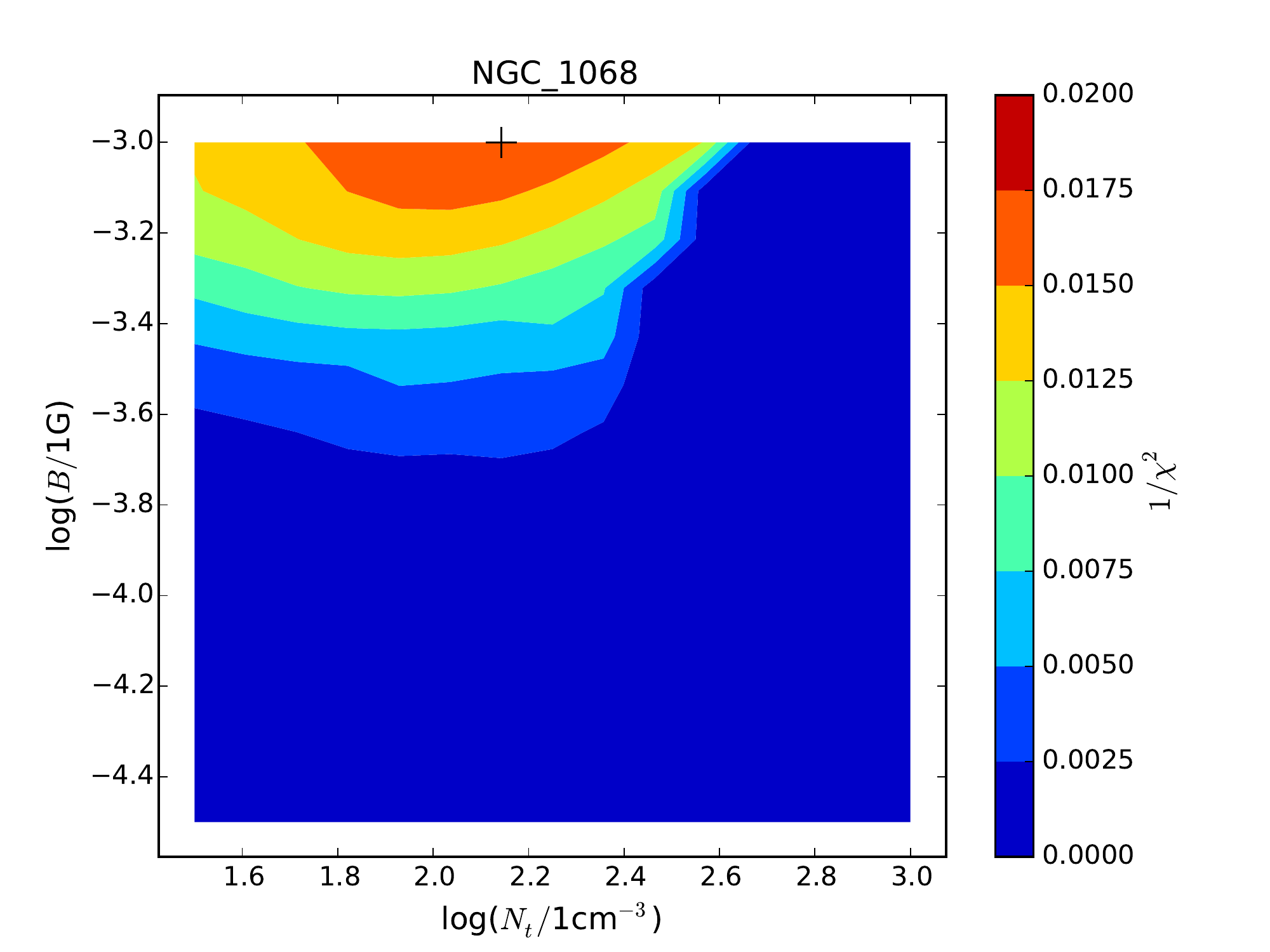}
\end{subfigure}
\caption{The inverse sum of the chi-squared test to the radio and gamma-ray data dependent on $l_e$ and $v_{adv}$ (left) as well as $B$ and $N_t$ (right) of NGC 253, M82, NGC 4945 and NGC 1068. The black cross marks the best-fit model parameter values.}
\label{Chi2test}
\end{figure}

\begin{figure}
  \centering
\begin{subfigure}[b]{0.4\textwidth}
    \includegraphics[width=0.9\textwidth]{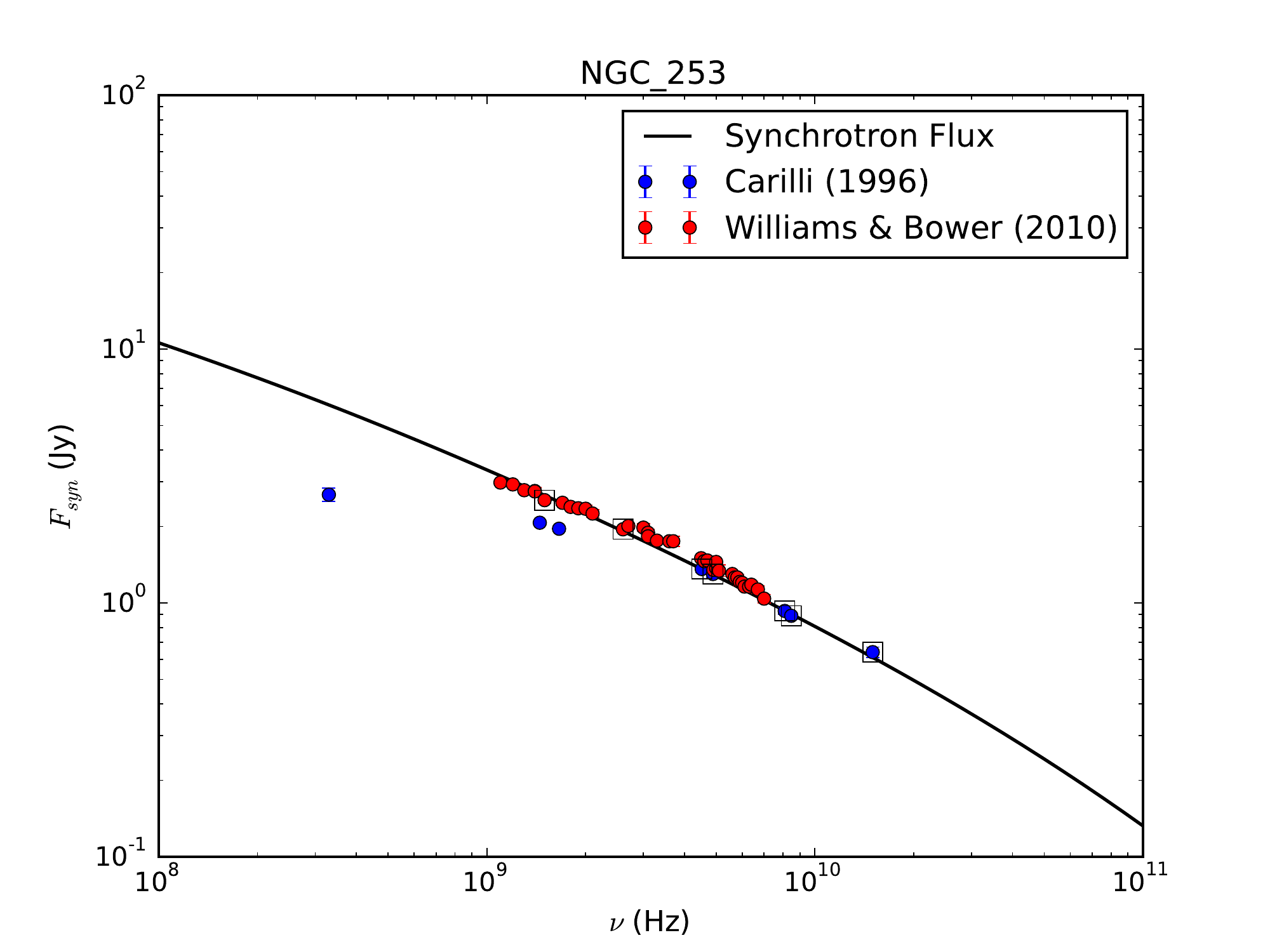}
\end{subfigure}
\begin{subfigure}[b]{0.4\textwidth}
    \includegraphics[width=0.9\textwidth]{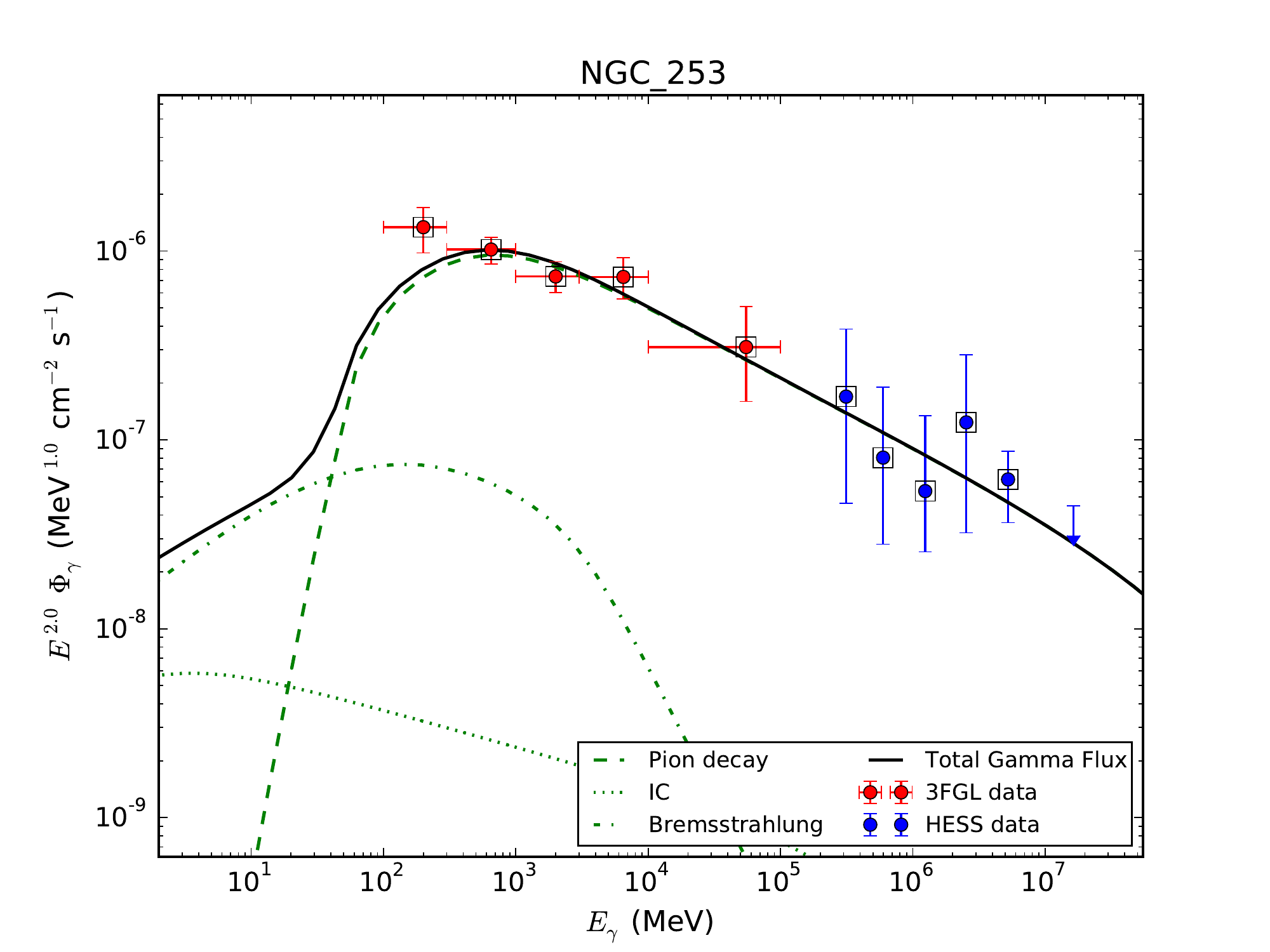}
\end{subfigure}
\begin{subfigure}[b]{0.4\textwidth}
    \includegraphics[width=0.9\textwidth]{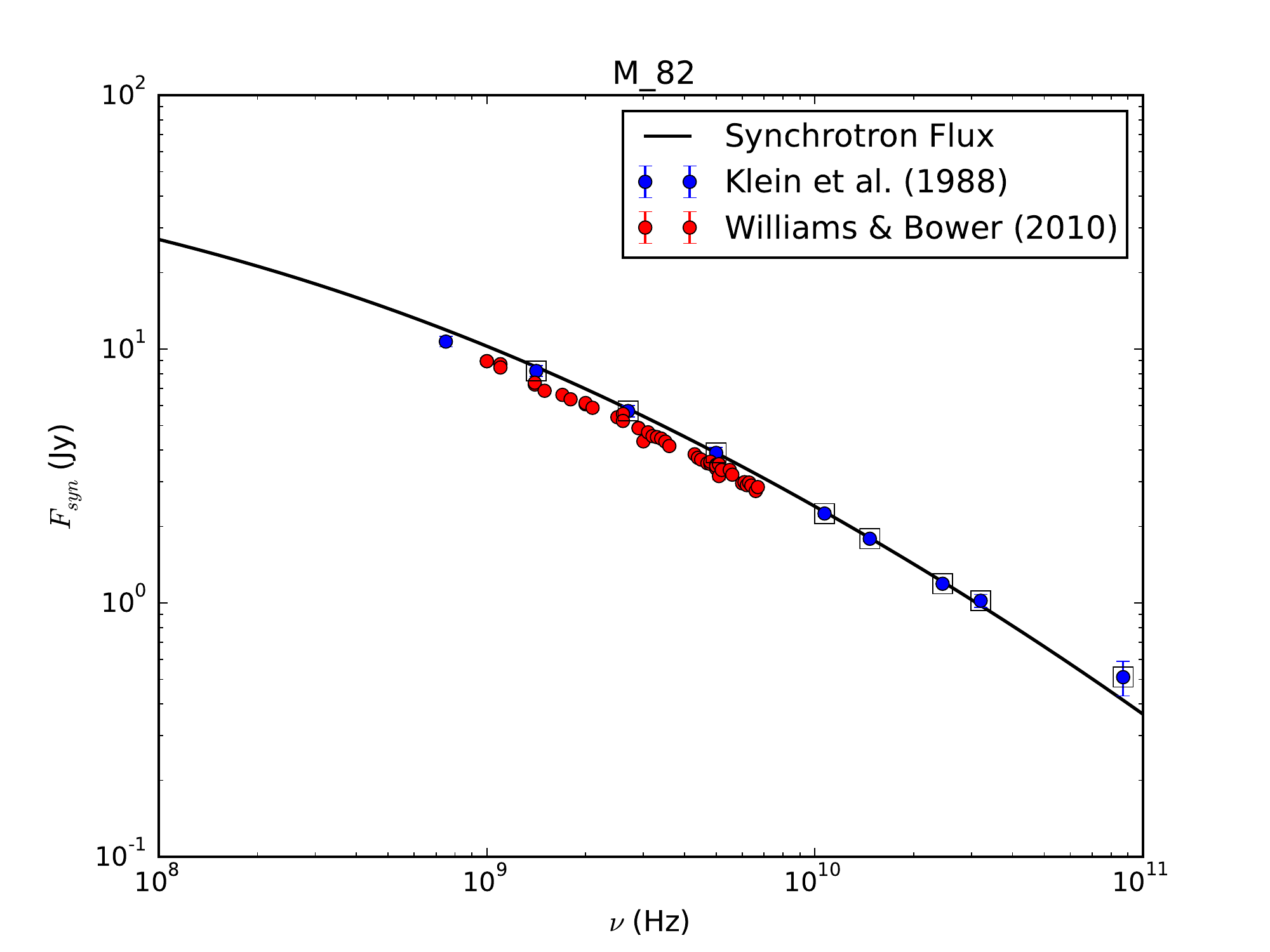}
\end{subfigure}
\begin{subfigure}[b]{0.4\textwidth}
    \includegraphics[width=0.9\textwidth]{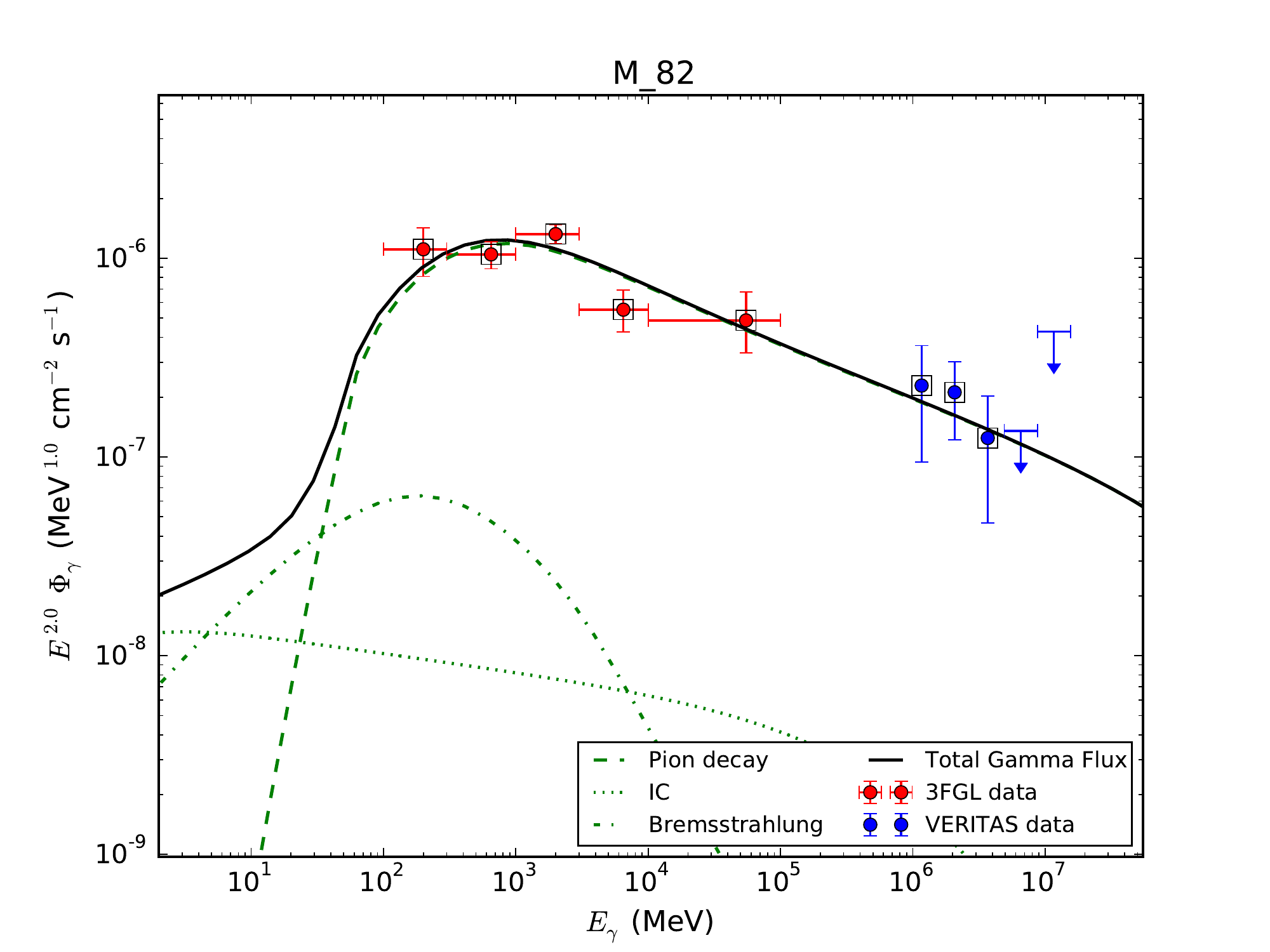}
\end{subfigure}
\begin{subfigure}[b]{0.4\textwidth}
    \includegraphics[width=0.9\textwidth]{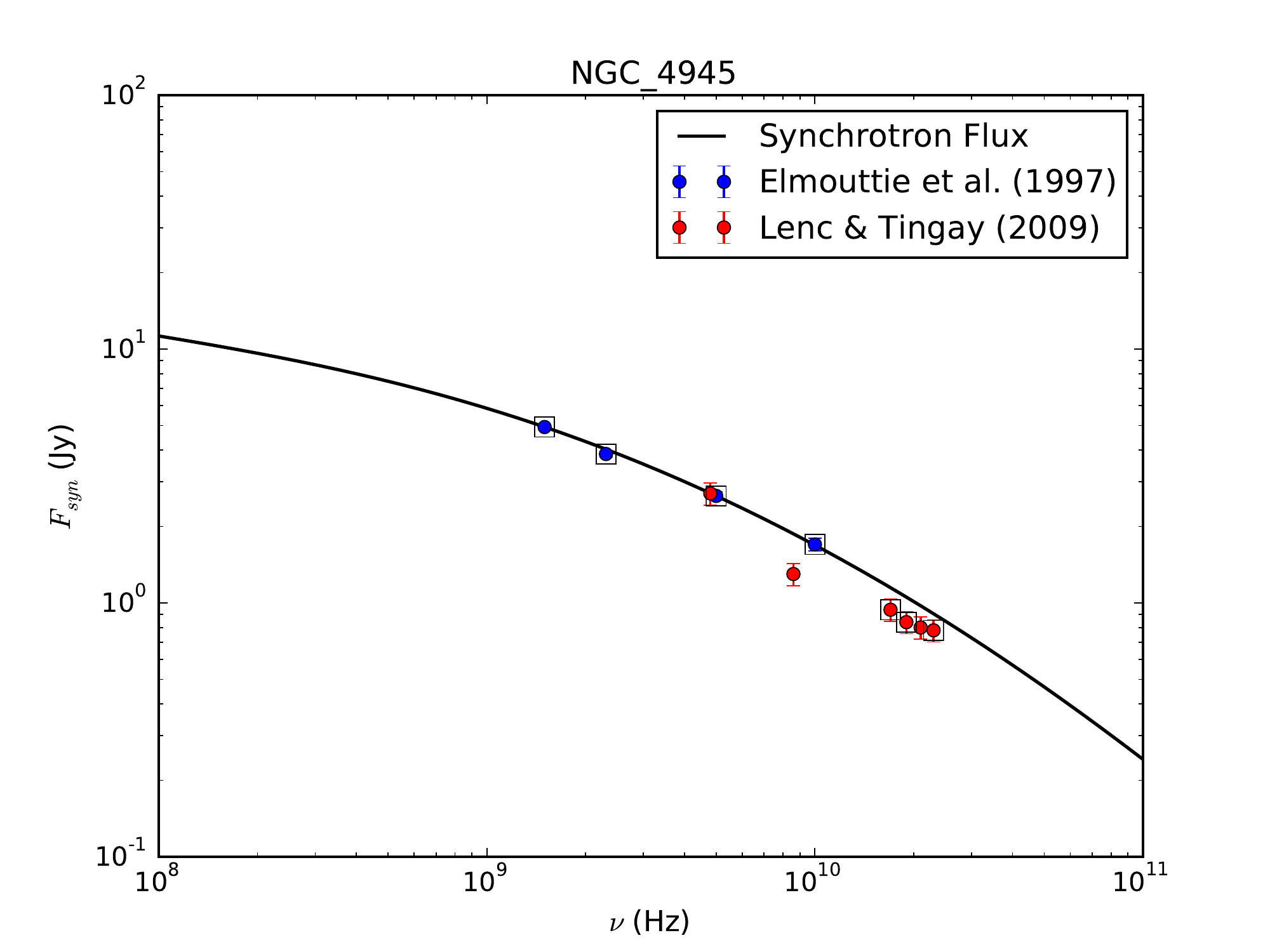}
\end{subfigure}
\begin{subfigure}[b]{0.4\textwidth}
    \includegraphics[width=0.9\textwidth]{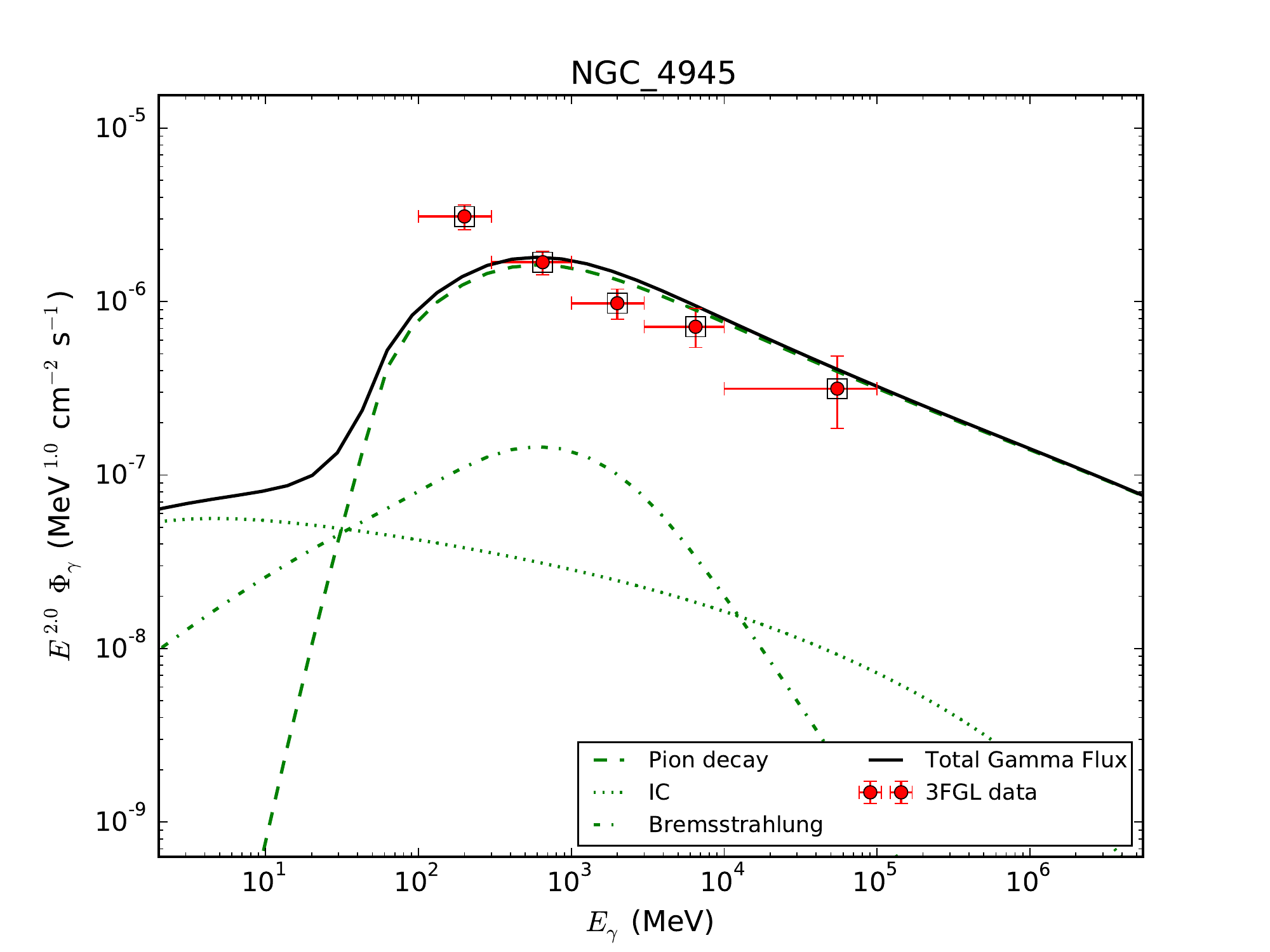}
\end{subfigure}
\begin{subfigure}[b]{0.4\textwidth}
    \includegraphics[width=0.9\textwidth]{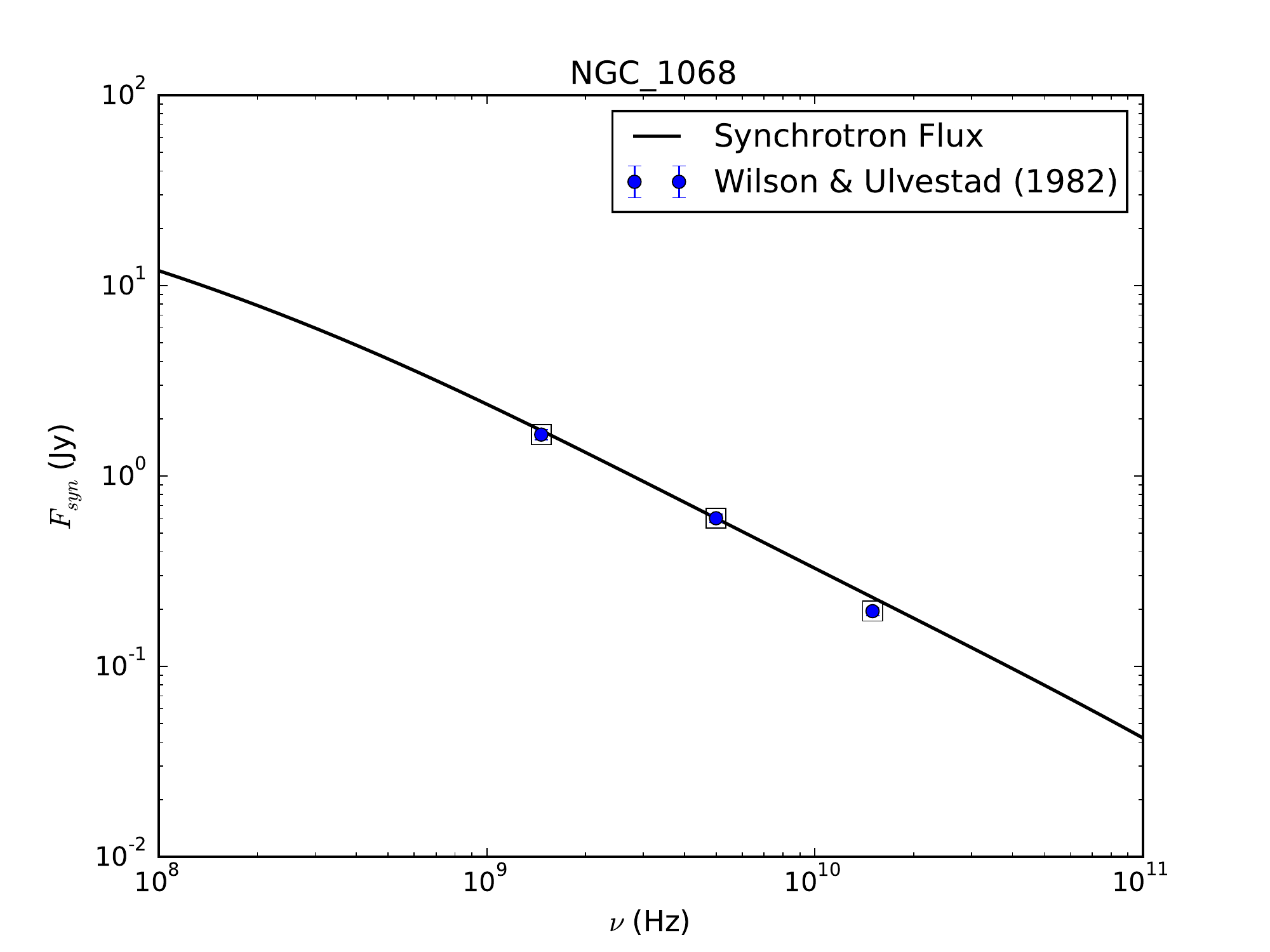}
\end{subfigure}
\begin{subfigure}[b]{0.4\textwidth}
    \includegraphics[width=0.9\textwidth]{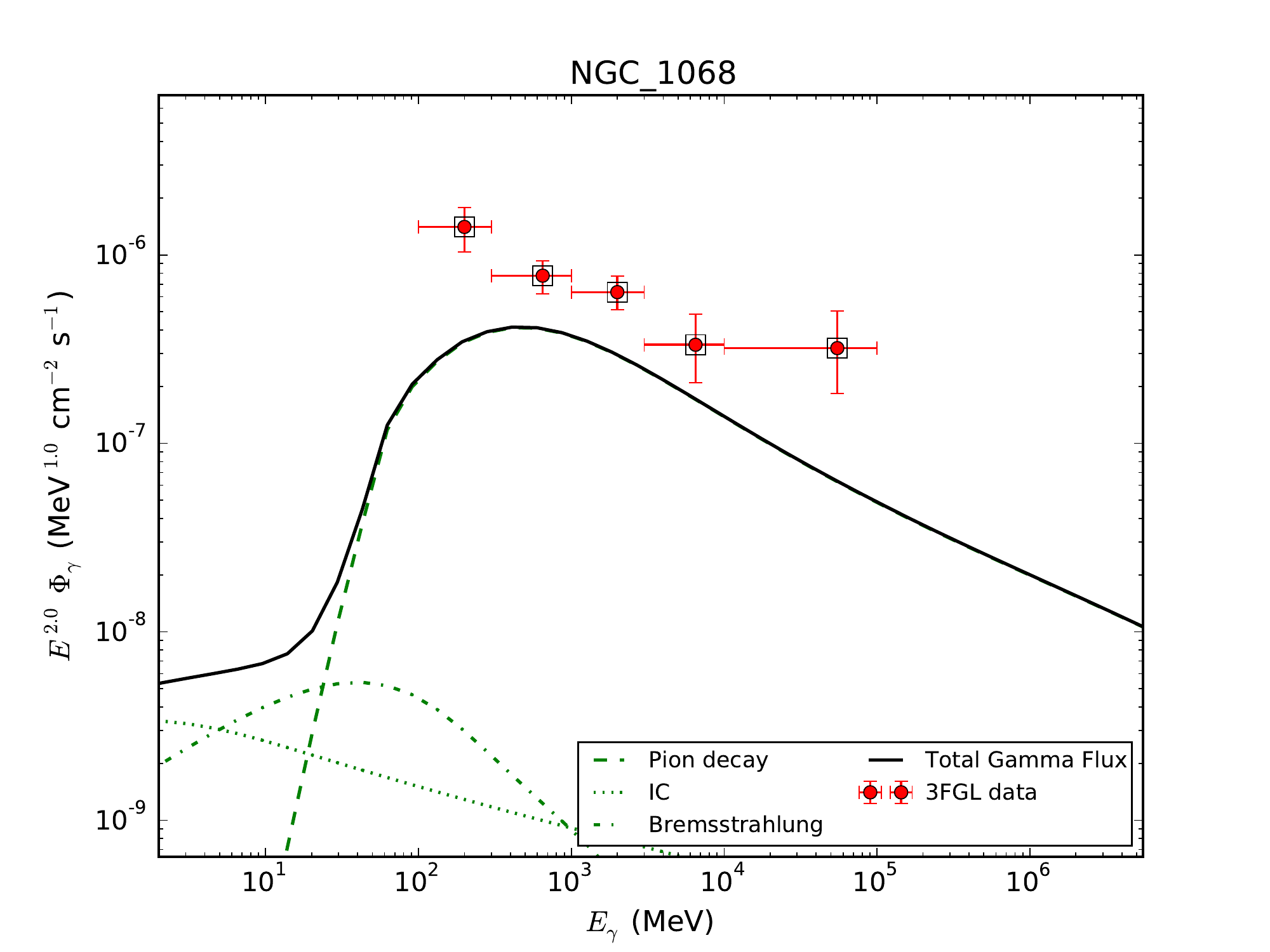}
\end{subfigure}
\caption{The radio (left) and gamma-ray (right) data as well as the best-fit model of NGC 253, M82, NGC 4945 and NGC 1068. Open squares represent the data set that is used for the $\chi^2$ test.}
\label{AllFits}
\end{figure}

In the following, the relation (\ref{protonToSNrate}) as well as the previously derived synchrotron and $\gamma-$ray flux are used to fit the observed SN rate (see table \ref{starburst_data_table}), the radio and gamma data, respectively, in order to determine $N_t$, $B$, $\alpha$, $\gamma_B$, $q_0^{e,p}$, $l_e$ and $v_{adv}$ of the four starburst galaxies.
Thereby, the ratio of the source rate $q_0^e/q_0^p$ is already fixed by the quasi-neutrality of the source plasma according to eq.\ (\ref{gen_source-ratio}) and the observed radio flux at a frequency $1\,\text{GHz}<\nu<10\,\text{GHz}$ is used to constrain $q_0^e$ for a given set of parameters. \\
In the case of $N_t$, $B$, $\alpha$ and $\gamma_B$ we test $11250$ $(= 15\times 15\times 10 \times 5)$ different value combinations within the reasonable parameter space of $B=[10^{-4.5};\, 10^{-3} ]\,\text{G}$, $N_t=[10^{1.5};\, 10^{3} ]\,\text{cm}^{-3}$, $\alpha=[1.9;\, 2.4 ]$ and $\gamma_B=[10^{4};\, 10^{8} ]$, respectively.
Due to the computational intense calculations, we test our model only for $l_e=\{10^{15},\,5\cdot 10^{15},\, 10^{16}, \, 2.5\cdot 10^{16}, \, 5\cdot 10^{16} \}\,\text{cm}$ as well as $v_{adv}=\{10^{7}, \,2.5\cdot 10^{7}, \,5\cdot 10^{7}, \,7.5\cdot 10^{7}, \, 10^{8}, 5\cdot 10^{8} \}\,\text{cm/s}$. \\
The goodness of the fits is expressed by the reduced chi-squared test $\chi^2 = (\tilde{\chi}^2_{SN} + \tilde{\chi}^2_{radio}+\tilde{\chi}^2_{\gamma})/\delta_{f}$ to the data with $\delta_{f}$ degrees of freedom. 
Here $\tilde{\chi}^2_{SN}$, $\tilde{\chi}^2_{radio}$ and $\tilde{\chi}^2_{\gamma}$ denote the chi-squared test to the observed SN rate, the radio data and the gamma-ray data, respectively. 
However, it needs to be taken into account, that due to the computational intense procedure the error of the considered energy of the FERMI data is not included. 
Therefore, $\tilde{\chi}^2_{\gamma}$ is more an upper limit of the chi-squared test to the gamma data and can not be used in order to determine the corresponding p-Value. 
In order to determine $\tilde{\chi}^2_{SN}$ a total proton energy output per SN of $\eta E_{SN}= 10^{50}\,\text{erg}$ is supposed as well as an error in the observed SN rate of $\sigma_{SN}=0.5\,\nu_{SN,obs}$ due to systematic errors in $\eta E_{SN}$. \\
Choosing a certain mean free path $l_e$ and $l_p$, respectively, as well as a galactic wind speed $v_{adv}$, we determine $\chi^2$ for $11250$ $(= 15\times 15\times 10 \times 5)$ different parameter sets of $(N_t,\,B,\,\alpha,\,\gamma_B)$. 
Afterward, the same procedure is done for 29 other parameter sets of $(l_e,\,v_{adv})$.
The resulting maximum of the derived $1/\chi^2$ dependent on $l_e$ and $v_{adv}$ (as displayed on the left hand side of fig.\ \ref{Chi2test}) yields the best-fit model parameters of the individual starburst galaxy as displayed in table \ref{starburst_bestfit_table} (the corresponding radio and gamma flux is shown in fig.\ \ref{AllFits}, respectively). 
The right hand side of fig.\ \ref{Chi2test} shows the values of $1/\chi^2$ dependent on $N_t$ and $B$ using the best-fit model values of $l_e$ and $v_{adv}$.
Consequently, the diffusion and advection have almost the same influence on the goodness of the fits and either a large diffusion length $l_e\sim 5\cdot 10^{16}\,\text{cm}$ or a strong galactic wind $v_{adv}\sim 5\cdot 10^{8}\,\text{cm/s}$ is needed to obtain an accurate fit to the data of NGC 253, M82 and NGC 4945. 
Here, only a small range of the escape parameters $l_e$ and $v_{adv}$ can be excluded. 
The major influence on the goodness of the fits is determined by the cooling parameters $B$ and $N_t$ as well as the initial source spectrum. 
Here a distinct range of possible magnetic field strengths is exposed that dependent on the target density according to $B\propto N_t$ in the case of NGC 253, M82 and NGC 4945. 
Thus, these three starburst galaxies show a distinct indication of an efficient particle escape as well as a certain correlation between the magnetic field strength and the target density, which is in good agreement with previous results from \citealt{2012ApJ...755..106P}. \\
However, for NGC 1068 there is neither an indication of efficient particle escape nor a hint of a certain relation between $B$ and $N_t$.
\begin{table}[h!]
\centering
  \caption{Best-fit Model Parameters}
  \begin{tabular}{ l || c | c | c | c }
    Physical parameters & NGC 253 & M82 & NGC 4945 & NGC 1068 \\ \hline \hline
    $q_0^e$ [$10^{-18}\,$cm$^{-3}\,$s$^{-1}$] & $11$ & $1.6$ & $3.9$ & $32$ \\ \hline
    $q_0^p$ [$10^{-20}\,$cm$^{-3}\,$s$^{-1}$] & $8.5$ & $2.6$ & $5.1$ & $20$ \\ \hline
    $\alpha$ & $2.3\pm 0.025$ & $2.1 \pm 0.025$ & $2.15 \pm 0.025$ & $2.35 \pm 0.025$  \\ \hline
    $\gamma_B$ & $10^{4\pm 0.5}$ & $10^{8 \pm 0.5}$ & $10^{6 \pm 0.5}$ &  $10^{4 \pm 0.5}$ \\ \hline
    $B$ [$\mu$G] & $373^{+52}_{-41}$ & $611^{+85}_{-67}$ & $178^{+25}_{-19}$ & $1000 \pm 109$  \\ \hline
    $N_t$ [cm$^{-3}$] & $477^{+67}_{-52}$ & $291^{+41}_{-32}$ & $178^{+25}_{-19}$ & $139^{+19}_{-15}$  \\ \hline
    $v_{adv}$ [$10^7\,$cm/s] & $50\pm 25$ & $5^{+2.5}_{-1.3}$ & $1\pm 1.3$ & $1\pm 1.3$  \\ \hline
    $l_e$ [$10^{16}\,$cm] & $5\pm 1.3$ & $5 \pm 1.3$ & $5 \pm 1.3$ &  $0.1 \pm 0.25$ \\ \hline
  \end{tabular}
  \begin{tablenotes}
  \item \emph{Note:} Errors refer to half of the distance between the adjacent parameters considered.
  \end{tablenotes}
  \label{starburst_bestfit_table}
\end{table}
 \\
Furthermore, the $\chi^2$ values expose that the best-fit model is able to describe the data of M82 and NGC 253 very well. 
Also the observations of NGC 4945 can be explained, although there is a discrepancy at low gamma-ray energies (between 100 and 300 MeV) and at radio frequencies $>10\,\text{GHz}$. 
The gamma-ray data of all three starburst galaxies is accurately described by hadronic pion production. 
Bremsstrahlung and IC scattering only affect the total $\gamma-$ray spectrum below a few tens of MeV. 
In the case of NGC 1068, the observed gamma-ray flux and the SN rate can not be described by our model. 
Here, a SN rate that is about an order of magnitude bigger than the observed one is need in order to accurately describe the gamma-ray data. 
\\
However, in the case of NGC 4945, M82 and NGC 253 it needs to be taken into account, that there are multiple different parameter models (subsequently called $2\sigma$ models) that fit within $2\sigma$ of our best-fit models (45 for NGC 4945, 298 for M82 and 1193 for NGC 253).
The large number of accurate models, in particular for M82 and NGC 253, is predominantly a consequence of the large uncertainties in the gamma-ray and SN rate data. 
Furthermore, the total minimum of the $\chi^2$ within the whole range of the 6-dimensional parameter space is most probably not covert, as $\chi^2$ is due to the computational intense procedure only determined for a relatively rough grid of parameters.
Since $\chi^2$ dependent on a certain parameter can have multiple local maximums the obtained best-fit parameter is not necessarily in the center of the distribution of $2\sigma$ models.
Hence, it is crucial to discuss also the models within $2\sigma$ of our best-fit model, in order to understand how well the astrophysical parameters are constrained. 
\begin{figure}
  \centering
\begin{subfigure}[b]{0.32\textwidth}
    \includegraphics[width=0.95\textwidth]{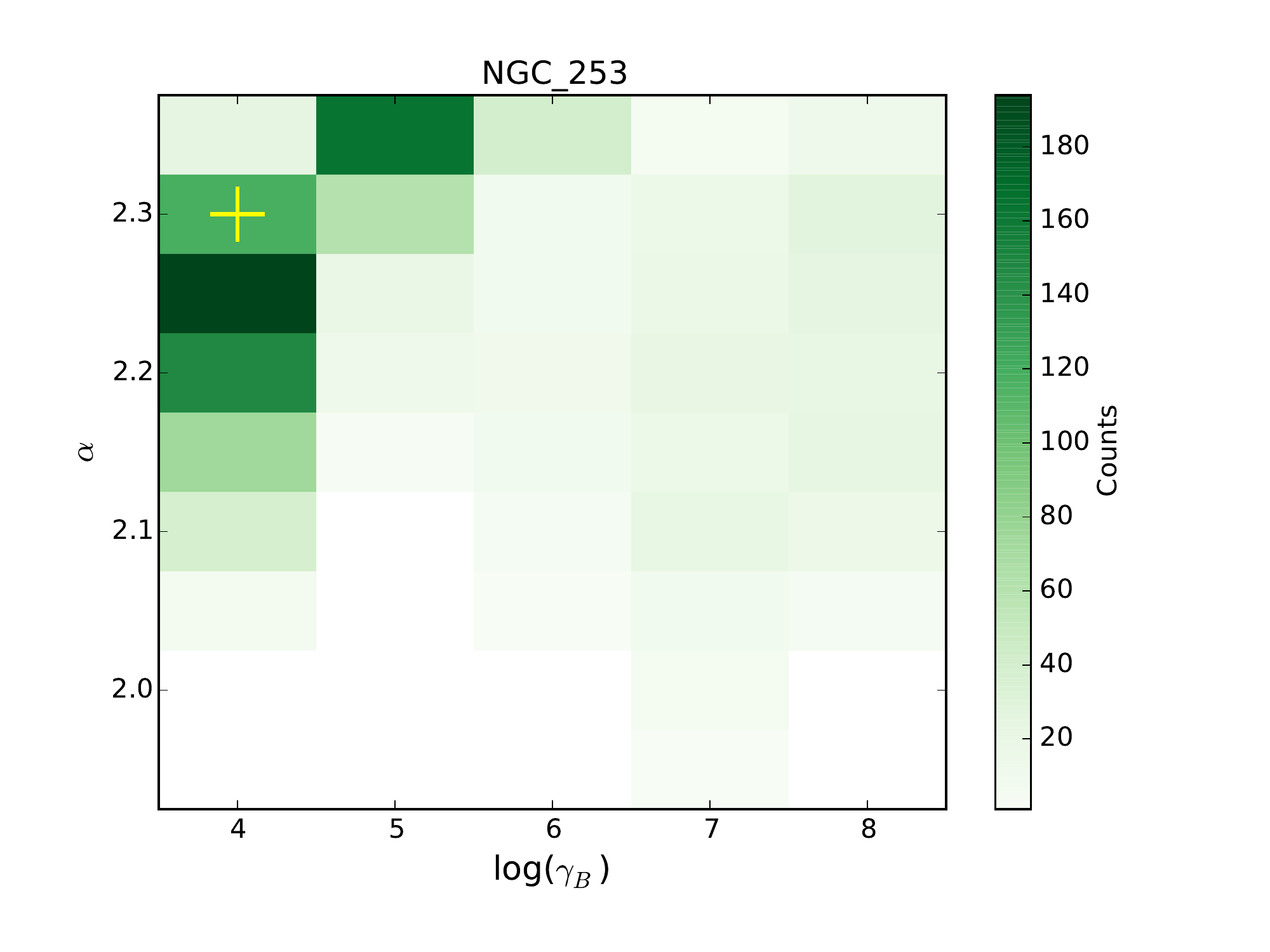}
\end{subfigure}
\begin{subfigure}[b]{0.32\textwidth}
    \includegraphics[width=0.95\textwidth]{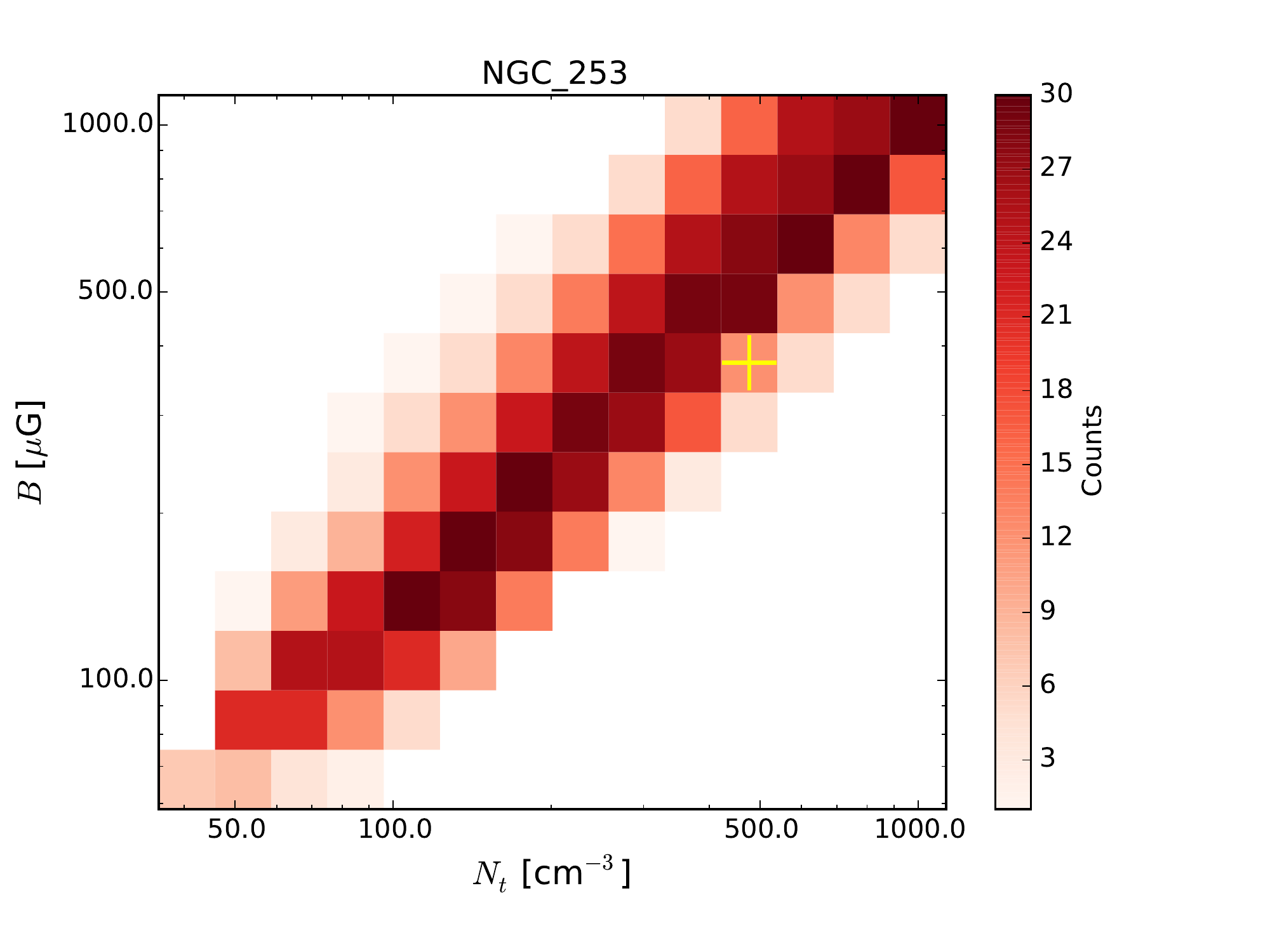}
\end{subfigure}
\begin{subfigure}[b]{0.32\textwidth}
    \includegraphics[width=0.95\textwidth]{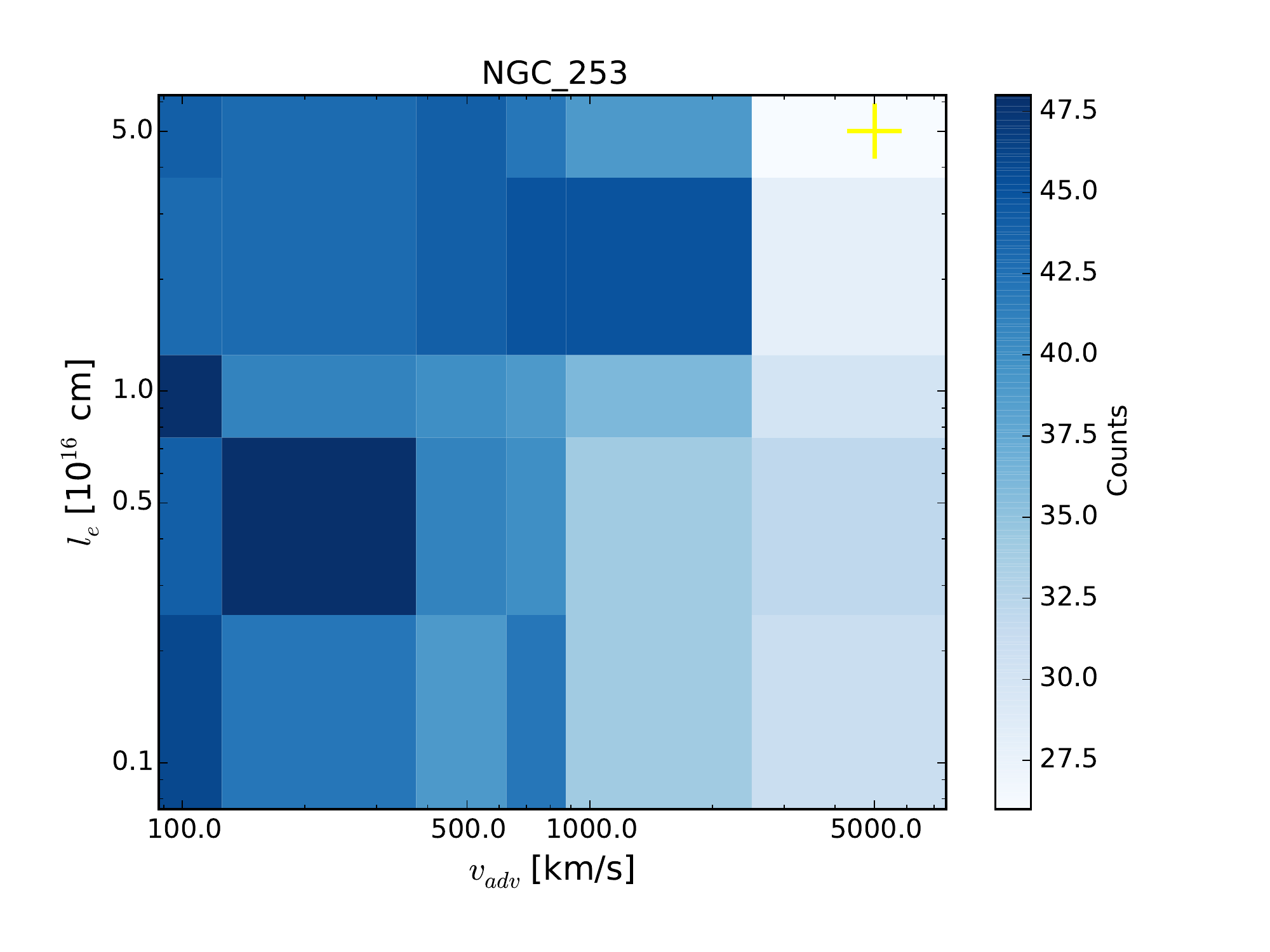}
\end{subfigure}
\begin{subfigure}[b]{0.32\textwidth}
    \includegraphics[width=0.95\textwidth]{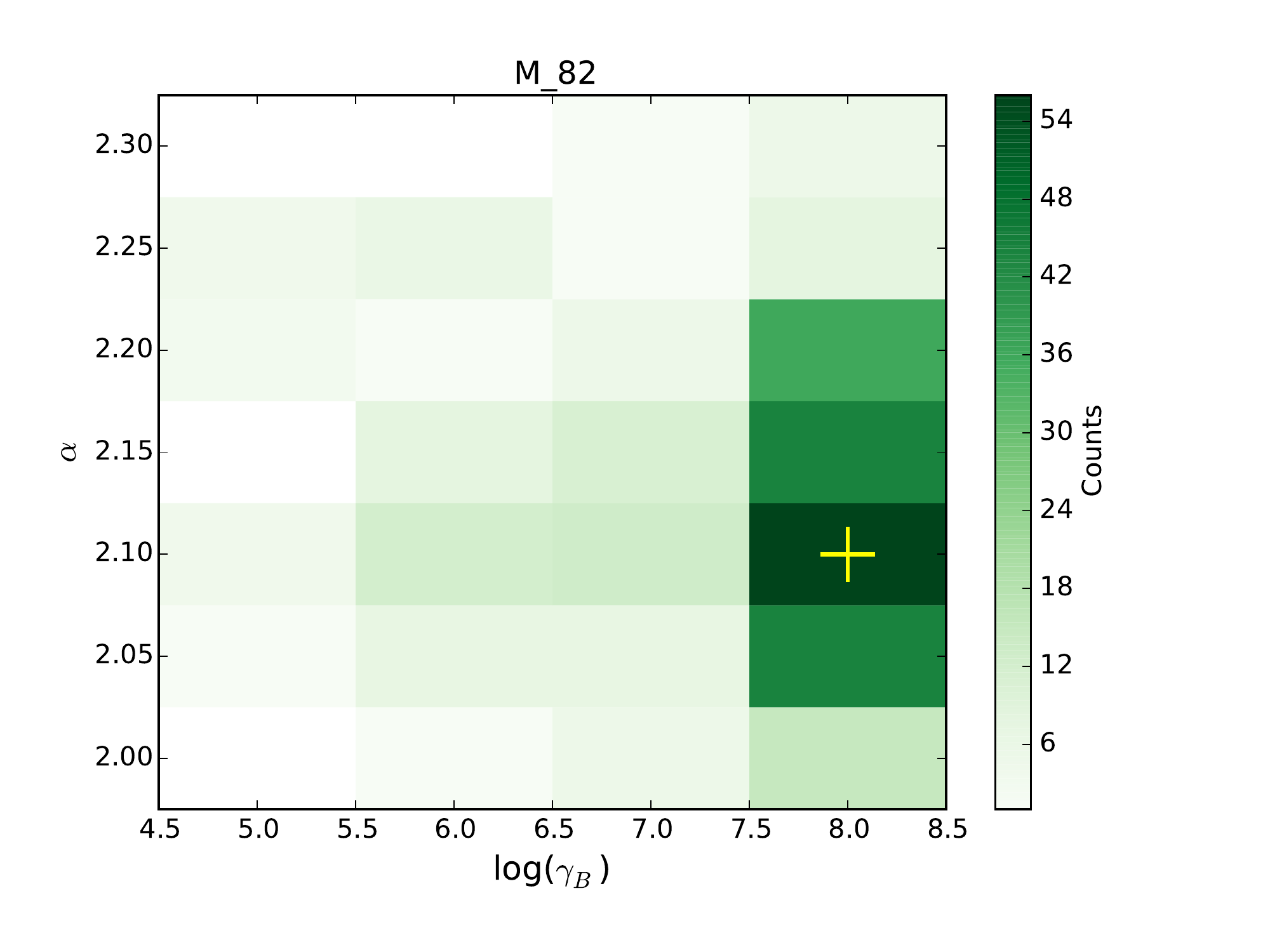}
\end{subfigure}
\begin{subfigure}[b]{0.32\textwidth}
    \includegraphics[width=0.95\textwidth]{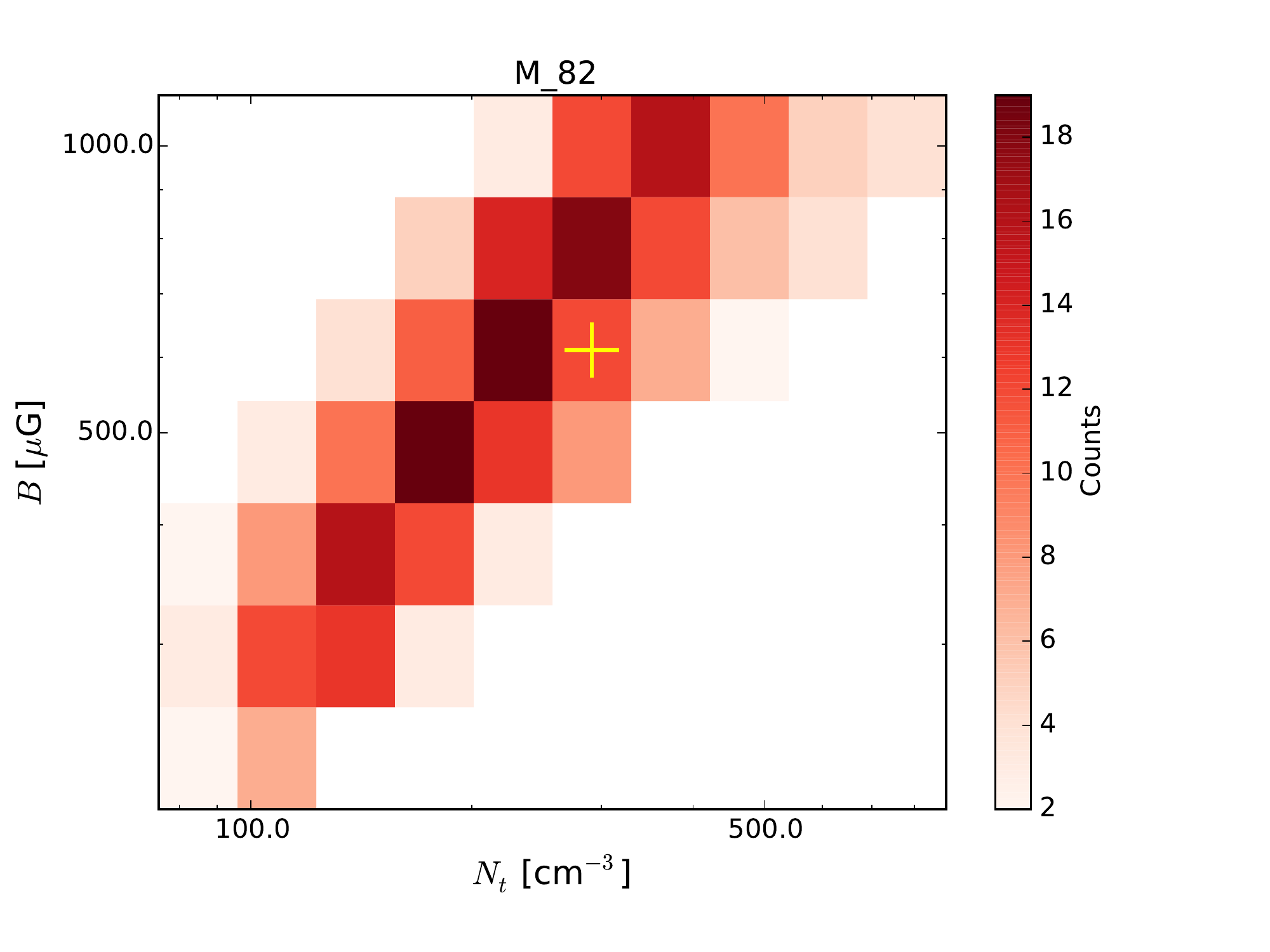}
\end{subfigure}
\begin{subfigure}[b]{0.32\textwidth}
    \includegraphics[width=0.95\textwidth]{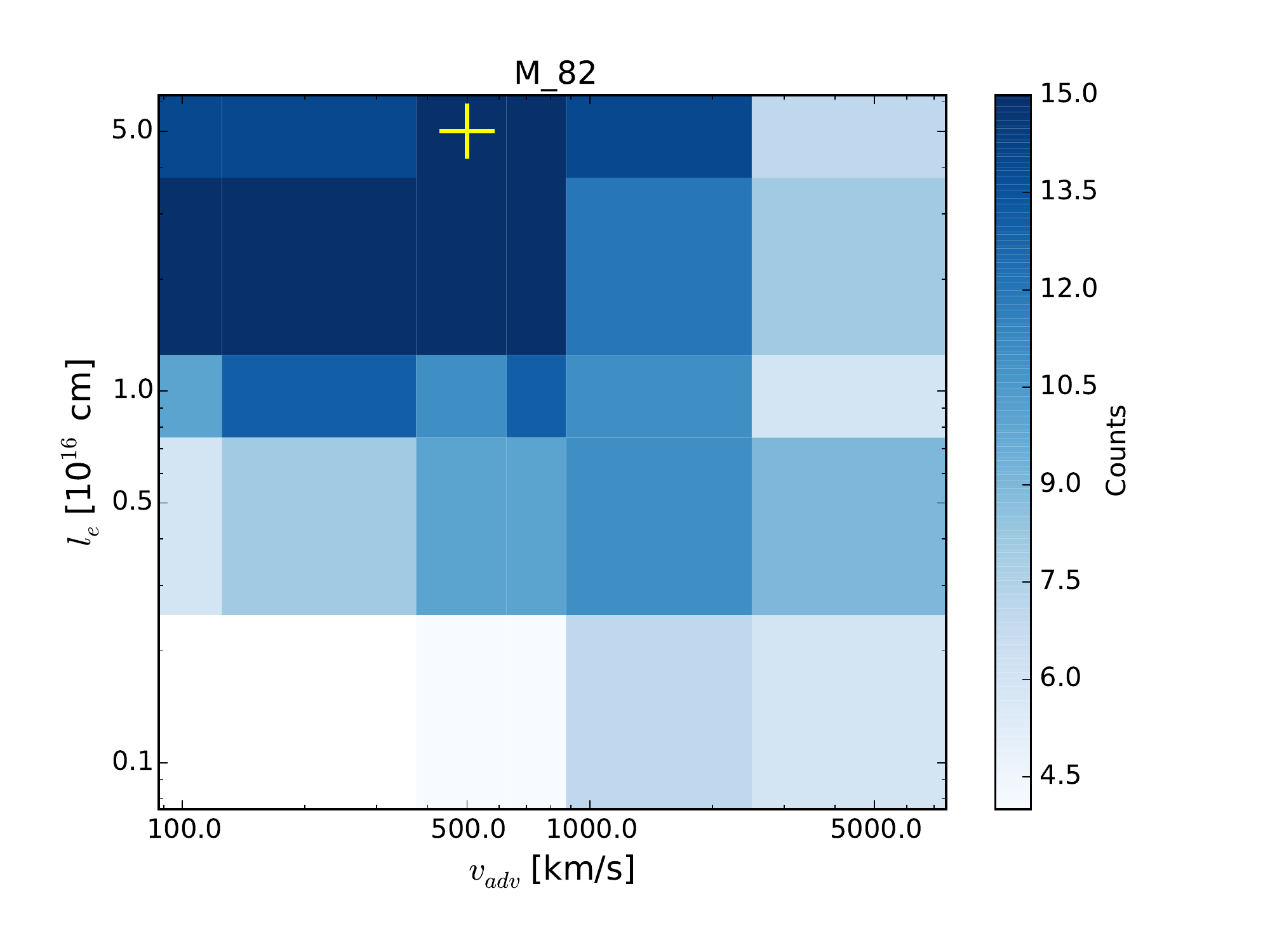}
\end{subfigure}
\begin{subfigure}[b]{0.32\textwidth}
    \includegraphics[width=0.95\textwidth]{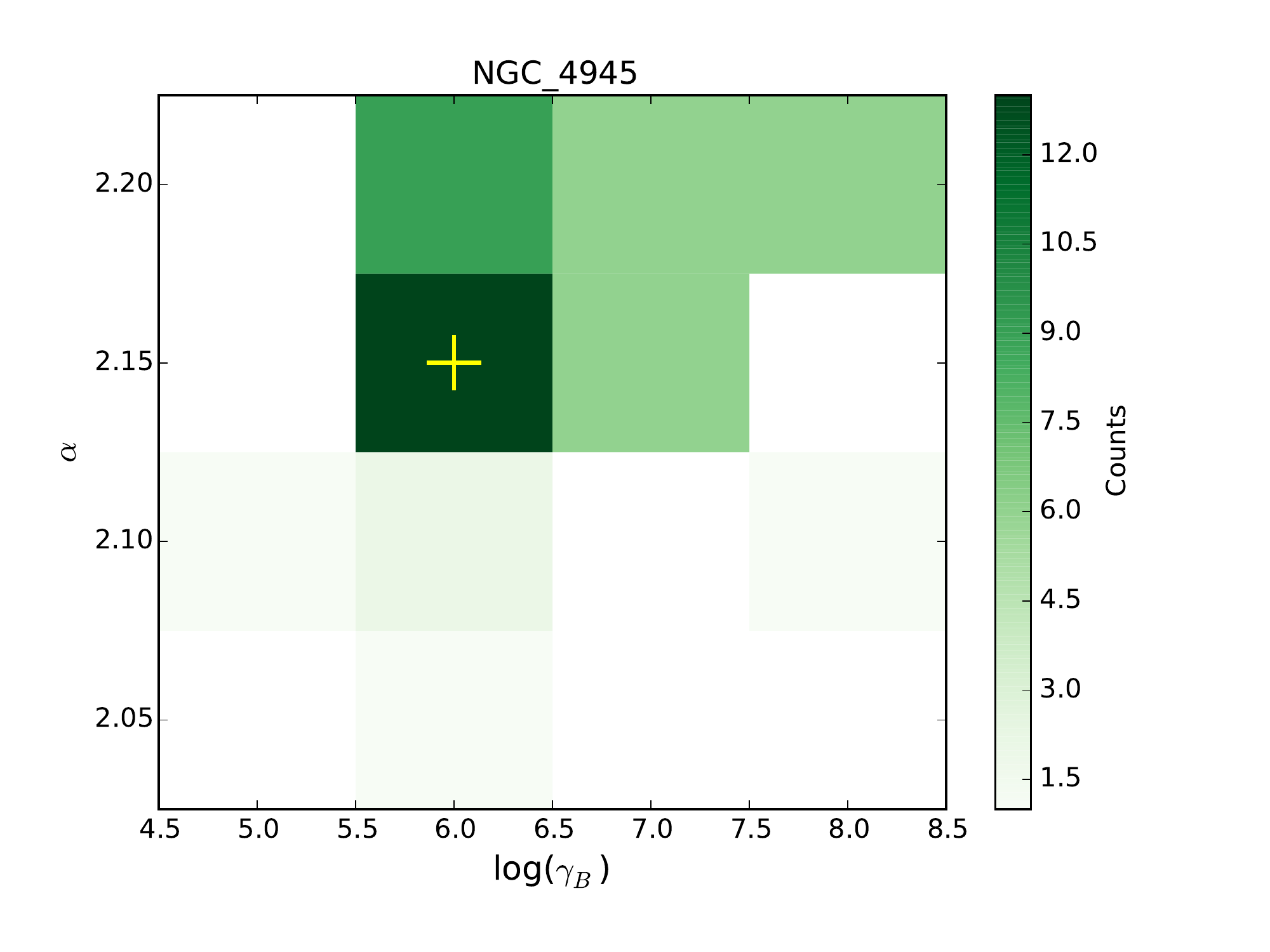}
\end{subfigure}
\begin{subfigure}[b]{0.32\textwidth}
    \includegraphics[width=0.95\textwidth]{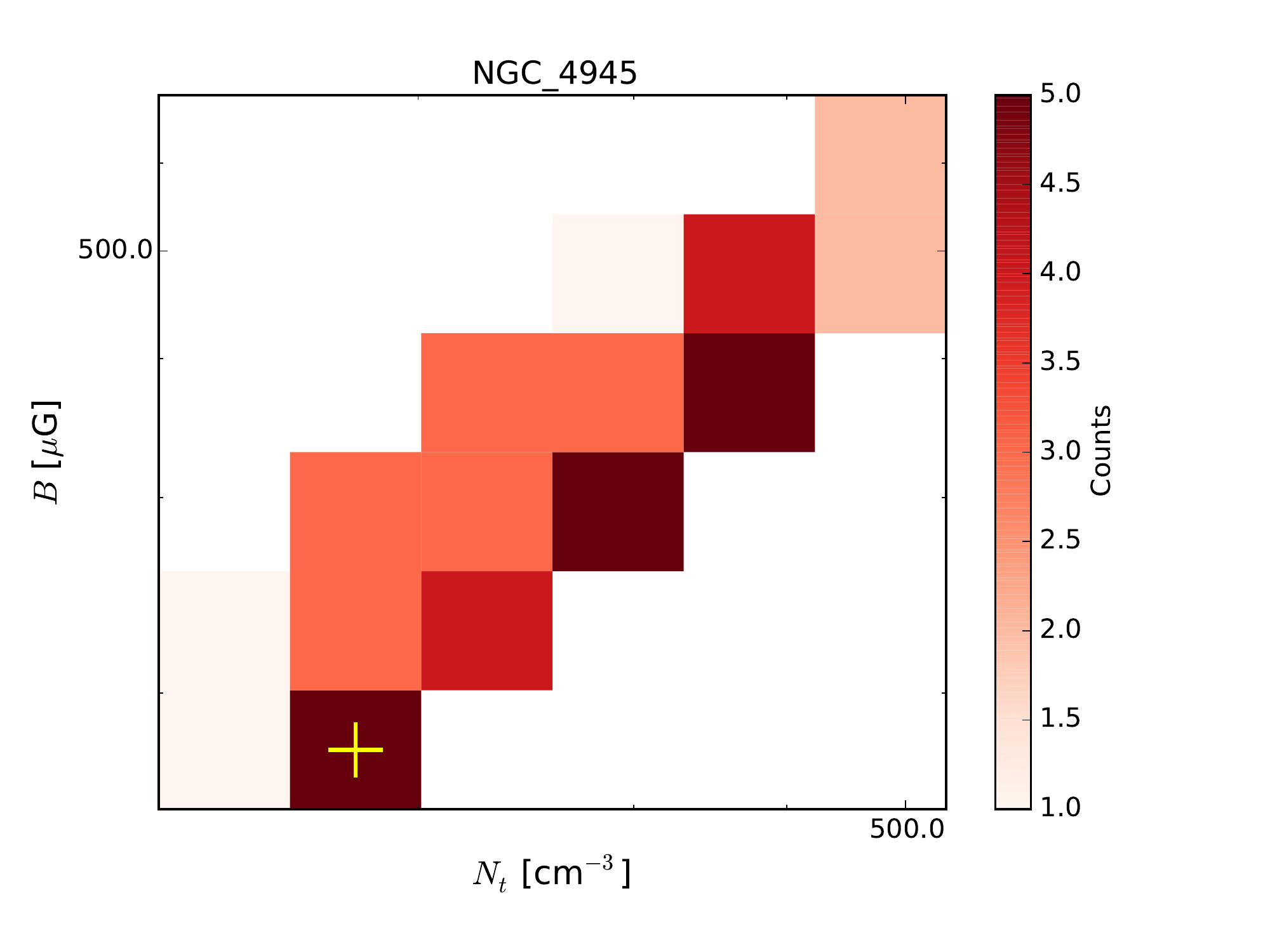}
\end{subfigure}
\begin{subfigure}[b]{0.32\textwidth}
    \includegraphics[width=0.95\textwidth]{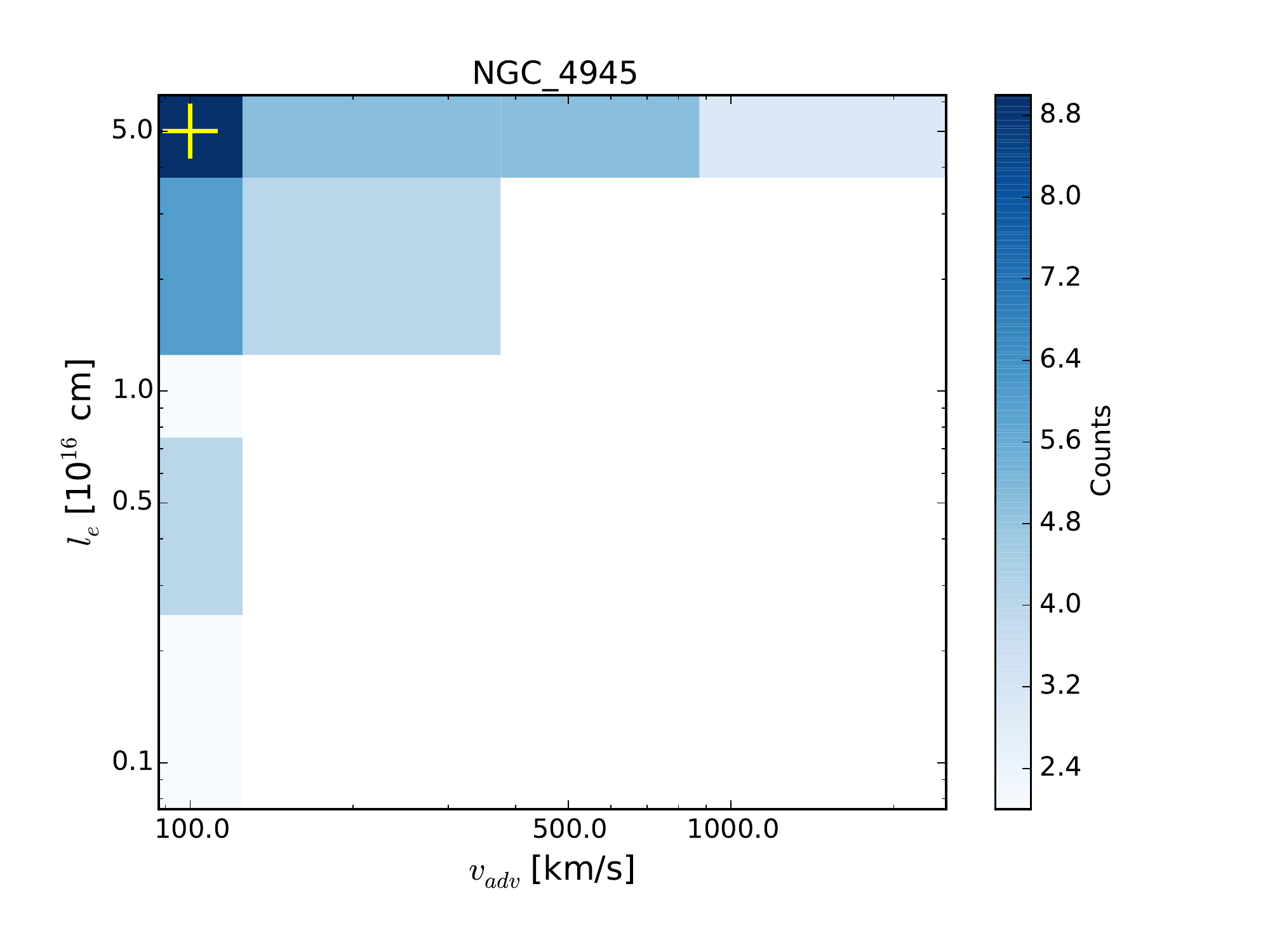}
\end{subfigure}
\caption{The distribution of $2\sigma$ models dependent on the source rate parameters (left), the cooling parameters (middle) and escape parameters (right) for NGC 253 (first line), M82 (second line) and NGC 4945 (third line). The yellow cross marks the best-fit model parameter values.}
\label{2sigmaModels}
\end{figure}\\
In doing so, fig.\ \ref{2sigmaModels} shows that apart from NGC 4945, the escape parameters are hard to constrain, whereas the source rate parameters and the cooling parameters constrain a distinct regime of accurate fit models (meaning those within $2\sigma$ of our best-fit model).
For NGC 4945 a huge range of escape parameters can be excluded, whereas for M82 we can only conclude that $v_{adv}\gtrsim 500\,\text{km/s}$ or $l_e\gtrsim 5\cdot 10^{15}\,\text{cm}$. 
An even less constraining picture is shown for NGC 253, where the distribution is almost flat with a slight tendency towards galactic wind speeds $< 1000\,\text{km/s}$. 
So the advection and diffusion timescales of NGC 253 and M82 do not necessarily deviate from each other. 
But, as the observed supernova rate of NGC 253 is slightly bigger than the supernova rate of M82, we also expect a shorter advection or diffusion timescale for NGC 253, since the gamma-ray flux of both galaxies is about the same. \\
The distribution of $2\sigma$ models dependent on the cooling parameters $B$ and $N_t$ do not show a certain maximum for any starburst, however, there is a clear linear relation between both parameters once again.
Only the distribution of $2\sigma$ models dependent on the source rate parameters indicate that $\alpha$ and $\gamma_B$ are rather well defined and close to the ones from the best-fit model. 
However, in the case of M82 and NGC 253 there are still a few accurate models for a wide range of source rate parameters.
In total, a more elaborated fit algorithm (with a higher resolution in certain parameter regimes) as well as less uncertainties within the observations are needed in order to better constrain the parameter space of the starburst galaxies. 
\\
Comparing the best-fit model parameters with previous examinations, we recognize that most of our results agree with previous ones quite well, although the starburst region of NGC 4945 is hardly observed.
The most noticeable differences of the best-fit model parameters are given by: (i) the high magnetic field strength of M82 (compared to \citealt{0004-637X-768-1-53}) and (ii) the high galactic wind speed in NGC 253 (compared to \citealt{2009A&A...494..563H}). \\
\subsubsection{Consequences on the supernova rate}
\begin{figure}[h!]
  \centering
\begin{subfigure}[b]{0.49\textwidth}
    \includegraphics[width=0.9\textwidth]{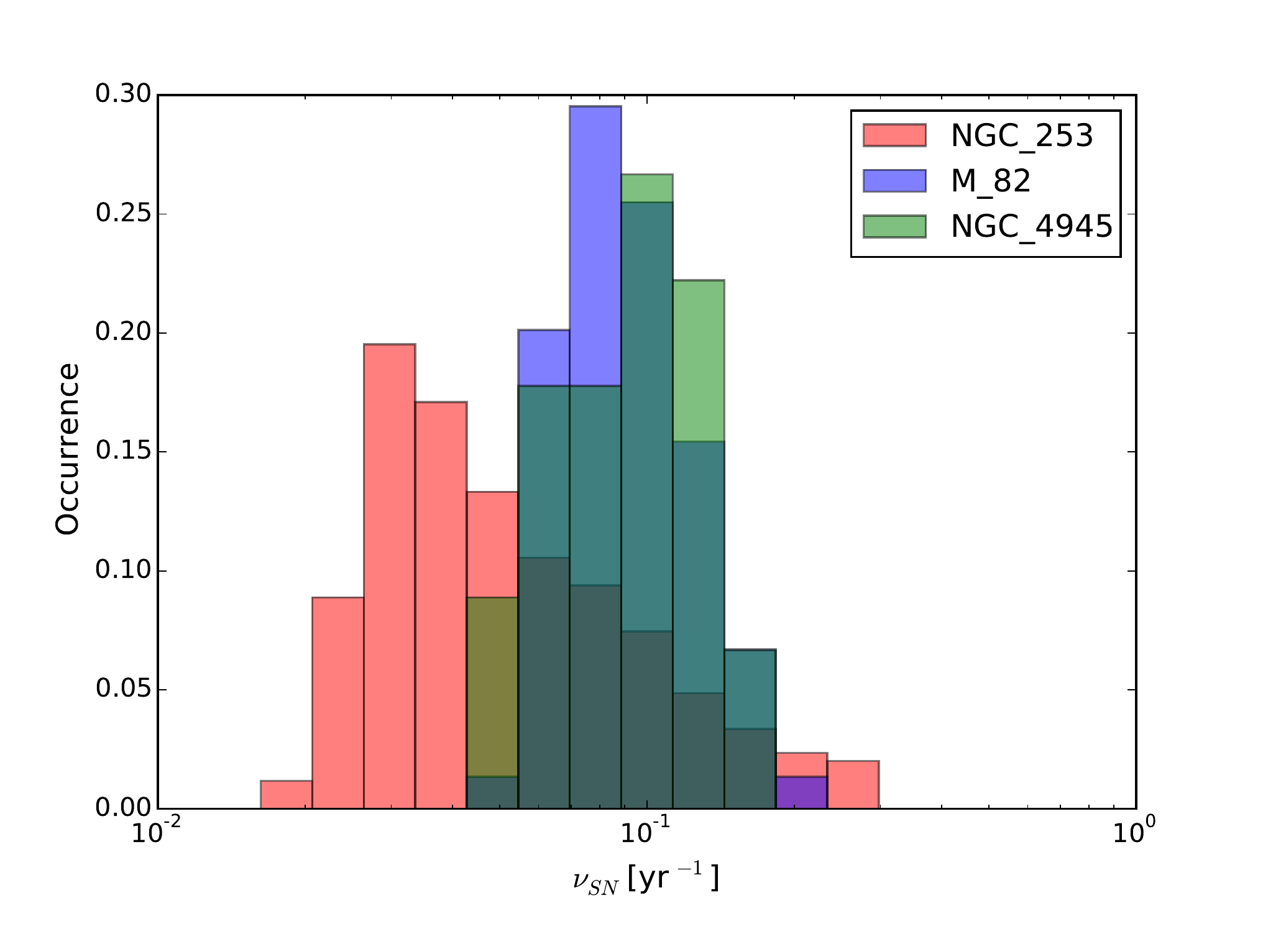}
\end{subfigure}
\begin{subfigure}[b]{0.49\textwidth}
    \includegraphics[width=0.9\textwidth]{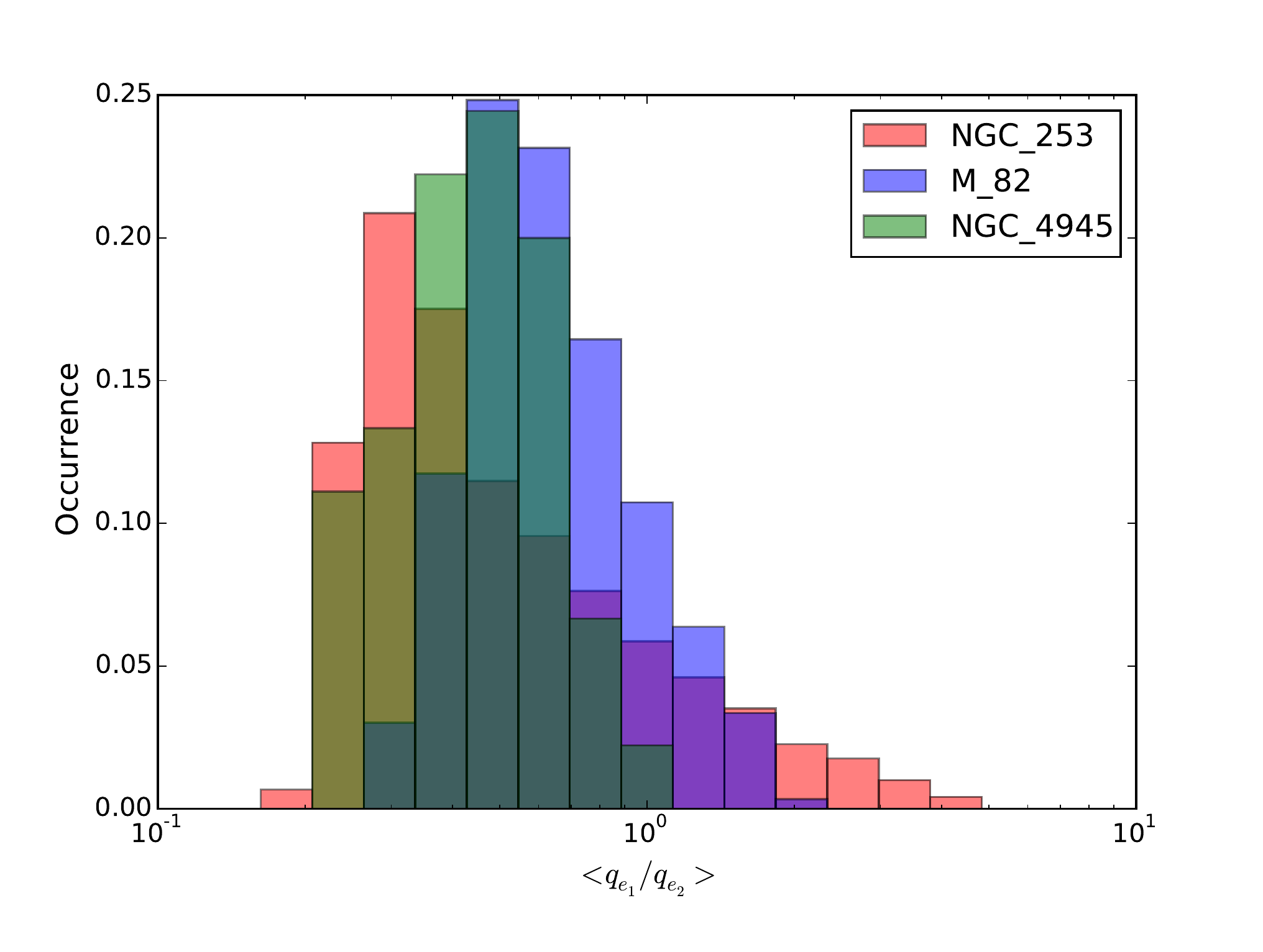}
\end{subfigure}
\caption{The Distribution of the resulting SN rate (left) and the mean of the ratio of primary to secondary electrons at $\gamma=\sqrt{\nu_{GHz}/\nu_s}$ with $\nu_{GHz}=(1-10)\,\text{GHz}$ (right) according to the $2\sigma$ models of NGC 253, M82 and NGC 4945.}
\label{SNrate_prim-sec_Distr}
\end{figure}

The fig.\ \ref{SNrate_prim-sec_Distr} displays the distribution of the obtained SN rates that refer to the $2\sigma$ models of NGC 253, M82 as well as NGC 4945 and the SN rate of the best-fit model is given in table \ref{starburst_rate_table}. 
In contrast to the best-fit model SN rates of NGC 253 most of the $2\sigma$ models of NGC 253 yield a SN rate of $\sim 0.03\,\text{yr}^{-1}$, whereas in the case of NGC 4945 and M82 the obtained SN rate from the best-fit model and the majority of $2\sigma$ models are about the same. 
\begin{table}[h!]
\centering
  \caption{Consequences of the Best-fit Model}
  \begin{tabular}{ l || c | c | c | c }
     Physical parameters & NGC 253 & M82 & NGC 4945 & NGC 1068 \\ \hline \hline
     $\nu_{SN}$ [yr$^{-1}$] & $0.11$ & $0.09$ & $0.13$ & $0.22$  \\ \hline
    $<q_{e_1}/q_{e_2}>(\gamma=\sqrt{\nu_{GHz}/\nu_s})$ & $1.0$ & $0.6$ & $0.6$ & -- \\ \hline 
    $\dot{E}_{p,loss}$ [$10^{40}\,$erg\,s$^{-1}$] & $7.5$ & $5.9$ & $3.0$ & -- \\  
    by advection [\%] & $98$ & $68$ & $44$ & -- \\ \hline
  \end{tabular}
  \begin{tablenotes}
  \item \emph{Note:} For the SN rate, the mean primary to secondary electron ratio and the total relativistic proton energy ($> 1$ GeV) leaving the starburst galaxy (as well as the proportion of advection).
  \end{tablenotes}
  \label{starburst_rate_table}
\end{table}
However, the best-fit model SN rates of all three starburst galaxies agree very well with the observed ones (see table \ref{starburst_data_table}). 
In the case of NGC 1068 we obtain a best-fit model SN rate $\nu_{SN}\gg \nu_{SN,obs}$, which gets even higher when we try to match the gamma-ray data. 
Thus, there needs to be an other relativistic proton source that dominates the acceleration of CRs within this starburst.
Here, the active galactic nucleus and the related jet-driven particle acceleration is a promising source candidate.
In the following, we will therefore focus on the starburst galaxies NGC 253, M82 and NGC 4945 and discuss some further consequences. 
\\
\subsubsection{Consequences on the primary to secondary electron ratio}
Since most synchrotron radiation is emitted around the characteristic frequency $\nu_s\,\gamma^2$, we evaluate whether the radio flux at $\nu_{GHz}=(1-10)\,\text{GHz}$ is mostly produced by primary or secondary electrons. 
Hence, the mean of the ratio $q_{e_1}/q_{e_2}$ at $\gamma=\sqrt{1\,\text{GHz}/\nu_s}$ and $\gamma=\sqrt{10\,\text{GHz}/\nu_s}$ is determined for the best-fit model parameters (see table \ref{starburst_rate_table}) as well as the $2\sigma$ model parameters (as displayed in fig.\ \ref{SNrate_prim-sec_Distr}).
Only the best-fit model for NGC 253 yields an almost equal ratio of primary and secondary electrons that generate the observed synchrotron flux at a few GHz.
However, the $2\sigma$ models as well as the best-fit models of the other starburst galaxies indicate that the number of secondary electrons is about two till three times higher than the number of primary electrons.
\subsubsection{Consequences on the particle accelerator}
An important consequence on the particle acceleration process is given by the initial steepening of the electron distribution at $\gamma_B$.
Here, the best-fit model indicates that the supposed influence of IC losses on a shock accelerated electron spectrum according to eq.\ (\ref{initial_spec-break}) is in good agreement with the observations for M82 and NGC 4945. 
In the case of NGC 253 the supposed shock velocity has to be about a magnitude smaller, i.e. $u_5=0.1$. 
However, the distribution of $2\sigma$ models has shown that an initial break in the source spectrum at $\gamma_B\gg 10^4$ can still result in an accurate fit to the data.
Thus, also for NGC 253 it is possible to obtain a good agreement with the shock acceleration approach supposing $u_5=1$.
Consequently, we approved for all three starburst galaxies that the common shock acceleration approach for a strong shock is in agreement with our model when the diffusion coefficient in the accelerator and in the starburst region hardly deviate. 
In doing so, the initial spectrum of accelerated electrons is not necessarily affected by IC (or synchrotron) losses.
\subsubsection{Consequences on the calorimetric behavior}
\begin{figure}[h!]
  \centering
\begin{subfigure}[b]{0.32\textwidth}
    \includegraphics[width=0.95\textwidth]{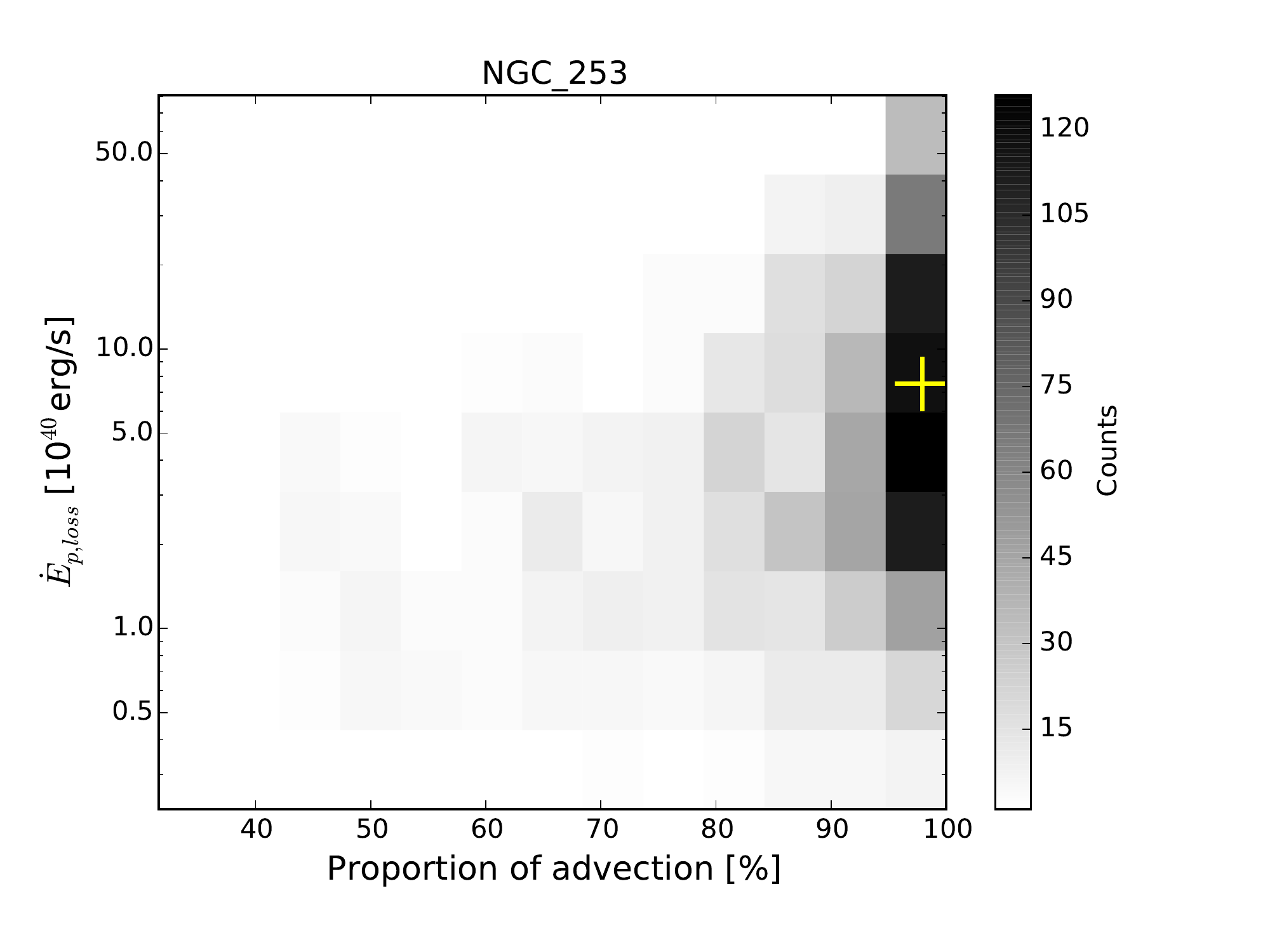}
\end{subfigure}
\begin{subfigure}[b]{0.32\textwidth}
    \includegraphics[width=0.95\textwidth]{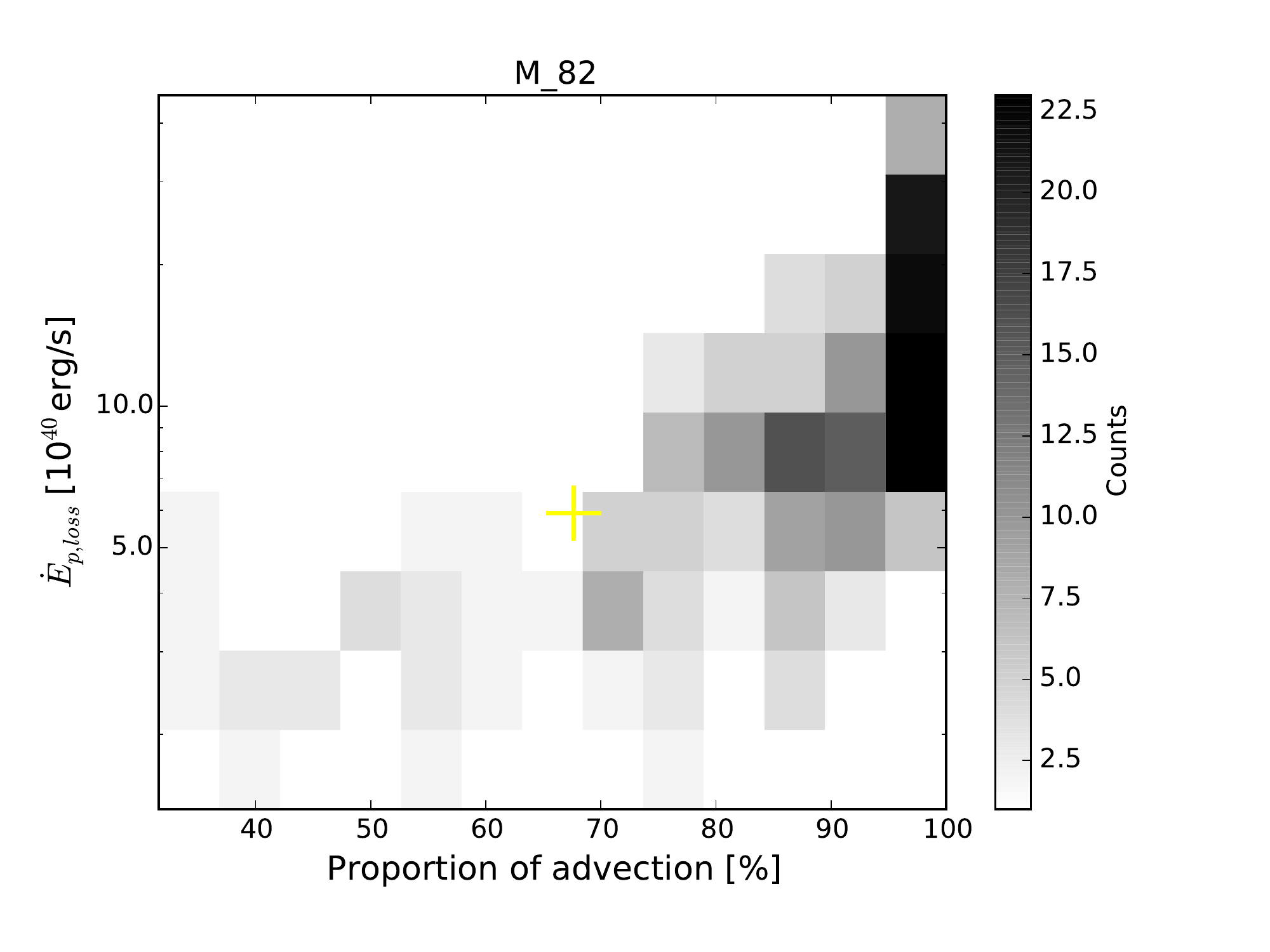}
\end{subfigure}
\begin{subfigure}[b]{0.32\textwidth}
    \includegraphics[width=0.95\textwidth]{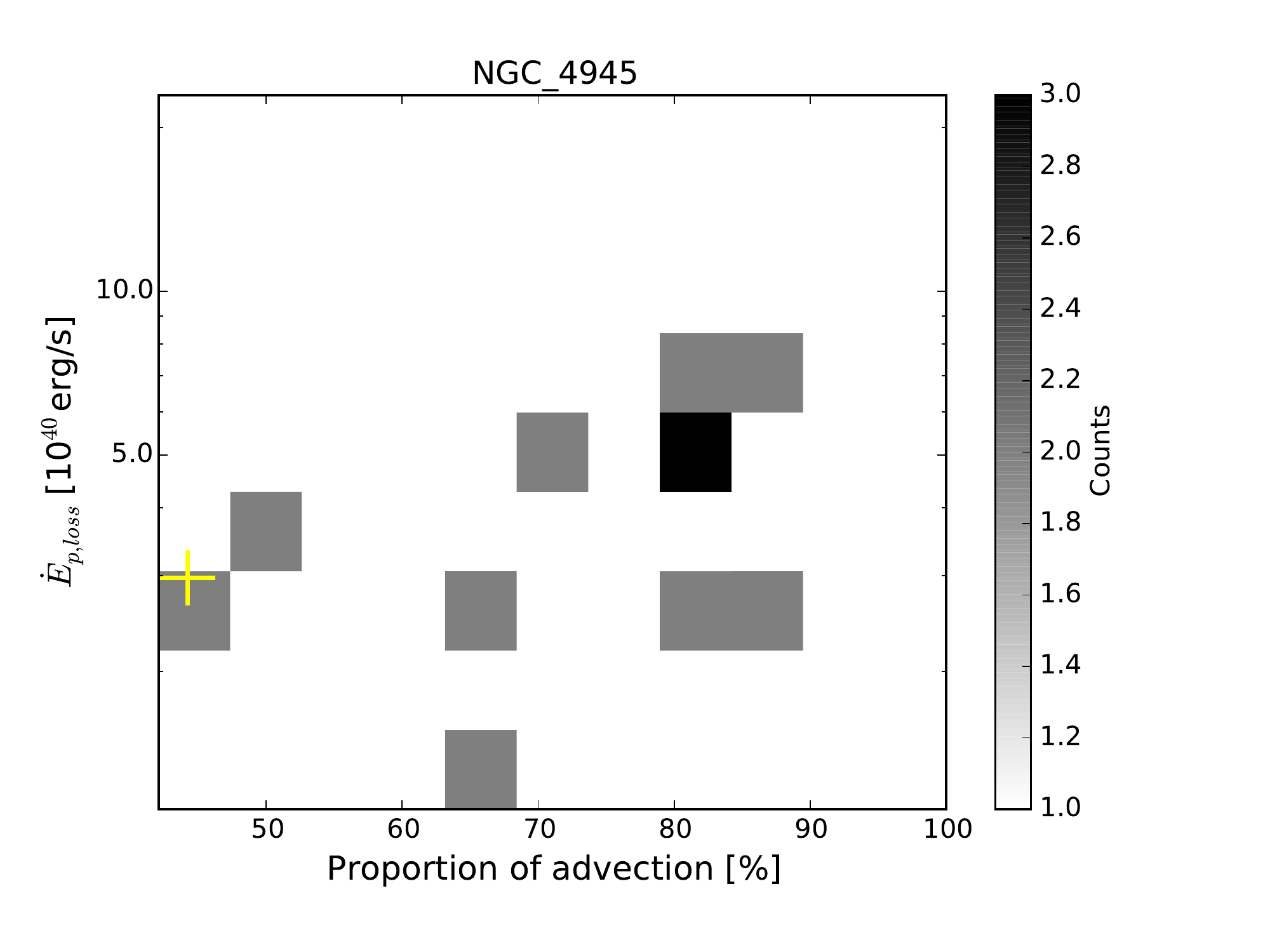}
\end{subfigure}
\caption{The distribution of $2\sigma$ models dependent on the total relativistic proton outflow and the proportion of advection for NGC 253 (left), M82 (middle) and NGC 4945 (right). The yellow cross marks the result of the best-fit model.}
\label{2sigmaModels2}
\end{figure}

Another important feature of starburst galaxies is their calorimetric behavior, i.e.\ whether or not the energy losses of CR protons or electrons dominate over advection or diffusion losses. 
However, the analysis of the $2\sigma$ models has shown that the relevant parameters $l_e$, $v_{adv}$, $B$ and $N_t$ are rather hard to constrain, especially for NGC 253 and M82.
\begin{figure}[h!]
  \centering
\begin{subfigure}[b]{0.45\textwidth}
    \includegraphics[width=1.05\textwidth]{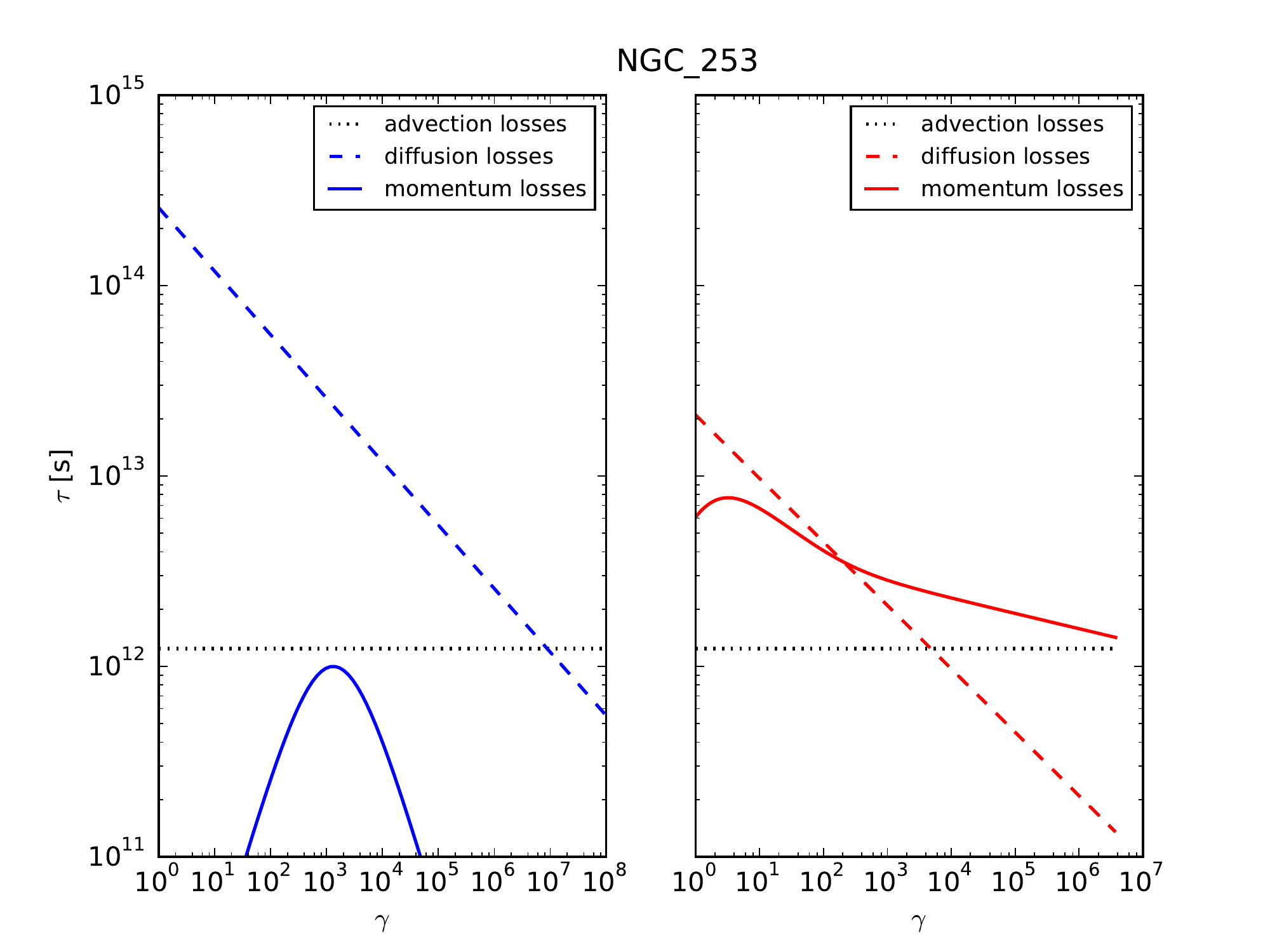}
\end{subfigure}
\begin{subfigure}[b]{0.45\textwidth}
    \includegraphics[width=1.05\textwidth]{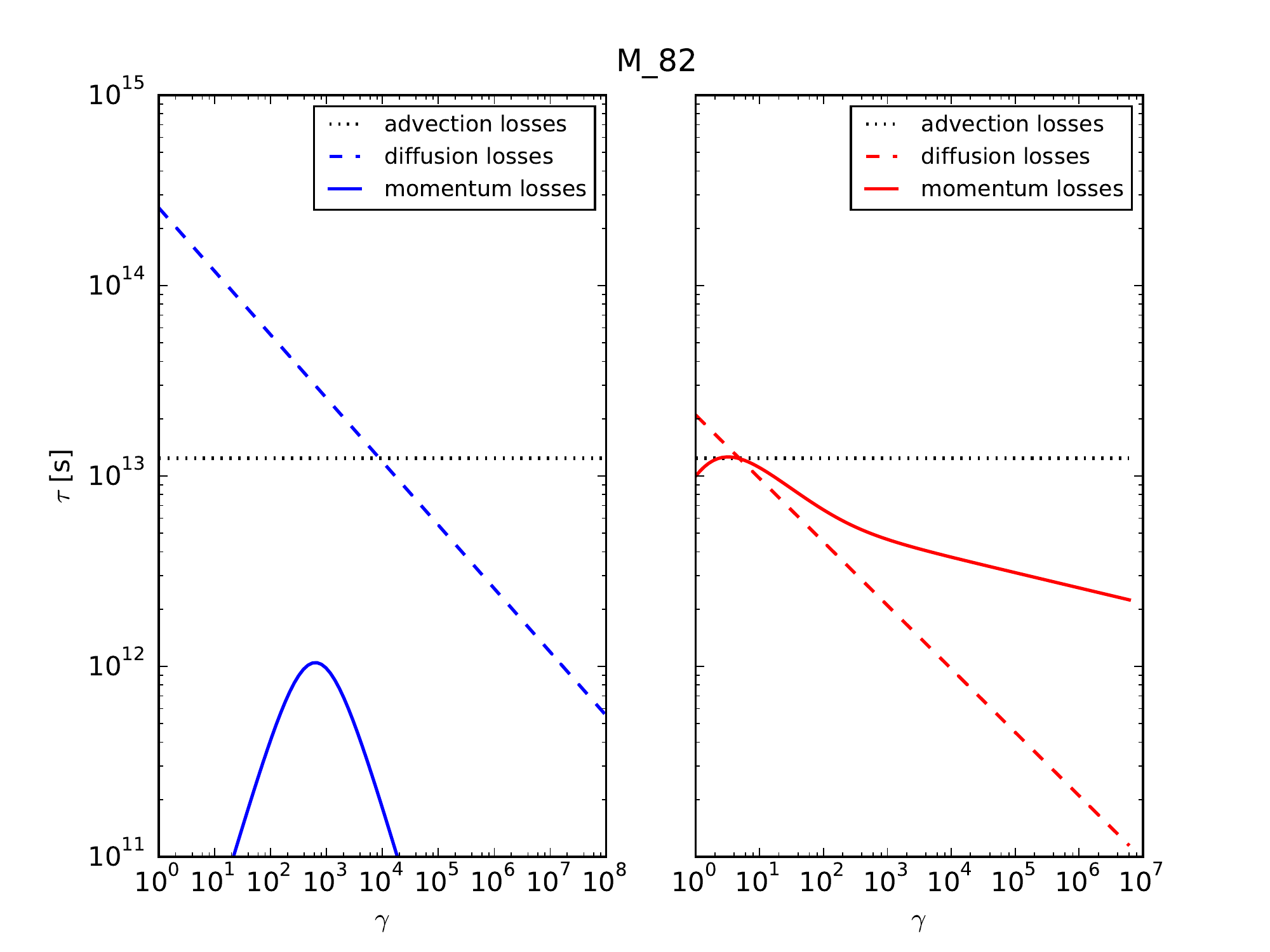}
\end{subfigure}
\begin{subfigure}[b]{0.45\textwidth}
    \includegraphics[width=1.05\textwidth]{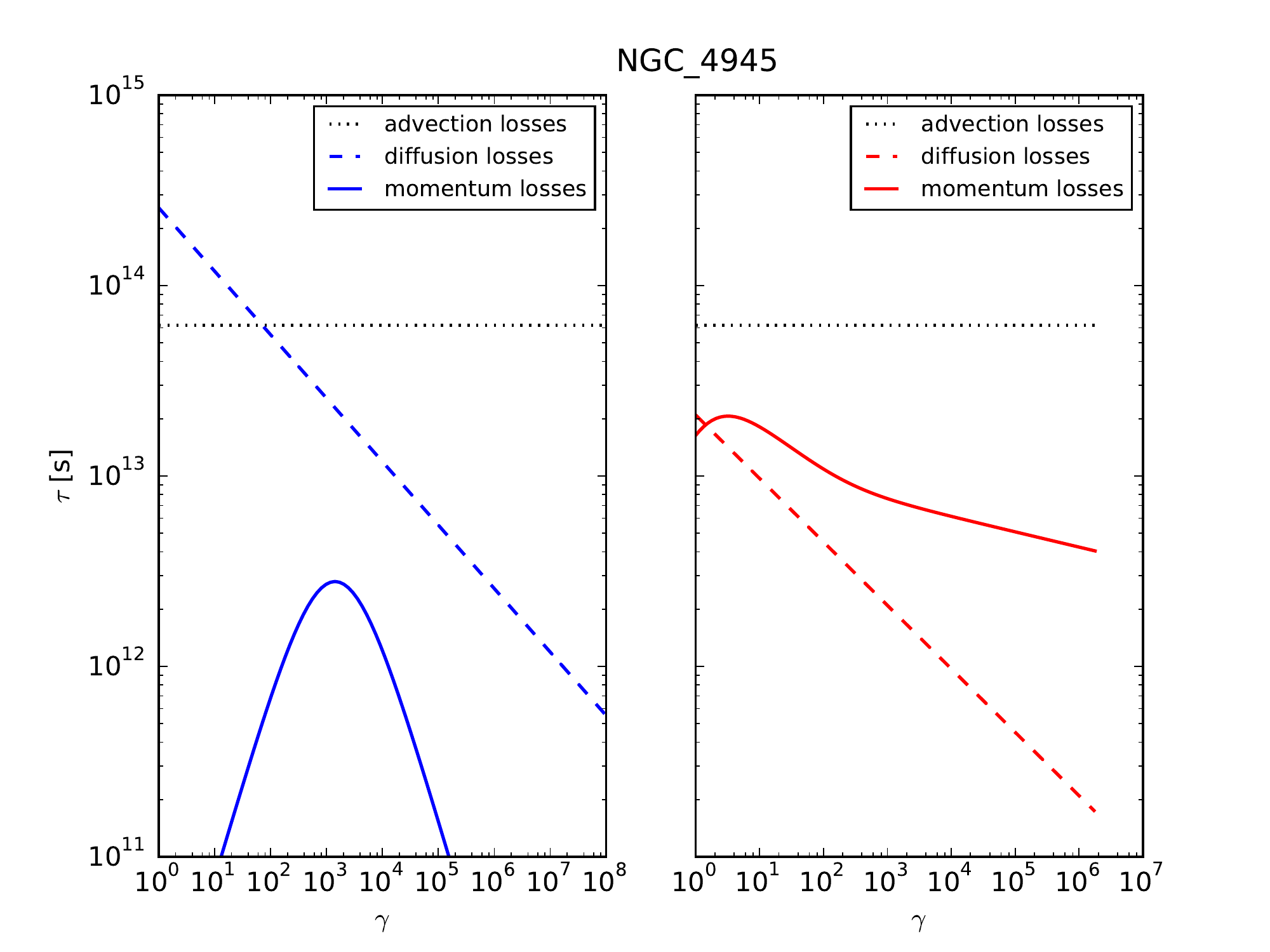}
\end{subfigure}
\begin{subfigure}[b]{0.45\textwidth}
    \includegraphics[width=1.05\textwidth]{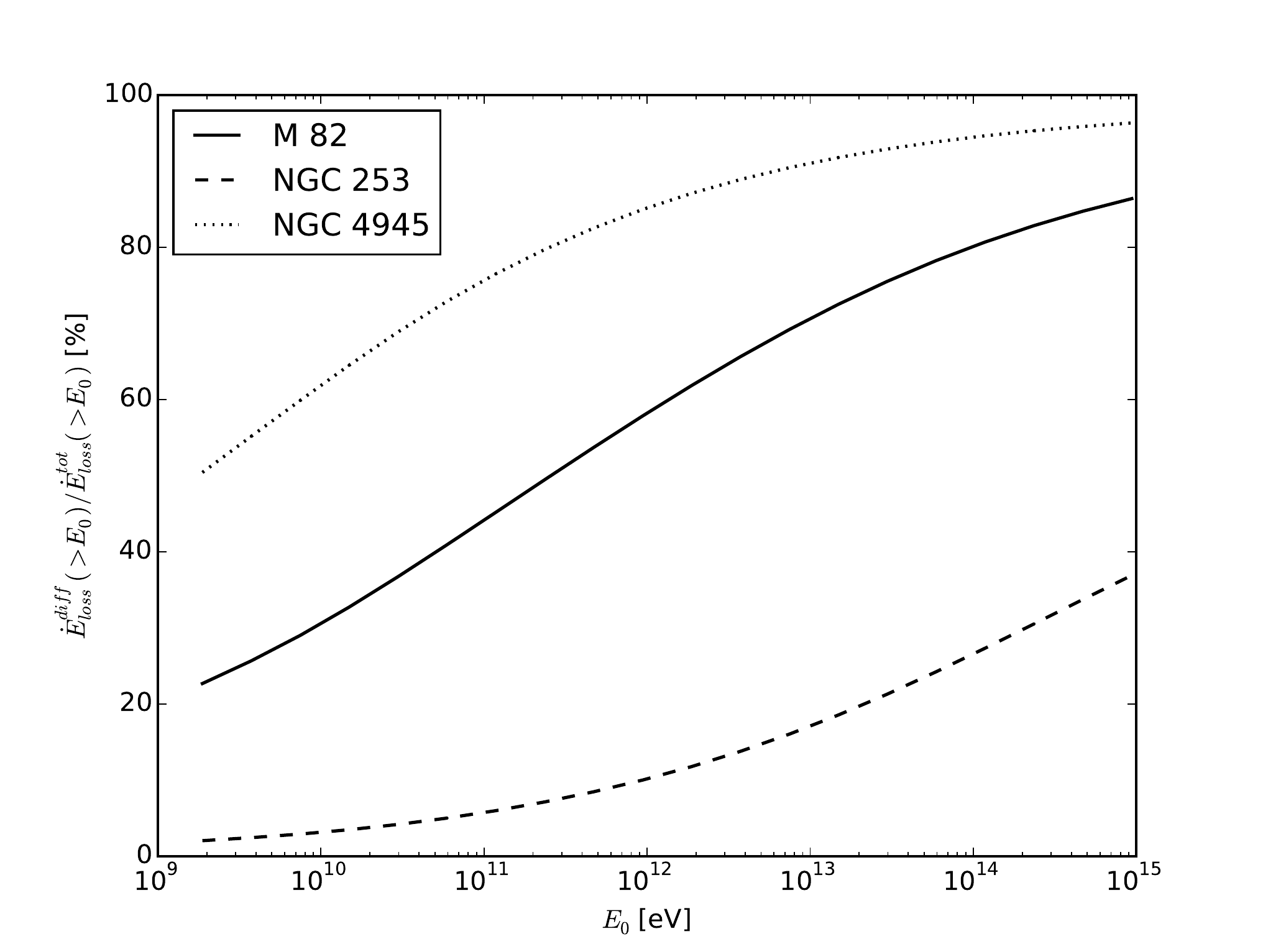}
\end{subfigure}
\caption{The resulting loss timescales of electrons (blue) and protons (red) of NGC 253, M82 and NGC 4945 that correspond to the best-fit model parameters (upper and lower left graphs). The lower right graph displays the proportion of diffusive proton losses dependent on the minimal energy of the protons.}
\label{LossTimes}
\end{figure} 
Nevertheless, we used the best-fit model parameters and fig.\ \ref{LossTimes} displays the resulting total energy loss timescale $\tau_{E}^{e,p} = \gamma/|\dot{\gamma}|_{e,p}$ as well as the catastrophic loss timescales $\tau_{adv}$ and $\tau_{diff}$ dependent on $\gamma$.
Here, some distinct characteristics of the calorimetric behavior are indicated, even though the timescales can change about an order of magnitude, when one of the $2\sigma$ model is considered instead of the best-fit model.
Hence, there is a crucial difference between the calorimetric behavior of relativistic protons and electrons. 
All considered starburst galaxies are a perfect electron calorimeter, whereas none of them is a proton calorimeter. 
Furthermore, the significance of diffusive particle losses increase with energy, so that especially for M82 and NGC 4945 diffusive particle escape is the dominant loss process at high proton energies. 
Thus, relativistic protons predominantly escape from the starburst region following the magnetic field lines (in the diffusive scenario) or the galactic wind (in the advective scenario). 
In doing so, inelastic collisions generate secondary electrons which can form a synchrotron halo as observed in the case of M82 (e.g.\ \citealt{1992A&A...256...10R}, \citealt{1994A&A...282..724R}) and NGC 253 (e.g.\ \citealt{1992ApJ...399L..59C}). 
Hence, the luminosity of the halo is not supposed to exceed the total energy outflow of the relativistic proton. 
According to the transport equation (\ref{teq0}) the total relativistic proton energy transported outwards by the galactic wind as well as the diffusion process is determined by
\be
\dot{E}_{p,loss} = {4 \over 3}\,\pi\,R^3\,m_pc^2\,\int_{\gamma_{0}}^{\gamma_{max}}\text{d}\gamma\,\, \gamma\left( {n_p(\gamma) \over \tau^p_{diff}(\gamma)} + {n_p(\gamma) \over \tau_{adv}(\gamma)}  \right)\,,
\ee
where $\gamma_0\geq \gamma_{rel}$ denotes the minimal Lorentz factor of the considered CRs. 
As already discussed, fig.\ \ref{LossTimes} exposes that the proportion of diffusive energy loss increases with the minimal Lorentz factor $\gamma_0$, so that for M82 and NGC 4945 high-energy CRs predominantly leave the starburst region by diffusion.
The total relativistic proton outflow with $E_0=1\,GeV$ as well as the corresponding proportion of advection is shown in table \ref{starburst_rate_table} for the best-fit model and in fig.\ \ref{2sigmaModels2} for the $2\sigma$ models.
\\
Here, the total energy loss rate of the best-fit model of M82 is close to the prediction from \citealt{1991ApJ...369..320S} of about $10^{40}$ erg/s due to the non-thermal radio halo. 
However, the distribution of $2\sigma$ models indicate that the majority of the parameter models yield a slightly higher total energy outflow as well as a higher proportion of advection. 
Since the proton outflow at low energies is mostly dominated by advection, it is not surprising that $\dot{E}_{p,loss}$ decreases with a decreasing influence of advection on the particle escape. 
\subsubsection{Consequences on the neutrino flux}
\begin{figure}[h!]
  \centering
\begin{subfigure}[b]{0.32\textwidth}
    \includegraphics[width=0.98\textwidth]{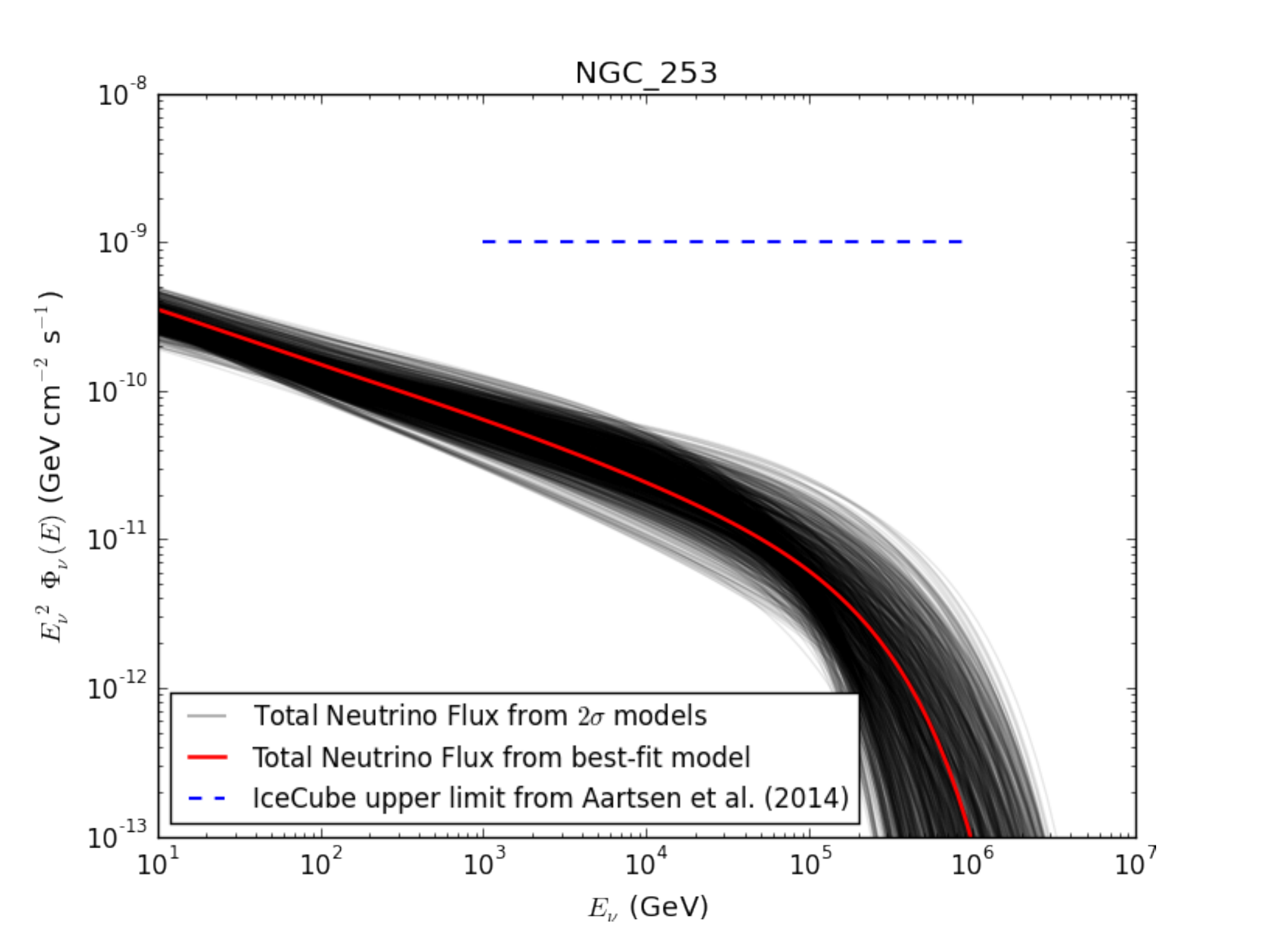}
\end{subfigure}
\begin{subfigure}[b]{0.32\textwidth}
    \includegraphics[width=0.98\textwidth]{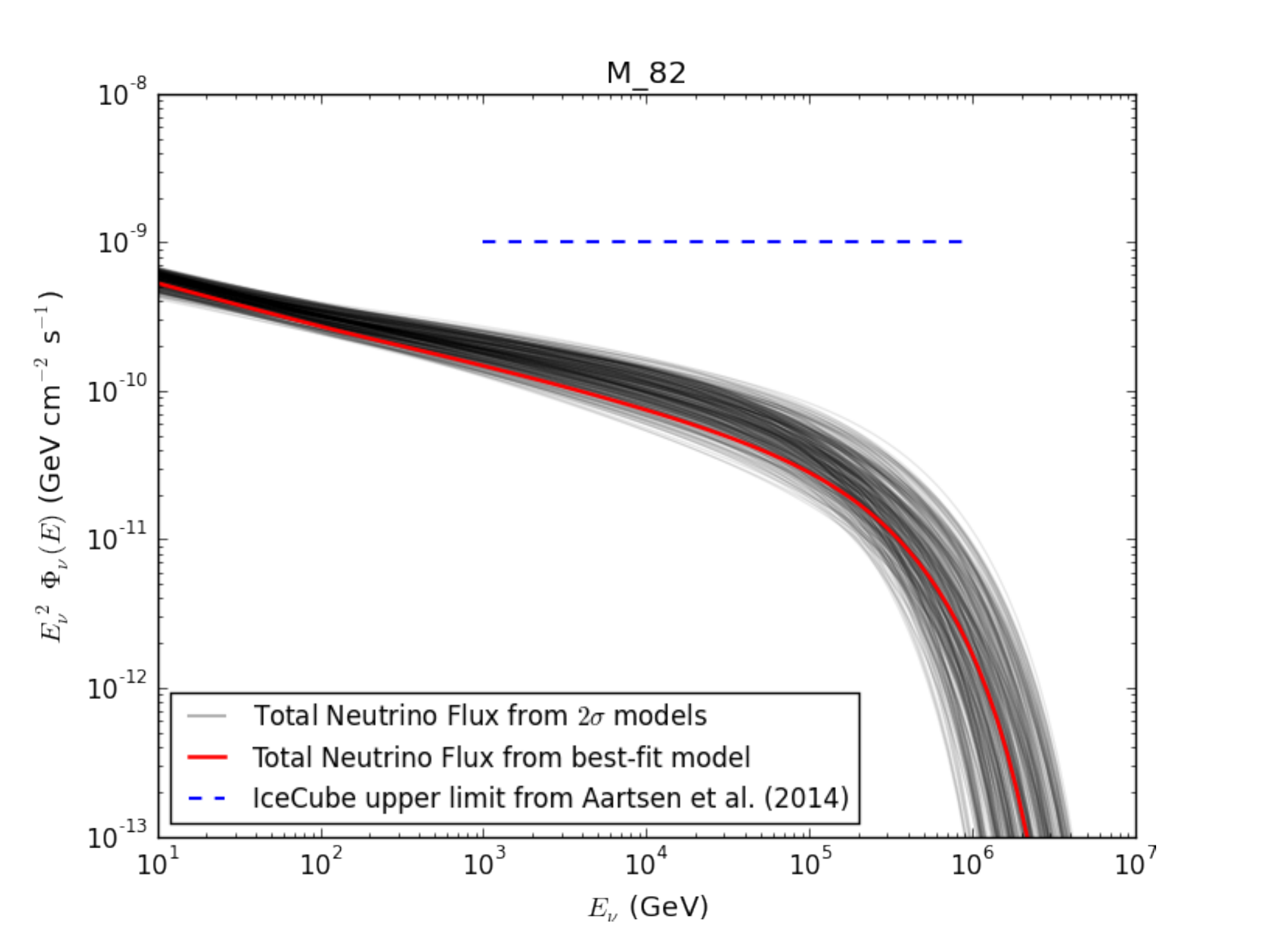}
\end{subfigure}
\begin{subfigure}[b]{0.32\textwidth}
    \includegraphics[width=0.98\textwidth]{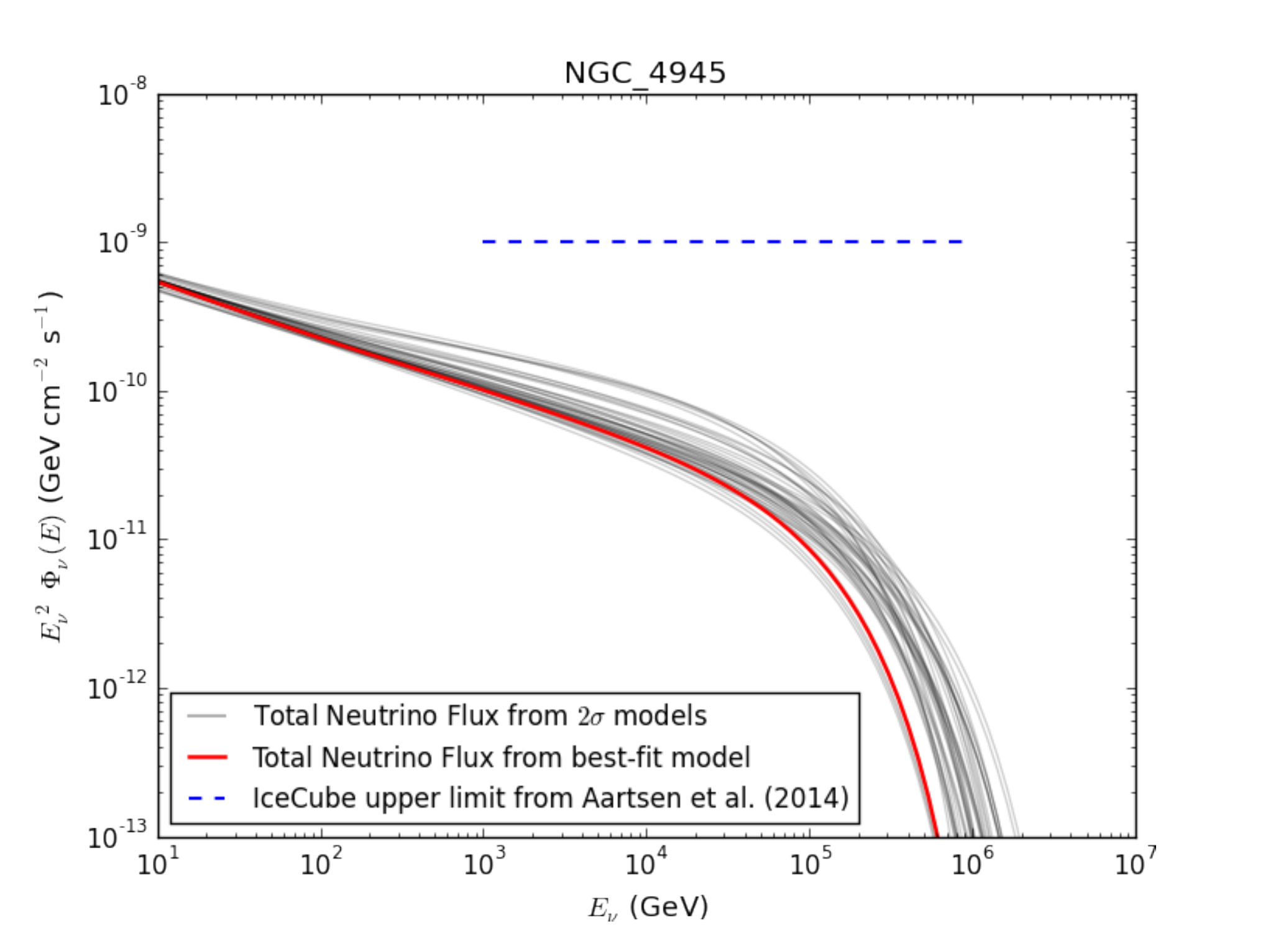}
\end{subfigure}
\caption{The differential neutrino flux of NGC 253, M82 and NGC 4945 that results from the parameters of the best-fit model (red line) and the $2\sigma$ models (grey lines). The blue dashed lines represent IceCube limits for the specific source from \citealt{2014ApJ...796..109A}.}
\label{neutrino-flux_1-4}
\end{figure}

A neutrino signal would be a clear indication of hadronic interactions within the starburst, however, no neutrino flux from an astrophysical source (apart from the Sun and the supernova SN1987A) has been observed so far.
Therefore, it is useful to determine the expected neutrino flux of the considered starburst galaxies based on the observed gamma-ray flux.
In doing so, the best-fit model as well as the $2\sigma$ models are used to determine the resulting neutrino flux.
Fig.\ \ref{neutrino-flux_1-4} shows that NGC 253, M82 and NGC 4945 are potential sources for high energy neutrinos.
However, the neutrino flux of all starburst galaxies is below the current observation limit of the \emph{IceCube} neutrino detector.
\\
\section{Conclusions}
The present paper examines the radiation processes in starburst galaxies. 
In doing so, a steady state description of the relativistic electrons and protons within the spherically symmetric inner hundreds of parsecs of the galaxy is used. 
We suppose homogeneously distributed sources of the CRs that generate a power-law distribution between a minimal and a maximal momentum.
Due to the strong cooling of the ultrarelativistic electrons (via IC or synchrotron losses) the electron source rate can steepen at a Lorentz factor $\gamma_B$ by one.
A general solution of the relativistic particle distribution is given and subsequently specified by the particle dependent source rate, cooling mechanism and diffusion loss rate. 
Here, the electron to proton source rate ratio $q_0^e/q_0^p$ is determined by the quasi-neutrality of the plasma, so that $q_0^e/q_0^p$ increases with decreasing $\gamma_B$, however, only for $\gamma_B\ll 10^2$ can the ratio become much bigger than the commonly used value of $\left(  m_p/ m_e  \right)^{(\alpha-1)/2}$.
\\ 
Using the primary electrons from the accelerator as well as the secondary electrons from hadronic pion production, we determine the resulting synchrotron radiation as well as the gamma radiation by Bremsstrahlung and IC collisions of the relativistic electrons.
However, the main constituent of the gamma ray flux of starburst galaxies is given by the gamma radiation from hadronic pion production of the relativistic protons as shown in Section \ref{sec:cons_to_obs}.
\\
Finally, we tested $337500$ $(= 15\times 15\times 10 \times 5 \times 5 \times 6)$ different combinations within a reasonable parameter range of $(B,\,N_t,\,\alpha,\,\gamma_B,\,l_e,\,v_{adv})$ in order to obtain the best-fit model parameters of four different starburst galaxies which are summarized in table \ref{starburst_bestfit_table}. 
As displayed in fig.\ \ref{AllFits}, the radio and gamma-ray flux of our best-fit model is in good agreement with the observations of NGC 4945 and especially M82 and NGC 253.
Only the data of NGC 1068 can hardly be described by our model, as a much higher source rate is needed to explain the gamma-ray data, however, this also leads to an increasing SN rate which become about an order of magnitude higher than observed. 
Therefore, we suggest that other/ additional particle accelerators (e.g.\ the active galactic nucleus) need to be at work, as already exposed by \citealt{2014ApJ...780..137Y}.
In order to determine the physical processes in NGC 1068 in more detail further examinations are needed that account for spatial inhomogeneities and the AGN nature within these objects as well as a bigger data set with less uncertainties.
Here, \citealt{2010A&A...524A..72L} have already shown that a leptonic model for the jet outflow of the AGN is able to describe the emission spectrum of NGC 1068 accurately. \\
In the case of NGC 253, M82 and NGC 4945 we obtain a good agreement with the observed SN rates. 
Here, the secondary electrons from hadronic pion production contribute about $50^{+30}_{-10}\,\%$ in NGC 253 and about $75^{+5}_{-25}\,\%$ in M82 and NGC 4945 to the observed radio flux via synchrotron radiation.
All starburst galaxies are potential high energy neutrino sources, but the expected neutrino flux is about $1-2$ orders of magnitude below the current observation limit of the IceCube detector (see fig.\ \ref{neutrino-flux_1-4}).  
\\
Furthermore, it has been shown that there is no mandatory need for a steepening of the initial electron source spectrum at $\gamma_B$ and the obtained parameters are in agreement with the common shock acceleration approach by supernovae. 
In addition, we reveal that the four starburst galaxies are perfect electron calorimeters but poor proton calorimeters which agrees with the conclusions by \citealt{2013ApJ...762...29L} and \citealt{0004-637X-768-1-53}. 
The total energy loss rate of the relativistic protons is dominated by the galactic wind, however, especially in the case of M82 and NGC 4945 the high energy protons ($>1\,\text{TeV}$) predominantly escape via diffusion (see fig.\ \ref{LossTimes} and \ref{2sigmaModels2}).
In doing so, the total energy loss rate of M82 is in agreement with the predictions from \citealt{1991ApJ...369..320S} with respect to the observed non-thermal radio halo. 
Thus, the observed synchrotron halo needs to be driven by the escaping relativistic protons that generate secondary electrons by interactions with its environment. 
%
%
%
%
%
%
%
%
%
%
%
%
%
%
%
%
%
%
%
%
%
%
%
%
%
%
%
%
%
%
%
%
%
%
%
%
%
%
%
%
%
%
%
%
%
%
%
%
%
%
%
%
%
%



\acknowledgments

We are grateful to Dominik Els\"asser, Reinhard Schlickeiser, Ralf-J\"urgen Dettmar, Sebastian Sch\"oneberg, Florian Schuppan and Lukas Merten for very useful comments that helped to improve the original version of the paper.
The authors especially thank the referee for the vital suggestions that greatly strengthened this work.
We acknowledge support from the DFG Research group FOR 1048 on ``Instabilities, Turbulence and Transport in Cosmic Magnetic Fields'' (project BE 3714/5-1) as well as from the MERCUR project St-2014-0040 (RAPP center) and the research department of plasmas with complex interactions (Bochum).





\appendix
%
%
%
%
%
%
%
%
%
%
%
%
%
%
%
%
\section{General solution of the steady state particle transport equation}
\label{app:pte_sol}
In order to solve the transport equation (\ref{teq0}) we substitute
\be
n_{e,p}(\gamma) = \tilde{n}_{e,p}(\gamma)\,\exp\left(  \int {\text{d}\gamma \over |\dot{\gamma}|_{e,p}\tau_{adv}}  \right)
\label{density-trafo}
\ee
so that
\begin{equation}
0=\partialgamma\left(|\dot{\gamma}|_{e,p}\,\tilde{n}_{e,p}\right)-{\tilde{n}_{e,p}\over \tau^{e,p}_{diff}(\gamma)}+\tilde{q}_{e,p}(\gamma)\,,
\label{teq}
\end{equation}
with
\be
\tilde{q}_{e,p}(\gamma) = q_{e,p}(\gamma)\,\exp\left( - \int {\text{d}\gamma \over |\dot{\gamma}|_{e,p}\tau_{adv}}  \right)\,.
\ee
For non-vanishing continuous energy losses, the transport Eq.\ (\ref{teq}) can easily be rearranged to
\be
0={\partial \tilde{n}_{e,p} \over \partial\gamma}+\left({1\over |\dot{\gamma}|_{e,p}}\,{\partial |\dot{\gamma}|_{e,p}\over \partial \gamma}-{1\over \tau_{diff}(\gamma)\,|\dot{\gamma}|_{e,p}}\right)\,\tilde{n}_{e,p}(\gamma)+{\tilde{q}_{e,p}(\gamma)\over |\dot{\gamma'}|_{e,p}} \,.
\ee
Using the method of variation of a constant the previous equation is solved by
\be
\tilde{n}_{e,p}(\gamma)=n^{e,p}_0\,\exp(-\zeta^{e,p}_0(\gamma))\,\left(1-{1\over n^{e,p}_0}\zeta^{e,p}_1(\gamma)\right)\,,
\label{gen_sol_part_dens0}
\ee 
where
\be
\zeta^{e,p}_0(\gamma)=\int d\gamma\,\left( {1\over |\dot{\gamma}|_{e,p}}\,{\partial |\dot{\gamma}|_{e,p}\over \partial \gamma}-{1\over \tau^{e,p}_{diff}(\gamma)\,|\dot{\gamma}|_{e,p}}\right)
\ee
and
\be
\zeta_1(\gamma)=\int_{\gamma_0}^\gamma d\gamma'\,\,{\tilde{q}_{e,p}(\gamma') \over  |\dot{\gamma'}|_{e,p}}\,\exp(\zeta^{e,p}_0(\gamma'))\,.
\label{zeta1}
\ee
With eq.\ (\ref{density-trafo}) the relativistic particle density is generally given by
\be
n_{e,p}(\gamma)=n^{e,p}_0\,\exp\left(  \int {\text{d}\gamma \over |\dot{\gamma}|_{e,p}\tau_{adv}}  \right)\exp(-\zeta^{e,p}_0(\gamma))\,\left(1-{1\over n^{e,p}_0}\zeta^{e,p}_1(\gamma)\right)\,,
\label{gen_sol_part_dens}
\ee 
In order to determine the constants $n^{e,p}_0$ and $\gamma_0$ we need appropriate boundary conditions.
The first condition (i) demands a vanishing particle density when there is no source rate, i.e. $n_{e,p}(\gamma)=0$ for $q_{e,p}(\gamma)=0$\,.
Consequently, $n_{0}^{e,p}=0$ so that the solution (\ref{gen_sol_part_dens}) simplifies to
\be
n_{e,p}(\gamma)=-\exp\left(  \int {\text{d}\gamma \over |\dot{\gamma}|_{e,p}\tau_{adv}}  \right)\exp(-\zeta^{e,p}_0(\gamma))\,\zeta^{e,p}_1(\gamma)\,.
\label{gen_sol_part_dens_i}
\ee
The second condition (ii) demands that there is a maximum particle Lorentz factor given by $q_{e,p}(\gamma)=f(\gamma)\,H[\gamma_{max}-\gamma]$.
Thus, $n_{e,p}(\gamma\geq \gamma_{max})=0$ yields that $\gamma_0=\gamma_{max}$, so that the differential particle density (\ref{gen_sol_part_dens_i}) yields
\be
n_{e,p}(\gamma)=\exp\left(  \int {\text{d}\gamma \over |\dot{\gamma}|_{e,p}\tau_{adv}}  \right)\text{e}^{-\zeta^{e,p}_0(\gamma)}\,\int_{\gamma}^{\gamma_{max}} d\gamma'\,\,{q_{e,p}(\gamma') \over  |\dot{\gamma'}|_{e,p}}\,\text{e}^{\zeta^{e,p}_0(\gamma')}\,\exp\left( - \int {\text{d}\gamma' \over |\dot{\gamma'}|_{e,p}\tau_{adv}}  \right)\,.
\label{gen_sol_part_dens_ii_app}
\ee

\end{document}